\documentclass[aps,onecolumn,superscriptaddress,showpacs]{revtex4}
\usepackage{amssymb,epsf}
\usepackage{amsmath}
\usepackage[dvips]{graphicx}
\usepackage[dvips]{color}

\def\iu{i_U}

\def\dualu{{\dual{U}}}
\def\Real{{\mathbb R}}

\def\man{{M}}
\def\Set#1{{\left\{#1\right\}}}
\def\dual#1{{\widetilde{#1}}}
\def\GamTM{{\Gamma T\man}}
\def\GamLamM#1{{\Gamma \Lambda^{#1}\man}}
\def\qquadand{\qquad\text{and}\qquad}

\newcommand\BE[1]{{\begin{equation}#1\end{equation}}}
\newcommand\BAE[1]{{\begin{equation}{\begin{aligned}#1\end{aligned}}\end{equation}}}

\def\iuF{\iu F}
\def\iustarF{\iu{{\star}}F}

\newcommand{\bfe}{ \TYPE 1    {\mathbf e}  {}   }
\newcommand{\bfb}{ \TYPE 1 {\mathbf b}   {}}
\newcommand{\bfd}{ \TYPE 1 {\mathbf d}   {} }
\newcommand{\bfh}{\TYPE 1  {\mathbf h}   {} }

\newcommand{\bfE}{ \TYPE 2    {\mathbf E}  {}   }
\newcommand{\bfB}{ \TYPE 2 {\mathbf B}   {}}
\newcommand{\bfD}{ \TYPE 2 {\mathbf D}   {} }
\newcommand{\bfH}{\TYPE 2  {\mathbf H}   {} }
\newcommand{\bfEdot}{ \TYPE 2    {\dot{\mathbf E}}  {}   }
\newcommand{\bfBdot}{ \TYPE 2 {\dot{\mathbf B}}   {}}
\newcommand{\bfDdot}{ \TYPE 2 {\dot{\mathbf D}}   {} }
\newcommand{\bfHdot}{\TYPE 2  {\dot{\mathbf H}}   {} }
\def\Me{{\mathbf e}}
\def\Mb{{\mathbf b}}
\def\Md{{\mathbf d}}
\def\Mh{{\mathbf h}}
\newcommand\TYPE[3]{ \underset {(#1)}{\overset{{#3}}{#2}}  }

\newcommand\EP{\epsilon_r\epsilon_0\,}
\newcommand\MU{\mu_r\mu_0\,}
\newcommand\ep{\varepsilon}
\newcommand{\hash}{\#}
\newcommand{\hashat}{\hat{\#}}
\newcommand{\xibar}{\underline\xi}
\newcommand{\D}{{\boldsymbol d}\,}

\newcommand{\ot}{\otimes}
\newcommand\PD[1]{\frac{\partial}{\partial #1}}

\newcommand\PSI[2]{\Psi_{#1}{#2}}
\newcommand\PHI[2]{\Phi_{#1}{#2}}

\newcommand{\DD}{{\cal D}}

\newcommand\NN[1]{{\cal N}_{#1}}
\newcommand\MM[1]{{\cal M}_{#1}}

\newcommand\union\cup

\newcommand\EPS{{\mu { Y}^2}}

\newcommand{\EM}{ electromagnetic }
\newcommand{\beq}{\begin{equation}}
\newcommand{\beqa}{\begin{eqnarray}}
\newcommand{\eeq}{\end{equation}}
\newcommand{\eeqa}{\end{eqnarray}}
\newcommand{\non}{\nonumber}
\newcommand{\fr}[1]{(\ref{#1})}

\newcommand{\ts}{\widetilde s}

\begin{document}

\pagestyle{myheadings} \markboth{Shin-itiro GOTO and Robin W Tucker}
{Electromagnetic Fields in a Curved Beam Pipe}

\title{\bfseries Electromagnetic Fields Produced by Moving Sources in a Curved Beam Pipe}

\author{Shin-itiro Goto}
\affiliation{  Department of Physics, Lancaster University 
and the Cockcroft Institute, Daresbury}
\author{Robin  W Tucker}
\affiliation{  Department of Physics, Lancaster University 
and the Cockcroft Institute, Daresbury}

\date{\today}






\begin{abstract}
\noindent 
A new geometrical perturbation scheme is developed
in order to calculate the electromagnetic fields produced by
charged sources in prescribed  motion moving in a
non-straight perfectly conducting beam pipe. The pipe
is  regarded as a perturbed infinitely long hollow
right-circular cylinder. The perturbation  maintains the pipe's
circular cross-section while deforming its axis into a planar
space-curve with, in general, non-constant curvature.
Various charged source models are considered  including a charged
bunch and an off-axis
point particle. In the ultra-relativistic limit this permits a calculation of the longitudinal wake
potential in terms of powers of the product of the pipe radius and the
 arbitrarily varying curvature of the axial space-curve.
 Analytic expressions to leading order are presented for beam pipes with piecewise
  defined constant curvature modelling pipes with straight segments linked by
  circular arcs of finite length.
The language of differential forms is used
throughout and to illustrate the power of this formalism a
pedagogical introduction is developed by deriving the theory
ab-initio from Maxwell's equations expressed intrinsically as a
differential system on (Minkowski) spacetime.
\end{abstract}
%
%
%
%
%
%
%
\maketitle
\section{Introduction}
Considerable activity is being devoted to the design of advanced
machines that can produce pulsed sources of intense focussed \EM
radiation. Such sources offer unprecedented opportunities for
probing the temporal and spatial microstructure of processes in
Nature. Many designs rely on being able to control the motion of
ultra-relativistic electron beams by external fields in beam pipes
with non-uniform spatial curvature. The production of femtosecond
radiation pulses requires high peak electric currents and the
maintenance of low emittance electron beams for the self-amplified
spontaneous emission of X-rays demands extreme design criteria in
order to sustain beam stability in the presence of radiation
back-reaction on accelerated sources.

A direct analytic approach to this electrodynamic problem via the
coupled system of Maxwell's field equations and the equations of
motion for the particle beams encounters difficult problems due to
non-linearities and retardation effects
\cite{GBR07,GBR06-proceedings,BGR06-asymptotic}.
A direct approach using
statistical methods suffers from similar complications. Numerical
approximations based on these equations exist but few are able to
address the full complexities encountered in a realistic situation.
In particular even when non-linearities are deemed
negligible,
 the effects of (conducting) boundaries on the
accelerating source via the radiation fields are often ignored in an
attempt to build tractable models and the effects of pipe curvature
are often restricted to those produced by motion in arcs of
circles \cite{saldin}.
Although insight can be gained from such modelling we feel that much
remains to be understood in more general scenarios.

In this paper we explore a new geometrical perturbation scheme that
addresses some aspects of the general electrodynamical problem of
charged sources in prescribed arbitrary motion moving in a
non-straight beam pipe. In the analysis  below the pipe
will be regarded as a perturbed infinitely long hollow
right-circular cylinder. The perturbation will maintain the pipe's
circular cross-section while deforming its axis into a planar
space-curve with, in general, non-constant curvature. It is assumed
that the curvature $\kappa(z)$  depends  on the arc-length $\vert
z\vert $ of the planar space-curve and tends to zero as $z \to
\pm\infty$. Furthermore if the cylinder has fixed radius $a$ we
require that $\vert \kappa(z) a\vert \ll 1$ for the perturbation
analysis to be effective. Under these conditions the
initial-boundary  value problem for the Maxwell fields in a
perfectly conducting hollow perturbed cylinder given prescribed
currents will be considered. The source currents will excite
superpositions of \EM modes of the empty perturbed cylinder as well
as generating their own \lq\lq self-fields\rq\rq.  The latter will
include {\it acceleration-fields} induced by the motion of the
sources in the curved regions of the beam pipe. A perturbation
scheme will be established to calculate all fields in the pipe as a
perturbation expansion in powers of $a\kappa$. Mode expansions based
on Dirichelet and Neumann eigen-functions of the Laplacian for a
circular disc domain can be used to reduce the general problem at
each order to  a two-dimensional linear telegraph-type equation with
prescribed sources. The general causal solution of this  equation is known
and from it the complete set of \EM fields can be constructed. Thus
the perturbed beam pipe impedances can be found. Furthermore the
fields arise naturally in a space-time description and thereby offer
direct input into \lq\lq leap-frog\rq\rq coding schemes that couple
the Maxwell sector to the equations of motion of the sources.

In section 2 the notation used in the paper is established and
illustrated by means of a pedagogic review of exterior methods used
for solving Maxwell boundary value problems in sections 3 and 4.
Section 5 deals with the introduction of Dirichelet and
Neumann modes used  to reduce the Maxwell system with perfectly
conducting boundaries to  a telegraph type equation in two
dimensions. Section 6 uses a geometric perturbation  approach
to explore the dependence of the radiated power
from a smooth longitudinal convective  current on
local beam-pipe curvature. Section 7 deals with a moving
point charge source and discusses in some detail
radiant instantaneous power.
This permits a perturbative calculation of
the ultra-relativistic wake potentials in a planar pipe
with arbitrary curvature and explicit analytic expressions can
be found in the case where the curvature is piecewise constant
modelling pipes with straight segments linked by
circular arcs of finite length. Finally section 8 considers
the radiation from smooth convected localised bunches
with fixed total charge while the Appendices tabulate coupling
coefficients and transfer kernels used in the main text.

\section{Notation}
The exterior
calculus of differential forms offers a versatile and powerful
means for analysing Maxwell's equations \cite{des,flanders}.
The notation used below
follows standard conventions for a manifold $\man$ with a metric
tensor field. Thus $\GamTM$ denotes the set of vector fields and
$\GamLamM{p}$ the set of $p$-form fields on $\man$. Metric duals
with respect to any metric tensor $g$ are written with a tilde so
that $\dual{X}=g(X,-)\in\GamLamM{1}$ for $X\in\GamTM$ and
$\dual{\alpha}=g^{-1}(\alpha,-)\in\GamTM$ for
$\alpha\in\GamLamM{1}$. The Hodge dual map associated with $g$ is
denoted by a star so that the canonical $n$-form measure (\lq\lq
volume element\rq\rq)  on an $n$-dimensional manifold $M$ is the
image of 1 under the  Hodge
map. In $4$-dimensional spacetime we adopt the flat Minkowski metric.
In  a $3$-dimensional space we adopt the Euclidean flat metric and
regard time as an evolution parameter for  forms in three
dimensions.
 In a  $2$-dimensional space we adopt the Euclidean flat metric and
regard time and a longitudinal coordinate as parameters for forms in
two dimensions.
 One must then distinguish notationally between the different
metrics introduced and their associated Hodge maps. However for any
manifold $M$ with Hodge map $\star$  one always has the standard
relations
\begin{align}
\Phi\wedge\star\Psi=\Psi\wedge\star\Phi, \qquad&\text{for}\quad
\Phi,\Psi\in\GamLamM{p}
\label{id_star_pivot}
\\
\nonumber \\[-1ex]
i_X\star \Phi=\star(\Phi\wedge\widetilde{X}), \qquad&\text{for}\quad
X\in\GamTM,\ \Phi\in\GamLamM{p} \label{id_iX_star}
\end{align}
where $i_X$ denotes the interior (contraction) operator on
forms.

Maxwell's equations find their most cogent formulation as a theory
of $2$-forms on spacetime modelled on a space and time oriented
$4$-dimensional manifold with a metric tensor field $g$ of
Lorentzian signature $(-,+,+,+)$. On such a spacetime $M$ the set
$\Set{e^0,e^1,e^2,e^3}$  will denote a local $g$-orthonormal coframe
(a linearly independent collection of $1$-forms).
The Hodge map associated with the Lorentzian metric $g$ will be
denoted by $\star$. Then \vspace{-0.3cm}
\begin{align}
\star\, i_X\Phi=-\star\Phi\wedge\widetilde{X}, \qquad&\text{for}\quad
X\in\GamTM,\ \Phi\in\GamLamM{p} \label{id_star_iX}
\\ \nonumber \\
\star\star\Phi=(-1)^{p+1}\Phi,\qquad&\text{for}\quad
\Phi\in\GamLamM{p} \label{id_star_star}
\end{align}

For manifolds with a Euclidean signature and different dimensions
these last two relations change as will be indicated for three and
two dimensional spaces below. Finally note that for all
$n$-dimensional manifolds of any signature one has the useful
results:
\begin{align}
i_{X}\Phi \wedge \Psi =(-1)^{p+1}\Phi \wedge i_{X}\Psi, \quad & \text{for}%
\quad \Phi \in \GamLamM{p},\   \nonumber \\
\text{ \ \ \ \ \ \ \ \ }\Psi & \in \GamLamM{q},\ p+q\geq n+1
\label{id_iX_move} \\  \nonumber \\
d\Phi \wedge \Psi =(-1)^{p+1}\Phi \wedge d\Psi + d(\Phi\wedge\Psi
),\qquad & \text{for}\quad \Phi \in \GamLamM{p},\   \nonumber \\
\Psi & \in \GamLamM{q}
 \label{id_d_move}
\end{align}

As a notational convenience the expression $\TYPE {q} \Psi {}  $
below implies that $\Psi\in \Gamma\Lambda^{q}N$ is a differential
$q$-form on $N$ where the manifold $N$ follows from the
context.
\def\ee{\,\epsilon_0\,}
\def\cc{\, c_0 \,}

\section{Electromagnetic Fields in Spacetime}
\label{ch_fields}

Maxwell's equations for an electromagnetic field in an arbitrary
medium can be written
\begin{align}
d\,F=0 \qquadand d\,\star\, G =j, \label{Maxwell}
\end{align}
where $F\in\GamLamM{2}$ is the Maxwell $2$-form, $G\in\GamLamM{2}$ is
the excitation $2$-form and $j\in\GamLamM{3}$ is the $3$-form electric
current source\footnote{All tensors in this article have dimensions
  constructed from the SI dimensions $[M], [L], [T], [Q]$ where $[Q]$
  has the unit of the Coulomb in the MKS system. We adopt $[g]=[L^2],
  [G]=[j]=[Q],\,[F]=[Q]/\ee$ where the permittivity of free space
  $\epsilon_0$ has the dimensions $ [ Q^2\,T^2 M^{-1}\,L^{-3}] $ and
  $c_0$ denotes the speed of light in vacuo.}.
  To close this system, ``electromagnetic constitutive
relations'' relating $G$ and $j$ to $F$ are necessary.

The electric $4$-current $j$  describes both  (mobile) electric charge
and effective (Ohmic) currents in a conducting medium.  The {\it
electric } field $\Me\in\GamLamM{1}$ and {\it magnetic induction
field $\Mb\in\GamLamM{1}$} associated with $F$ are defined with
respect to an arbitrary {\it unit} future-pointing timelike
$4$-velocity vector field $U\in\GamTM$ by
\begin{align}
\Me = \iuF \qquadand \cc\Mb = \iustarF. \label{intro_e_b}
\end{align}
Thus $i_U\Me=0$ and $i_U\Mb=0$.

Since $g(U,U)=-1$
\begin{equation}
F=\Me\wedge \dualu - \star\,(\cc\Mb\wedge \dualu). \label{intro_F}
\end{equation}

The field $U$ may be used to describe an {\it observer frame} on
spacetime and its integral curves model idealised  observers.

Likewise the {\it displacement} field $\Md\in\GamLamM{1}$ and the
{\it magnetic} field $\Mh\in\GamLamM{1}$  associated with $G$ are
defined with respect to $U$ by
\begin{align}
\Md = \iu G\,, \qquadand \Mh/\cc = \iu\star G\,. \label{Media_d_h}
\end{align}
Thus
\begin{equation}
G=\Md\wedge \dualu - \star\,((\Mh/\cc)\wedge \dualu), \label{Media_G}
\end{equation}
and $i_U\Md=0$ and $i_U\Mh=0$. It may be assumed that a material
medium has associated with it a future-pointing timelike unit vector
field $V$  which may be identified with the  bulk $4$-velocity field
of the medium in spacetime. Integral curves of $V$ define the
averaged world-lines of identifiable constituents of the medium. A
{\it comoving observer frame with $4$-velocity $U$} will have
\footnote{If $U\ne V$ but at an event $p$ in spacetime  their
integral curves share the same tangent vector then it is sometimes said
that $V$ is {\it instantaneously} at rest at $p$ with respect to the
timelike frame $U$.} $U=V$.

\section{Time dependent Maxwell Systems in Space}
\label{ch2}
On any $n$-dimensional  manifold a chart  sets up a correspondence
between points on some region (patch) on the manifold and a set on
$\Real^n$. Thus in a $2$-dimensional patch let
$\hat\xibar=(\xi^1,\xi^2)$ be a generic set of coordinates.
Similarly let $\xibar=(\xi^1,\xi^2,\xi^3)$ denote coordinates on a
patch of a $3$-dimensional manifold and $ \xi=(\xibar,\xi^0)$ denote
coordinates on a patch of  $4$-dimensional spacetime.

Let $\D$ denote exterior differentiation in any domain of a
Euclidean space with coordinates $\xibar$.
Similarly let $\hat\D$
denote exterior differentiation in a patch with coordinates
$\hat\xibar$.
 A \lq\lq moving\rq\rq
orthonormal (Cartan) coframe in flat spacetime with Minkowski metric
$g$ is a set of (independent \footnote{i.e. $e^0\wedge e^1\wedge
e^2\wedge e^3 \neq 0$.}) $1$-forms $\{ e^0, e^1, e^2,e^3\}$ with
$e^0$ timelike. In general this will depend on the choice of
coordinates $\xi$ in the sense that its exterior derivative will not
be zero. In the following we adopt an inertial frame with laboratory
time $\xi^0=t$ and $e^0=c_0\,d t$ with $\{ e^1, e^2,e^3\}$
independent of $t$. Thus in general the coframe \lq\lq moves\rq\rq
as a function of $\xibar$. If $\beta$ is any form on spacetime it
will be convenient to adopt the abbreviation $\dot\beta$ for ${\cal
L}_{\PD t}\beta$, where ${\cal L}_X $ denotes the Lie derivative
\cite{rwt,frankel},
 with respect to $X$. Thus $\dot e^{k}=0$
for $k=1,2,3$. Within this framework introduce the tensor fields:
$$\hat {\underline g}= e^1\ot e^1 + e^2\ot
e^2,\quad
 {\underline g}= \hat {\underline g} + e^3 \ot
e^3,\quad g= -e^0\ot e^0+ \underline{g},
$$
where $e^0=c_0\,d t$ and $g$ is the metric tensor field on
Minkowski spacetime. At each instant ($t=$constant), $\underline g$
is the induced metric tensor on Euclidean space and $\hat
{\underline g}$ is the induced metric tensor on the $2$-dimensional
submanifolds (leaves) where $\xi_3=$ constant. Denote the Hodge map
associated with $\hat {\underline g}$ by $\hashat$ with
$$
\hashat 1 = e^1 \wedge  e^2,
$$
and that associated with $\underline g$ by $\hash$ with
$$
\hash 1= \hashat 1 \wedge e^3.
$$
Then
$$
\star 1=\hash 1\wedge e^0 := e^1\wedge e^2\wedge e^3\wedge e^0.
$$

To accommodate the effects of signature it is convenient to
introduce the  involution operator $\eta$ on $p$-forms $\Phi$  by
$\eta\Phi=(-1)^p\Phi$. Then
\BE{\star\star=-\eta,\qquad \hash \hash = 1,\qquad \hashat \hashat =
\eta.}

By linearity the action of the   Hodge map on an arbitrary form in
Euclidean $3$-space readily  follows by expanding it in an
orthonormal basis and using the relations
$$
\hash e^1=e^2\wedge e^3,
$$
$$
\hash e^2=e^3\wedge e^1,
$$
$$
\hash e^3=e^1\wedge e^2,
$$
on the basis forms. Furthermore in a $2$-dimensional
Euclidean space
$$
\hashat e^1=e^2,
$$
$$
\hashat e^2=-e^1.
$$

If $\TYPE p \beta {} (\xi)$ is a $p$-form {\it on spacetime} but
generated by forms in the exterior algebra generated by
$\{e^1(\xibar) ,e^2(\xibar) ,e^3(\xibar)\}$ then at any event with
coordinates $\xi$ one has
$$
\TYPE p \beta {}(\xi)= \sum _I \beta_I(\xi) e^I( \xibar),
$$
where,  for each multi-index $I$, the set of exterior
$p$-forms $\{e^I(\xibar)\}$ denotes a basis for $p$-forms generated
from the set $\{e^1(\xibar), e^2(\xibar), e^3(\xibar)\}$. One refers
to the functions  $ \beta_I  $ as the components of $ \TYPE p \beta
{} $ in the $e^I$ basis. With this notation
$$
\TYPE
p {\dot \beta} {}(\xi) := \sum _I \frac{\partial}{\partial
\xi^0}\beta_I(\xi) e^I( \xibar).
$$

Define the $2+1$ split of $\TYPE p { \beta} {}(\xi)$   into the pair
$\{ \TYPE {p-1} {\hat \beta} {}(\xi) , \TYPE p {\hat \beta} {}(\xi)
\}   $ by the unique decomposition with respect to $\D \xi^3$:
\BE{
\TYPE p { \beta} {}(\xi) = \TYPE {p-1} {\hat \beta} {}(\xi)
\wedge \D\xi^3 +
 \TYPE p {\hat \beta} {}(\xi),
}
where $\TYPE {p-1} {\hat \beta} {}(\xi)$ and $ \TYPE p
{\hat \beta} {}(\xi) $ are   $p-1$  and $p$-forms respectively,
generated from the $1$-forms in $\{ \D\xi^1,\,\D\xi^2\}$  satisfying
$i_{\PD {\xi^3}} \TYPE {p-1} {\hat \beta} {}(\xi)=0$ and $i_{\PD
{\xi^3}} \TYPE p { \hat\beta} {}(\xi)=0 $. Thus $ \TYPE {p-1} {\hat
\beta} {}  $ and $ \TYPE {p} {\hat \beta} {}  $ are forms that do
not contain $\D \xi^3$.

It follows that for $q=0,1,2$:
\BE{ \hash(
 {\TYPE q {\hat\beta}  {}}  \wedge e^3
 ) =
\hashat (\eta {\TYPE q {\hat\beta} {}} ),
}
\BE{\hash(  {\TYPE q
{\hat\beta} {}} ) = \hashat  ({\TYPE q{\hat\beta}{}}) \wedge e^3.
}
For any $0$-form ${\TYPE 0 {\hat\beta} {}}$
$$
\D {\TYPE 0 {\hat\beta} {}}= \hat\D {\TYPE 0 {\hat\beta} {}}
 + ({\cal L}_{{\frac{\partial}{\partial\xi^3}}} {\TYPE 0 {\hat\beta}
   {}})\, \D\xi^3,
$$
where
$$
\hat\D {\TYPE 0 {\hat\beta} {}} := \PD {\xi^1}
{\TYPE 0 {\hat\beta} {}}\,\D\xi^1 +
 \PD  {\xi^2} {\TYPE 0 {\hat\beta} {}}\,\D\xi^2.
$$

From this it follows that,  for $q=0,1,2$:
$$
\D {\TYPE q {\hat\beta} {}}= \hat\D {\TYPE q {\hat\beta} {}}
 +  \D\xi^3 \wedge ({\cal L}_{{\frac{\partial}{\partial\xi^3}}}
{\TYPE q {\hat\beta} {}}),
$$
where $ \hat\D$
acts\footnote{When ${\hat\beta}$ 
is independent of $\xi^3$
(i.e.    $  {\cal L}_{{\frac{\partial}{\partial\xi^3}}}  
{\hat\beta} 
$
it is unnecessary to
distinguish notationly between $ \D $  and $\hat\D. $  }
on exterior forms generated by $\{ \D\xi^1,\D\xi^2 \}$. Note that
for all $2$-forms ${\TYPE 2 {\hat\beta} {}} $ one has $ \hat\D {\TYPE
2 {\hat\beta} {}} =0$. Let the $3+1$ split of the 4-current $3$-form
be
\BE{
{\TYPE 3 j {}} (\xi)= - {\TYPE 2 J {}}(\xi) \wedge d\,t + \TYPE
0 \rho {} (\xi) \hash 1,
}
with $i_{\PD t} {\TYPE 2 J {}}=0 $. Then, from
(\ref{Maxwell})
\BE{
d\, j=0,\label{dj}
}
yields
$$
\D {\TYPE 2 J {}}(\xi) + \dot{\underset{(0)}\rho}(\xi) \hash 1=0.
$$

It is convenient to introduce the (Hodge) dual forms:
$$
\bfE:=\hash \bfe, \quad \bfD:=\hash\bfd,\quad \bfB:=\hash \bfb, \quad
\bfH:=\hash \bfh,
$$
so that the $3+1$ split of the spacetime covariant Maxwell
equations (\ref{Maxwell})  with respect to $d\, t$ becomes
\BE{\D\bfe=-\bfBdot,\label{M1}}
\BE{\D\bfB=0,\label{M2}}
\BE{\D\bfh={\TYPE 2 J {}}+\bfDdot,\label{M3}}
\BE{\D\bfD=\underset{(0)}\rho\hash 1.\label{M4}}

All $p$-forms ($p\ge1$) in these equations are independent of $e^0$
but may depend on $t$. Furthermore they are independent of the
choice of (stationary) spatial co-frame constructed from $\{\D
\xi^1, \D \xi^2,\D \xi^3\}$, in any chart with local coordinates $
\xi^1,\xi^2,\xi^3$.

In the following it is assumed that  $\bfb=\mu \bfh$ and $ \bfd=\ep
\bfe$  (with constant $\ep, \mu $ ) where $\ep=\EP$, $\mu=\MU$. Thus
in terms of $\bfe,\bfh, \bfE, \bfH$:
\BE{\D   \bfe=-\mu\bfHdot,\label{M11}}
\BE{\D   \bfH=0,\label{M12}}
\BE{\D   \bfh=\ep \bfEdot + {\TYPE 2 J {}},   \label{M13}}
\BE{\ep\D\bfE=\underset{(0)}\rho \hash 1.\label{M14}}


\newcommand{\DDHAT}{\D\,}

\newcommand\PSII[2]{{\Psi}_{#1}{#2}}
\newcommand\PHII[2]{{\Phi}_{#1}{#2}}

\newcommand\PSIB[2]{\overline{\Psi_{#1}}{#2}}
\newcommand\PHIB[2]{\overline{\Phi_{#1}}{#2}}

\newcommand\U{{\cal U}}

\newcommand\basiceqns{(\ref{eqn:M1}), (\ref{eqn:M2}),
 (\ref{eqn:M3}), (\ref{eqn:M4}) }
\section{The Maxwell System with Sources in a Curved Beam Pipe}
\label{sec:generic-internal-sources}
In terms of time dependent $1$-form \EM fields with general sources
$(\TYPE 0 {\rho} {} , {\TYPE 2 J {} {}}) $ the Maxwell's equations
in a medium with scalar permeability $\mu$  can be written
 \beqa
&&\D \bfe +\mu\hash\dot{\bfh}=0,\label{eqn:M1}\\
&&\D \bfh -\EPS\hash\dot{\bfe}-{\TYPE 2 J {} {}}=0,\label{eqn:M2}\\
&&\D \hash\bfh =0,\label{eqn:M3}\\
&&\EPS\D\hash\bfe-\TYPE 0 {\rho}{}\hash1=0,\label{eqn:M4}
\eeqa
where the admittance $Y= 1/(\mu c)$ with $c$ being the speed of
light in the medium.

These equations involve time dependent forms and are independent of
particular local spatial coordinates. They depend explicitly on the Euclidean
metric and for conserved sources define a well posed
initial-boundary value problem. We now choose a coordinate system
adapted to the interior $\U$ of a beam pipe with a circular disc
cross-section of fixed radius $a$ at every point and an axis given
by a planar space-curve with, in general,  non-constant curvature $\kappa$ and
$\vert\kappa a \vert \ll 1$. At each point on this
curve one may erect a triad of orthogonal vectors
in space, one member of
which is tangent to the curve. The remaining vectors define a
transverse plane. All points in the interior $\U$ of the beam pipe
lie on some transverse plane associated with such a triad with
origin at some point on the axial space-curve.
 Let the region
$\U\subset {\mathbb R}^3$ inside the beam pipe be described in terms of
coordinates $(\xi^1,\xi^2,\xi^3) := (r,\theta, z)$ adapted to
the central space-curve with curvature $\kappa(z)$ such that
$$
0\leq r\leq a,\qquad 0 < \theta \leq 2\pi,\qquad -\infty\leq z \leq
\infty.
$$
A convenient field of orthonormal coframes \cite{sagnac} on $\U$ is given
in these coordinates by
\BE{
\{ e^1=\D r,\quad  e^2=r\D\theta,\quad
e^3 =\left(1-\epsilon\kappa_0(z) x_1\right)\D z\},
\label{coframes}
}
with $x_1=r\cos\theta$. Thus the Euclidean metric tensor $\underline
g$ on ${\cal U}$ is given by
$$
\underline g=
e^1 \otimes e^1 + e^2 \otimes e^2 +
  e^3\otimes e^3.
$$
In these coordinates the pipe boundary is the surface $r=a$, the
coordinate $z$ measures arc-length along the space-curve and on the
space-curve $r=0$.

It proves convenient in the following to write
$\kappa(z)=\epsilon\kappa_0(z)$ and use $\epsilon$ as a book-keeping
device in order to keep track of different orders of $\kappa$. In
terms of adapted coordinates
$$
\underline g=\underline{\hat{g}}+
(1-\epsilon\kappa_0(z) x_1)^2\D z\otimes\D z,
$$
where for each
cross-section at constant $z$ one has the induced metric tensor
${\hat{\underline g}}$ on the $2$-dimensional disc ($0\leq r \leq a,
0<\theta \leq 2\pi $):
$$
\underline{\hat{g}}= \D r\otimes \D r
+ r^2\D\theta\otimes\D\theta.
$$
The associated  contravariant
tensors  are
\beqa
\underline{\hat{g}}^{-1}&=&\frac{\partial}{\partial r}
  \otimes\frac{\partial}{\partial r}
+\frac{1}{r^2}\frac{\partial}{\partial\theta}
  \otimes\frac{\partial}{\partial\theta}, \non\\
\underline{g}^{-1}&=&\underline{\hat{g}}^{-1}
+\frac{1}{(1-\epsilon\kappa_0(z) x_1)^2}\frac{\partial}{\partial z}
 \otimes \frac{\partial}{\partial z}.
\non
\eeqa
The source forms will be expressed in terms of the scalar functions
$J_r, J_{\theta},J_0$ and ${\rho}$  of $(\epsilon,t,z,r,\theta)$. We
choose to write $\TYPE 2 J {}{}$ as
\beqa
&&{\TYPE 2 J {}{(\epsilon,t,z,r,\theta)}}\non\\
&&=
  \bigg(J_{\theta}(\epsilon,t,z,r,\theta)\D r
+ r J_{r}(\epsilon,t,z,r,\theta)\D\theta
  \bigg)\wedge\D z(1-\epsilon\kappa_0(z) r \cos\theta)\non\\
&&\quad + J_0(\epsilon,t,z,r,\theta)r\D r\wedge\D\theta,
\label{eqn:S2-general}
\eeqa
from which it immediately follows that
$$
\hash {\TYPE 2 J {}{}}=-J_\theta r\D \theta +J_r\D r
+J_0(1-\epsilon\kappa_0(z) x_1)\D z,
$$
with orthogonal components $(-J_\theta, J_r,J_0 )$ of
$ \hash {\TYPE 2 J {}{}} $.
The associated electric current vector field is
$\widetilde{\hash {\TYPE 2 J {}{}} }$.

The objective is to solve (\ref{eqn:M1}), (\ref{eqn:M2}),
 (\ref{eqn:M3}),(\ref{eqn:M4}) for the fields $\bfe$ and $\bfh$  on
$\U$ in terms of  prescribed sources and initial data as a
perturbative expansion in the axial curvature of the beam pipe. The
strategy will be to project the field system into suitable modes
that ensure that perfectly conducting boundary conditions are
satisfied at the pipe boundary. In the adapted coordinate system
this is achieved with the aid of complex Dirichelet and Neumann
eigen-modes of the two-dimensional Laplacian associated with each
transverse plane in the beam pipe.

\subsection{Dirichelet Modes}
\label{ch5}
Let ${\cal D}$ be the  smooth $2$-dimensional submanifold
($z$=constant) with boundary $\partial{\cal D}$, embedded in
Eucldean $\Real^3$. The tensor $\hat {\underline g}$  on  ${\cal D}$
is that induced from the Euclidean metric $\underline g$ in
$\Real^3$. A {\it complex Dirichelet mode set} $\{ \Phi_N\}$ is a
collection of complex eigen $0$-forms of the Laplacian operator
$-\DDHAT \hashat \DDHAT $ on ${\cal D}$
(associated with the metric $\hat g$ and Hodge operator $\hashat $)
that vanishes on $\partial{\cal D} $.
This boundary condition and the nature of the domain determine the
associated (positive non-zero real) eigenvalues $\beta_N^2$. The
label $N$ here consists of an ordered pair of real numbers. Thus
\BE{ \PHII N {}  :
{\cal D} \to \Real,\quad  {r,\theta} \mapsto \PHII N {(r,\theta)},}
 satisfies \BE{ \DDHAT \hashat \DDHAT \PHII N {} + \beta_N^2 \PHII N {}
\hashat 1=0,
\label{dmode}}
with $\PHII N {} \vert_{\partial{\cal
D}}=0.$ It is straightforward to show from these properties that if
$\beta_N^2\ne \beta_M^2\neq 0$, $(N\neq M)$
 then
$$
\int_{\cal D}\PHIB M {}\, \PHI N {} \,\hashat 1 =0,
$$
where the bar denotes complex conjugation.
If one normalises these modes so that
\beq
\int_{{\cal D}}\PHIB
 M {}\, \PHI N {} \,\hashat 1=
{\cal N}_N^2\,\delta_{NM},
\label{eqn:Dirichelet-orthogonal}
\eeq
then it is also easy to show that
$$
\int_{\cal D} \DDHAT \PHIB N {}\wedge \hashat \DDHAT \PHII M {} =
\beta^2_N{\,\cal N}_N^2 \,\delta_{NM}.
$$
An explicit form for
$\Phi_N$ is for $n\in\mathbb{Z}$
\beq
\Phi_N(r,\theta)=J_n\left(x_{q(n)}\frac{r}{a}\right) e^{in\theta},
\label{eqn:Phi-basic}
\eeq
where $J_n(x)$ is the $n$-th Bessel function
and the numbers $\{x_{q(n)}\}$ are defined by $J_n(x_{q(n)})=0$ and
$N:=\{n,q(n)\}$. The eigenvalues are given by
$\{\beta_N=x_{q(n)}/a\}$. It follows from the integral
\cite{abramowitz}
$$
\int_0^a dr r J_m\left(x_{q(m)}\frac{r}{a}\right)
J_m\left(x_{q'(m)}\frac{r}{a}\right)
=\frac{a^2}{2}J_{m+1}^2(x_{q(m)})\delta_{q(m),q'(m)},
$$
that
 $\NN{N}^2=\pi a^2 J^2_{n+1}(x_{q(n)})$.

\subsection{Neumann Modes}
\label{ch5}
In a similar manner one defines  a {\it Neumann mode set}
$\{\Psi_N\}$ as a collection of eigen $0$-forms of the Laplacian
operator  on ${\cal D}$ such that $\hashat \DDHAT \PSII N {}$
vanishes on $\partial{\cal D} $.
This alternative boundary condition and the nature of the domain determine the
associated (positive non-zero real) eigenvalues $\alpha_N^2$ where
again the label $N$ consists of an ordered pair of real
numbers. Thus
$$
\PSII N {}  :
{\cal D} \to \Real, \quad{r,\theta} \mapsto \PSII N {(r,\theta)},
$$
satisfies
\beq
\DDHAT \hashat \DDHAT \PSII N {} + \alpha_N^2 \PSII N {}
\hashat 1=0,
\label{nmode}
\eeq
with $ \hashat \DDHAT \PSI N {}
\vert_{\partial{\cal D}}=0$ It is straightforward to show from these
properties that if
$\alpha_N^2\ne \alpha_M^2\neq 0$, ($N\neq M$)
then
\beq
\int_{\cal D}\PSIB M {}\, \PSII N {} \,\hashat 1 =0.
\label{eqn:Neumann-orthogonal}
\eeq
If one
normalises these modes so that
$$
\int_{{\cal D}}\PSIB M {}\, \PSII
N {} \,\hashat 1={\cal M}_N^2\,\delta_{NM},
$$
then it is also easy to
show that
$$
\int_{\cal D} \DDHAT \PSIB N {}\wedge \hashat \DDHAT
\PSII M {} = \alpha^2_N{\,\cal M}_N^2 \,\delta_{NM}.
$$
An explicit
form for  $\Psi_M$ is  for $m\in\mathbb{Z}$
\beq
\Psi_M(r,\theta)=J_m\left(x'_{p(m)}\frac{r}{a}\right) e^{im\theta},
\label{eqn:Psi-basic}
\eeq
where the numbers  $\{x'_{p(m)}\}$ are defined by $J_m'(x'_{p(m)})=0
$ and $M:=\{m,p(m)  \}$. The eigenvalues are given by
$\{\alpha_M=x_{p(m)}'/a\}$ and
 $\MM{M}^2=\pi a^2 J^2_{m+1}(x'_{p(m)})$.

\subsection{Mode Decompositions}
\label{sec:1-forms}
Since ${\cal U}$ is simply connected one can represent the
electromagnetic $1$-forms  $\bfe=\bfe(\epsilon,t,z,r,\theta)$ and
$\bfh=\bfh(\epsilon,t,z,r,\theta)$ as \cite{RWT-theoret}
\beqa
\bfe(\epsilon,t,z,r,\theta)&=& \sum_N V_N^E(\epsilon,t,z)\D\Phi_N+
\sum_M V_M^H(\epsilon,t,z)\hash(\D z\wedge\D\Psi_M)\non\\
&&+\sum_N\gamma_N^E(\epsilon,t,z)\Phi_N(r,\theta)\,\D z,
\label{eqn:e1-basic}\\
\bfh(\epsilon,t,z,r,\theta)&=& \sum_N I_N^E(\epsilon,t,z)\hash(\D
z\wedge\D\Phi_N)
+\sum_MI_M^H(\epsilon,t,z)\D\Psi_M\non\\
&&+\sum_M\gamma_M^H(\epsilon,t,z)\Psi_M(r,\theta)\,\D z.
\label{eqn:h1-basic} \eeqa Here, for any scalars $f_M$ with
$M=(m,p(m))$ and $f_N$ with
$N=(n,q(n))$, 
the summations above  are  abbreviations for:
$$
\sum_M f_M\cdots=\sum_{m\in\mathbb{Z}}\sum_{p(m)\in\mathbb{N}}
f_{m,p(m)}\cdots,\quad \sum_N
f_N\cdots=\sum_{n\in\mathbb{Z}}\sum_{q(n)\in\mathbb{N}}
f_{n,q(n)}\cdots.
$$
For future convenience the further abbreviation:
$$
{\sum_N}' f_N\cdots=\sum_{n=m\pm 1}\sum_{q(n)\in\mathbb{N}}
f_{n,q(n)}\cdots.
$$
will be used.
 The expansions above in terms of  $\Phi_N$ and
$\Psi_M$ and their derivatives  ensure that {\it the \EM fields
satisfy perfectly conducting boundary conditions at the surface}
$r=a$.

Since $\Phi\vert_{r=a}=0$ we note that \BAE{
\int_\DD\D\Psi_M\wedge\D\Phi_N= \int_{\partial\DD} \Phi_N\,\D\Psi_M
=0, \label{eqn:basic-int} } and for $m,n \in\mathbb Z $
$$
\delta_{m,n}=\int_0^{2\pi}\frac{d\theta}{2\pi} e^{i(m-n)\theta}.
$$

Furthermore with $n\in\mathbb{Z}$,
$$
J_{-n}(x)=(-1)^nJ_n(x),
$$
so
$$
p(m)=p(-m),\quad \mbox{and}\quad q(n)=q(-n).
$$
Hence \beq \alpha_{n,q(n)}=\alpha_{-n,q(-n)},\quad
\beta_{n,q(n)}=\beta_{-n,q(-n)}, \label{eqn:symmetry-alpha-beta}
\eeq and \beq
\D\Phi_{n,q(n)}=(-1)^n\D\overline{\Phi_{-n,q(-n)}},\quad
\D\Psi_{m,p(m)}=(-1)^m\D\overline{\Psi_{-m,p(-m)}}.
\label{eqn:symmetry-dphi-dpsi} \eeq

From Eq. \fr{eqn:symmetry-dphi-dpsi} and the  reality conditions,
$$
\bfe={\overline\bfe},\qquad \mbox{and}\qquad \bfh={\overline\bfh},
$$
one has\BAE{ V_{n,q(n)}^{E}=(-1)^n{\overline{ V_{-n,q(-n)}^E}},~
V_{m,p(m)}^{H}=(-1)^m{\overline{ V_{-m,q(-m)}^H}},\\
\gamma_{n,q(n)}^{E}=(-1)^n{\overline{ \gamma_{-n,q(-n)}^E}},\\
I_{n,q(n)}^{E}=(-1)^n{\overline{ I_{-n,q(-n)}^E}},~
I_{m,p(m)}^{H}=(-1)^m{\overline{ I_{-m,q(-m)}^H}},\\
\gamma_{m,p(m)}^{H}=(-1)^m{\overline{ \gamma_{-m,p(-m)}^H}}.
\label{eqn:symmetry-e-h} } Similarly from the relations $$
zJ_n'(z)=nJ_n(z)-zJ_{n+1}(z),
$$
it follows that
\beq
\NN{-n,q(-n)}^2=\NN{n,q(n)}^2,\qquad\mbox{and}\qquad
\MM{-m,p(-m)}^2=\MM{m,p(m)}^2.
\label{eqn:symmetry-NN-MM}
\eeq
 These
relations enable one to pass simply from complex to real
representations of the mode summations   for the fields above.



\subsection{Perturbation Expansions}
\label{sec:perturbation_expansions}
Since for small $\vert\kappa a\vert$ the
beam pipe approximates a
straight cylinder we adopt the perturbative  field-mode expansions

\beq
V_N^E(\epsilon,t,z)=V_N^{E(0)}(t,z)+\epsilon V_N^{E(1)}(t,z)
+{\cal O}(\epsilon^2). \label{eqn:expans-example1}
\eeq
\beq
I_N^E(\epsilon,t,z)=I_N^{E(0)}(t,z)+\epsilon I_N^{E(1)}(t,z)
+{\cal O}(\epsilon^2). \label{eqn:expans-example2}
\eeq
\beq
\gamma_N^E(\epsilon,t,z)=\gamma_N^{E(0)}(t,z)+\epsilon
\gamma_N^{E(1)}(t,z) +{\cal O}(\epsilon^2),
\label{eqn:expans-example3}
\eeq
with analogous expansions for the magnetic modes $ V_M^H,
I_M^H,\gamma_M^H$ and express the sources as a power series in
$\epsilon$:
\BAE{ J_{\theta}(\epsilon,t,z,r,\theta)=
J_{\theta}^{(0)}(t,z,r,\theta) + \epsilon
J_{\theta}^{(1)}(t,z,r,\theta)
+{\cal O}(\epsilon^2),\\
J_{r}(\epsilon,t,z,r,\theta)=J_{r}^{(0)}(t,z,r,\theta)
+ \epsilon J_{r}^{(1)}(t,z,r,\theta) +{\cal O}(\epsilon^2),\\
J_{0}(\epsilon,t,z,r,\theta)=J_{0}^{(0)}(t,z,r,\theta)
+ \epsilon J_{0}^{(1)}(t,z,r,\theta) +{\cal O}(\epsilon^2),\\
{\rho}(\epsilon,t,z,r,\theta)={\rho}^{(0)}(t,z,r,\theta)
 + \epsilon {\rho}^{(1)}(t,z,r,\theta) +{\cal O}(\epsilon^2).
\label{eqn:source-expansions} }

These expansions are then inserted into \basiceqns and ${\cal
O}(\epsilon^n)$ systems extracted for $n=0,1$. For general sources
it is somewhat tedious to project out the equations for the
perturbative field coefficients above. This is achieved using the
orthogonality relations between the different Dirichelet and Neumann
mode sets and the explicit relations listed in Appendix A
to integrate \basiceqns over 
the domain ${\cal D}$.

\def\F{{\cal F}}
\def\G{{\cal G}}

In order to express the resulting equations for the mode amplitudes
in a unified way  we define
$$
\Xi_{M,N}^k
(\F,\G,x,y):=\int_0^a dr  r^k
\F_m\left(x_{p(m)}\frac{r}{a}\right)
\G_n\left(y_{q(n)}\frac{r}{a}\right),
$$
and note
$$
\Xi_{M,N}^k(\F,\F,x,x)=\Xi_{N,M}^k(\F,\F,x,x).
$$
Furthermore if $\F$ is a Bessel function $J$ or its derivative, $J'$:
\beq
\Xi_{M,N}^k(\F,\F,x,x)=(-1)^{m+n}\Xi_{-m,p(-m),-n,q(-n)}^k(\F,\F,x,x).
\label{eqn:symmetry-Xi}
\eeq
 Thus
\beqa
\int_0^a dr r
J_m\left(x'_{p(m)}\frac{r}{a}\right) J'_n\left(x'_{q(n)}\frac{r}{a}\right)&=&
\Xi_{M,N}^1(J,J',x',x'),\non\\
\int_0^a dr  J_m\left(x'_{p(m)}\frac{r}{a}\right)
J_n\left(x'_{q(n)}\frac{r}{a}\right)&=&
\Xi_{M,N}^0(J,J,x',x').\non
\eeqa
The symbol $\Xi_{M,N}^k({\cal F},{\cal G},x,y)$
enables one to write more compactly certain overlap coefficients
that arise in the projections of \basiceqns over $\DD$. These
coefficients are given in Appendix A.

Using these definitions the projection of \fr{eqn:M1} to lowest
order
 ${\cal O}(\epsilon^0)$ is
\beqa
&&V_M^{E(0)\prime} +\mu{\dot
I}_M^{E(0)}-\gamma_M^{E(0)}=0,
\label{eqn:E1:0}\\
&&V_M^{H(0)\prime}-\mu{\dot I}_M^{H(0)}=0,
\label{eqn:H1:0}\\
&&\mu{\dot\gamma}_M^{H(0)}-\alpha_M^2 V_M^{H(0)}=0,
\label{eqn:H2:0}
\eeqa
and to first order  ${\cal O}(\epsilon^1)$ is
\BAE{
(V_M^{E(1)\prime}+\mu{\dot I}_M^{E(1)}
-\gamma_M^{E(1)})\beta_M^2\NN{M}^2\\
-\kappa_0(z){\sum_N}'(V_N^{H(0)\prime}+\mu\dot{I}_N^{H(0)})
G_{M,N}^{\bar{\Phi},\Psi}
-\kappa_0'(z){\sum_N}'V_N^{H(0)}G_{M,N}^{\bar{\Phi},\Psi}
=0,
\label{eqn:E1:1}
}
\BAE{
(V_M^{H(1)\prime}-\mu{\dot I}_M^{H
(1)})\alpha_M^2\MM{M}^2 +\kappa_0(z){\sum_N}'
(V_{N}^{H(0)\prime}+\mu{\dot I}_{N}^{H(0)}) F_{M,N}^{\Phi}\\
+\kappa_0'(z){\sum_N}'V_N^{H(0)}F_{M,N}^{\Phi}
=0,
\label{eqn:H1:1}
}
\BAE{
(\mu{\dot\gamma}_{M}^{H(1)}-\alpha_M^2
V_M^{H(1)})\MM{M}^2 +\kappa_0(z){\sum_N}'
(\mu{\dot\gamma}_{N}^{H(0)}-\alpha_{N}^2
V_{N}^{H(0)}) E_{M,N}^{\Psi}\\
+\kappa_0(z){\sum_N}'V_N^{H(0)}(C_{M,N}^{\Psi}-D_{M,N}^{\Psi})=0.
\label{eqn:H2:1}
}
where a superscript prime attached to a function
denotes its partial derivative with respect to $z$ and
all non-constant functions in
\fr{eqn:E1:1},\fr{eqn:H1:1},\fr{eqn:H2:1}
depend on $z$ or $z,t$.

The projection of \fr{eqn:M2} to lowest order
 ${\cal O}(\epsilon^0)$ is
\BAE{
-(I_M^{H(0)\prime}-\EPS{\dot V}_M^{H(0)}
-\gamma_M^{H(0)})\alpha_M^2\MM{M}^2\\
+\int_\DD
J_\theta^{(0)}({\hashat}\D\overline{\Psi_M})
\wedge\D r +\int_\DD
rJ_r^{(0)}({\hashat}\D
\overline{\Psi_M})\wedge\D\theta=0,
\label{eqn:H3:0}
}
\BAE{
(I_M^{E(0)\prime}+\EPS{\dot V}_M^{E(0)})\beta_M^2\NN{M}^2\\
+\int_{\DD}J_{\theta}^{(0)}\D\overline{\Phi_M}\wedge\D r + \int_{\DD} r
J_{r}^{(0)}\D\overline{\Phi_M}\wedge\D \theta=0,
\label{eqn:E2:0}
}
\BAE{
(-\beta_M^2
I_M^{E(0)}-\EPS{\dot\gamma}_M^{E(0)})\NN{M}^2-\int_{\DD}J_0^{(0)}
\overline{\Phi_M}
{\hashat}1=0. \label{eqn:E3:0}
}
and  to first order ${\cal O}(\epsilon^1)$ is
\BAE{
-(I_M^{H(1)\prime}-\EPS{\dot V}_M^{H(1)}-\gamma_M^{H(1)})
\alpha_M^2\MM{M}^2\\
+\kappa_0(z){\sum_N}'(I_N^{E(0)\prime}-\EPS\dot{V}_N^{E(0)})
G_{M,N}^{\bar{\Psi},\Phi}\\
+\kappa_0'(z){\sum_N}'I_N^{E(0)}G_{M,N}^{\bar{\Psi},\Phi}
+\int_\DD(J_{\theta}^{(1)} - J_{\theta}^{(0)}\kappa_0(z) r\cos\theta)
({\hashat}\D\overline{\Psi_M})\wedge\D r\\
+\int_\DD r(J_{r}^{(1)} - J_{r}^{(0)}\kappa_0(z) r\cos\theta)
({\hashat}\D\overline{\Psi_M})\wedge\D\theta
=0,
\label{eqn:H3:1}
}
\BAE{
(I_M^{E(1)\prime}+\EPS{\dot V}_M^{E(1)})\beta_M^2\NN{M}^2
+\kappa_0(z){\sum_N}' (I_{N}^{E(0)\prime}
-\EPS{\dot V}_{N}^{E(0)})
F_{M,N}^{\Phi}\\
+\kappa_0'(z){\sum_N}' I_{N}^{E(0)}F_{M,N}^{\Phi}
+\int_\DD (J_{\theta}^{(1)}-J_{\theta}^{(0)}\kappa_0(z) r\cos\theta)
\D\overline{\Phi_M}\wedge\D r\\
+\int_\DD r(J_r^{(1)}-J_r^{(0)}\kappa_0(z) r\cos\theta)
\D\overline{\Phi_M}\wedge\D\theta =0, \label{eqn:E2:1}
}
\BAE{
-(I_M^{E(1)}\beta_M^2+\EPS{\dot\gamma}_M^{E(1)})\NN{M}^2
-\kappa_0(z){\sum_N}'
(\beta_{N}^2I_{N}^{E(0)}+\EPS{\dot\gamma}_{N}^{E(0)})
E^{\Phi}_{M,N}\\
+\kappa_0(z){\sum_N}' I_{N}^{E(0)}(C^{\Phi}_{M,N}-D^{\Phi}_{M,N})
-\int_\DD J_0^{(1)}\overline{\Phi_M}{\hashat}1
=0. \label{eqn:E3:1}
}


The projection of \fr{eqn:M3} to lowest order
 ${\cal O}(\epsilon^0)$ is
\BAE{ \gamma_M^{H(0)\prime}-\alpha_M^2 I_M^{H(0)}=0.
\label{eqn:H4:0}
}
and to first order ${\cal O}(\epsilon^1)$,
\BAE{
(\gamma_M^{H(1)\prime}-\alpha_M^2I_M^{H(1)})\MM{M}^2
+\kappa_0(z){\sum_N}'(\gamma_{N}^{H(0)\prime}+\alpha_{N}^2
I_{N}^{H(0)})
E^{\Psi}_{M,N}\\
+\kappa_0(z){\sum_N}'I_N^{H(0)}(-C^{\Psi}_{M,N}+D^{\Psi}_{M,N})=0.
\label{eqn:H4:1}
}

Finally the projection of \fr{eqn:M4} to lowest order
 ${\cal O}(\epsilon^0)$ is
\BAE{
\EPS(\gamma_M^{E(0)\prime}-\beta_M^2 V_M^{E(0)})\NN{M}^2-
\int_\DD\rho^{(0)}\overline{\Phi_M}{\hashat}1=0. \label{eqn:E4:0}
}
and to first order  ${\cal O}(\epsilon^1)$,
\BAE{
\EPS(\gamma_M^{E(1)\prime}-\beta_M^2V_M^{E(1)})\NN{M}^2 -\int_\DD
({\rho}^{(1)}-{\rho}^{(0)}\kappa_0(z) r\cos\theta)
\overline{\Phi_M}\hashat 1\\
+\EPS\kappa_0(z){\sum_N}'
\bigg\{(V_{N}^{E(0)}\beta_{N}^2+\gamma_{N}^{E(0)\prime})E^{\Phi}_{M,N}
+V_{N}^{E(0)}(-C^{\Phi}_{M,N}+D^{\Phi}_{M,N})
\bigg\}=0.
\label{eqn:E4:1}
}


\subsubsection{Decoupling to lowest order}
\label{subsec:GEN-0-derive-eqs}

The equations above to lowest order describe the fields that can be
excited by sources in a perfectly conducting straight beam pipe. They are
readily decoupled by substituting 
\fr{eqn:H2:0} and
\fr{eqn:H4:0} into 
\fr{eqn:H3:0} yielding an inhomogeneous
telegraph-type equation for $\gamma_N^{H(0)}$:
\beqa
&&\ddot{\gamma}_N^{H(0)}-c^2\gamma_N^{H(0)\prime\prime}
+c^2\alpha_N^2\gamma_N^{H(0)}\non\\
&&=\frac{-c^2}{\MM{N}^2}\bigg\{ \int_\DD
J_\theta^{(0)}(\hashat\overline{\Psi_N})\wedge\D r + \int_\DD
rJ_r^{(0)}(\hashat\overline{\Psi_N})\wedge\D\theta \bigg\}.
\label{eqn:GEN-Sol-gammaH:0}
\eeqa
 Similarly the equation for $\gamma_N^{E(0)}$ follows by substituting
\fr{eqn:E3:0} and \fr{eqn:E4:0} into 
\fr{eqn:E1:0}:
\beqa
&&\ddot{\gamma}_N^{E(0)}-c^2\gamma_N^{E(0)\prime\prime}
+c^2\beta_N^2\gamma_N^{E(0)}\non\\
&&= -\frac{1}{\EPS\NN{N}^2}\frac{\partial}{\partial t}\bigg(
\int_\DD J_0^{(0)}\overline{\Phi_N}\hashat 1 \bigg)
-\frac{c^4\mu}{\NN{N}^2}\frac{\partial}{\partial z} \bigg( \int_\DD
\rho^{(0)}\overline{\Phi_N}\hashat 1 \bigg).
\label{eqn:GEN-Sol-gammaE:0}
\eeqa

Once these equations are solved  the other
field components can be calculated
from $\gamma_N^{H(0)}$ and $\gamma_N^{E(0)}$ and their derivatives.
Thus
\fr{eqn:E4:0},\fr{eqn:H2:0},\fr{eqn:E3:0} and \fr{eqn:H4:0} yield
\beqa
V_N^{E(0)}&=&\frac{1}{\beta_N^2}\bigg(
\gamma_N^{E(0)\prime}-\frac{1}{\NN{N}^2\EPS} \int_\DD
\rho^{(0)}\overline{\Phi_N}\hashat 1 \bigg),
\label{eqn:GEN-Sol-VE:0}\\
V_N^{H(0)}&=&\frac{\mu}{\alpha_N^2}\dot{\gamma}_N^{H(0)},
\label{eqn:GEN-Sol-VH:0}\\
I_N^{E(0)}&=&-\frac{1}{\beta_N^2}\bigg( \EPS\dot{\gamma}_N^{E(0)}
+\frac{1}{\NN{N}^2} \int_\DD J_0^{(0)}\overline{\Phi_N}\hashat
1\bigg),
\label{eqn:GEN-Sol-IE:0}\\
I_N^{H(0)}&=&\frac{1}{\alpha_N^2}\gamma_N^{H(0)\prime}.
\label{eqn:GEN-Sol-IH:0}
\eeqa
These solutions are consistent for sources satisfying (\ref{dj}).

\subsubsection{ Decoupling to First Order }
\label{subsec:GEN-1-derive-eqs}

The  first order equations
involve the solutions to the
lowest order fields. Similar to the case above the equation for
$\gamma_M^{H(1)}$ follows by substituting
\fr{eqn:H2:1} and
\fr{eqn:H4:1} into
\fr{eqn:H3:1}:
\beqa
&&\ddot{\gamma}_M^{H(1)}-c^2\gamma_M^{H(1)\prime\prime}
+c^2\alpha_M^2\gamma_M^{H(1)} \non\\
&&= -\frac{c^2\kappa_0(z)}{\MM{M}^2}
{\sum_N}' (I_N^{E(0)\prime}-\EPS\dot{V}_N^{E(0)})
G_{M,N}^{\bar{\Psi},\Phi}
-\frac{c^2}{\MM{M}^2}\kappa_0'(z){\sum_N}'I_N^{E(0)}G_{M,N}^{\bar{\Psi},\Phi}
\non\\
&& -\frac{c^2}{\MM{M}^2}\int_\DD
(J_\theta^{(1)}-J_\theta^{(0)}\kappa_0(z) r\cos\theta)
(\hashat\D\overline{\Psi_M})\wedge\D r\non\\
&& -\frac{c^2}{\MM{M}^2}\int_\DD r(J_r^{(1)}-J_r^{(0)}\kappa_0(z)
r\cos\theta)
(\hashat\D\overline{\Psi_M})\wedge\D\theta\non\\
&&+\frac{c^2}{\MM{M}^2}\frac{\partial}{\partial z}
{\sum_N}' \kappa_0(z)
\bigg\{ (\gamma_N^{H(0)\prime}+\alpha_N^2
I_N^{H(0)})E_{M,N}^{\Psi} -I_N^{H(0)}(C_{M,N}^{\Psi}-D_{M,N}^{\Psi})
\bigg\}\non\\
&&-\frac{\kappa_0(z)}{\mu\MM{M}^2}\frac{\partial}{\partial t}
{\sum_N}'\bigg\{
V_N^{H(0)}(C_{M,N}^{\Psi}-D_{M,N}^{\Psi})\bigg\},
\label{eqn:GEN-Sol-gammaH:1}
\eeqa
where \fr{eqn:H2:0} has been used.
The equation for
$\gamma_M^{E(1)}$ follows by substituting
\fr{eqn:E3:1} and
\fr{eqn:E4:1} into
\fr{eqn:E1:1}:
 \beqa
&&\ddot{\gamma}_M^{E(1)}-c^2\gamma_M^{E(1)\prime\prime}
+c^2\beta_M^2\gamma_M^{E(1)} \non\\
&&
=-\frac{c^2\kappa_0(z)}{\NN{M}^2}
{\sum_N}' (V_N^{H(0)\prime}+\mu\dot{I}_N^{H(0)})
G_{M,N}^{\bar{\Phi},\Psi}
-\frac{c^2\kappa_0'(z)}{\NN{M}^2}
{\sum_N}' V_N^{H(0)}G_{M,N}^{\bar{\Phi},\Psi}
\non\\
&&-\frac{\kappa_0(z)}{\EPS\NN{M}^2}
\frac{\partial}{\partial t}{\sum_N}'\bigg\{
(\beta_N^2I_N^{E(0)}+\EPS\dot{\gamma}_N^{E(0)})E_{M,N}^{\Phi}
-I_N^{E(0)}(C_{M,N}^{\Phi}-D_{M,N}^{\Phi})
\bigg\}\non\\
&&+\frac{c^2}{\NN{M}^2} \frac{\partial}{\partial z}
{\sum_N}'\kappa_0(z)
\bigg\{ (\beta_N^2 V_N^{E(0)}+\gamma_N^{E(0)\prime})
E_{M,N}^{\Phi}-V_N^{E(0)}(C_{M,N}^{\Phi}-D_{M,N}^{\Phi})\bigg\}\non\\
&&-\frac{c^2}{\EPS\NN{M}^2}\frac{\partial}{\partial z} \int_\DD
({\rho}^{(1)} -{\rho}^{(0)}\kappa_0(z) r\cos\theta)
\overline{\Phi_M}{\hashat}1\non\\
&&-\frac{\mu c^2 }{\NN{M}^2}\frac{\partial}{\partial t} \int_\DD
J_0^{(1)}\overline{\Phi_M}\hashat 1. \label{eqn:GEN-Sol-gammaE:1}
\eeqa

Expressions for the other field components can be expressed in terms
of $\gamma_M^{H(1)}$ and $\gamma_M^{E(1)}$, their derivatives and
lower order fields. From
\fr{eqn:E4:1}, \fr{eqn:H2:1}, \fr{eqn:E3:1} and \fr{eqn:H4:1} one
has
\beqa
&&V_M^{E(1)}=\frac{1}{\beta_M^2}\bigg[
\gamma_M^{E(1)\prime}-\frac{1}{\EPS\NN{M}^2}
\int_{\DD}({\rho}^{(1)}-{\rho}^{(0)}\kappa_0(z) r\cos\theta)
\overline{\Phi_M}{\hashat}1\non\\
&&\qquad +\frac{\kappa_0(z)}{\NN{M}^2}{\sum_N}' \bigg\{
(V_N^{E(0)}\beta_N^2+\gamma_N^{E(0)\prime})E_{M,N}^{\Phi}
-V_N^{E(0)}(C_{M,N}^{\Phi}-D_{M,N}^{\Phi}) \bigg\}
\bigg],\non\\
&&V_M^{H(1)}=\frac{1}{\alpha_M^2}\bigg\{
\mu\dot{\gamma}_M^{H(1)}+\frac{\kappa_0(z)}{\MM{M}^2} {\sum_N}'
V_N^{H(0)}(C_{M,N}^{\Psi}-D_{M,N}^{\Psi})
\bigg\},\non\\
&&I_M^{E(1)}=-\frac{1}{\beta_M^2}\bigg[
\EPS\dot{\gamma}_M^{E(1)}+\frac{\kappa_0(z)}{\NN{M}^2}{\sum_N}' \bigg\{
(\beta_N^2I_N^{E(0)}+\EPS\dot{\gamma}_N^{E(0)})E_{M,N}^{\Phi}\non\\
&&\qquad -I_N^{E(0)}(C_{M,N}^{\Phi}-D_{M,N}^{\Phi})
\bigg\}+\frac{1}{\NN{M}^2}\int_\DD J_0^{(1)}\overline{\Phi_M}\hashat
1
\bigg],\non\\
&&I_M^{H(1)}=\frac{1}{\alpha_M^2}\bigg[
\gamma_M^{H(1)\prime}+\frac{\kappa_0(z)}{\MM{M}^2}{\sum_N}' \bigg\{
(\gamma_N^{H(0)\prime}+\alpha_N^2 I_N^{H(0)})E_{M,N}^{\Psi}\non\\
&&\qquad
-I_N^{H(0)}(C_{M,N}^{\Psi}-D_{M,N}^{\Psi}) \bigg\} \bigg].\non
\eeqa
\subsubsection{General Solutions}
In \S\ref{subsec:GEN-0-derive-eqs} and
\S\ref{subsec:GEN-1-derive-eqs} the problem of solving \basiceqns
has been reduced to solving an initial-value problem for the
decoupled fields
$\gamma_N^{H(0)},\gamma_N^{H(1)},\gamma_N^{E(0)}$ and
$\gamma_N^{E(1)}$. For some real constant $\sigma>0$ each satisfies a
second-order hyperbolic partial differential equation in the
independent variables $(t,z)$, of the form:
\beq
\ddot{f}-c^2
f^{\prime\prime}+c^2\sigma^2 f=g,
\label{eqn:typical-diff-in-z}
\eeq
for some prescribed source
function $g$. The causal solution of this partial differential
equation  for $t>0$, with prescribed values of $f(0,z)$ and
$\dot{f}(0,z)$, has been exhaustively studied in the
literature, see e.g. \cite{pinsky}. If the data and sources are
sufficiently smooth the general solution may be expressed in the
form
\beq
f(t,z)={\cal H}_{\sigma}[f^{init}](t,z)
+{\cal I}_{\sigma}[g](t,z),
\label{eqn:Pinsky}
\eeq
where
\beqa
&&{\cal H}_{\sigma}[f^{init}](t,z):=
\frac{1}{2}\bigg\{f(0,z-ct)+f(0,z+ct)\bigg\}\non\\
&& +\frac{1}{2c}\int_{z-ct}^{z+ct}d\zeta\,
\dot{f}(0,\zeta)J_0(\sigma\sqrt{c^2t^2-(z-\zeta)^2})\non\\
&& -\frac{ct\sigma}{2}\int_{z-ct}^{z+ct}d\zeta\, f(0,\zeta)
\frac{J_1(\sigma\sqrt{c^2t^2-(z-\zeta)^2}\,)}{\sqrt{c^2t^2-(z-\zeta)^2}},
\label{eqn:def-H}
\eeqa
and
\beq
{\cal I}_{\sigma}[g](t,z):=\frac{1}{2c}\int_0^t
dt'\int_{z-c(t-t')}^{z+c(t-t')}d\zeta\,
g(t',\zeta)J_0(\sigma\sqrt{c^2(t-t')^2-(z-\zeta)^2}),
\label{eqn:def-I}
\eeq
The functions $f(0,z), \dot{f}(0,z)$
constitute the initial $t=0$ Cauchy data in this solution and
determine the ${\cal H}_\sigma$ contribution above. Typically, in an
accelerating device, lowest order contributions  include externally
applied piecewise established magnetostatic and RF fields that are
together used to guide and accelerate charges along the beam tube.
In the following we assume that all ${\cal H}_\sigma$ contributions
to the field solutions arise in lowest order.

\subsection{Electromagnetic Power from Smooth Sources}
\label{sec:em-power}

In the general situation it is seen that all zero and first order
fields can be calculated in terms of finite range integrals
involving Bessel functions. It is of some interest to calculate how
the instantaneous  \EM power flux depends on the first order
curvature correction to that in a straight cylinder with smooth sources. This is
obtained by integrating the Poynting vector field over the
cross-section $\DD$ at an arbitrary point with coordinate $z$. In
terms of the Poynting $2$-form
\beq
{\TYPE 2 {\mathbf S} {} {}}
(\epsilon,t,z,r,\theta):=
\bfe(\epsilon,t,z,r,\theta)\wedge\bfh(\epsilon,t,z,r,\theta),
\label{eqn:def-Poyinting}
\eeq
such instantaneous   power
$w(\epsilon,t,z)$ is obtained by integrating ${\TYPE 2 {\mathbf S}
{} {}}$ over  $\DD$:
$$
w(\epsilon,t,z):=\int_\DD {\TYPE 2 {\mathbf S} {} {}}
(\epsilon,t,z,r,\theta) = \int_\DD {\TYPE 2 {\mathbf S^{\perp}} {}
{}} (\epsilon,t,z,r,\theta),
$$
where  ${\TYPE 2 {\mathbf S^{\perp}} {} {}}$ is the $2$-form that
does not contain $\D z$ in ${\TYPE 2 {\mathbf S} {} {}}$. From the
electromagnetic $1$-forms, \fr{eqn:e1-basic} and \fr{eqn:h1-basic},
it follows that
\beqa
&&w(\epsilon,t,z)\non\\
&&=\sum_N\sum_{N'}V_N^E(\epsilon,t,z) I_{N'}^E(\epsilon,t,z)
\int_\DD \D\Phi_N\wedge\hash(\D z\wedge\D\Phi_{N'})\non\\
&&\quad +\sum_M \sum_{M'}V_M^H(\epsilon,t,z) I_{M'}^H(\epsilon,t,z)
\int_\DD (\hash(\D z\wedge\D\Psi_M))\wedge\D\Psi_{M'}.
\label{eqn:allorder-Poyinting}
\eeqa
Taking into account the symmetries given by 
\fr{eqn:symmetry-dphi-dpsi} and
\fr{eqn:symmetry-e-h}, and the fact that the sums over
$n',m'\in\mathbb{Z}$ are from $-\infty$ to $\infty$,  Eq.
\fr{eqn:allorder-Poyinting} can be  re-written
\beqa
w&=&\sum_N\sum_{N'}V_N^E(-1)^{n'}\overline{I_{-n',q(-n')}^E}\non\\
&&\qquad\times\int_\DD
\D\Phi_N\wedge\hash(\D z\wedge(-1)^{-n'}
\D\overline{\Phi_{-n',q'(-n')}}~)\non\\
&&
+\sum_M \sum_{M'}V_M^H (-1)^{m'}\overline{I_{-m',p'(-m')}^H}\non\\
&&\qquad
\times \int_\DD (\hash(\D z\wedge\D\Psi_M))\wedge
(-1)^{m'}\D\overline{\Psi_{-m',p'(-m')}}\non\\
&=&\sum_N\sum_{N'}V_N^{E}\overline{I_{N'}^E}
\int_{\DD}\D\Phi_N\wedge\hash(\D
z\wedge\D\overline{\Phi_{N'}})\non\\
&&+\sum_M \sum_{M'}V_M^H \overline{I_{M'}^H} \int_\DD (\hash(\D
z\wedge\D\Psi_M))\wedge\D\overline{\Psi_{M'}}.\non
\eeqa
Writing
$$
w(\epsilon,t,z)=w^{(0)}(t,z)+\epsilon w^{(1)}(t,z)+{\cal
O}(\epsilon^2),
$$
one has
\beqa
&&w^{(0)}(t,z)=\sum_N\sum_{N'}V_N^{E(0)}(t,z)I_{N'}^{E(0)}(t,z)
\int_\DD\D\Phi_N\wedge{\hashat}\D\Phi_{N'}\non\\
&&\qquad +\sum_M\sum_{M'}V_M^{H(0)}(t,z)I_{M'}^{H(0)}(t,z)
\int_\DD({\hashat}\D\Psi_M)\wedge\D\Psi_{M'},
\label{eqn:pow-0th}
\eeqa
and
\beqa
&&w^{(1)}(t,z)=\sum_N\sum_{N'} \bigg\{
V_N^{E(0)}(t,z)I_{N'}^{E(0)}(t,z)
\kappa_0(z) \non\\
&&\qquad\qquad\times
\int_\DD r\cos\theta \D\Phi_N\wedge{\hashat}\D\Phi_{N'}\non\\
&& + \bigg( V_N^{E(0)}(t,z)I_{N'}^{E(1)}(t,z) +
V_N^{E(1)}(t,z)I_{N'}^{E(0)}(t,z) \bigg) \int_\DD
\D\Phi_N\wedge{\hashat}\D\Phi_{N'}\bigg\}\non\\
&&+\sum_M\sum_{M'} \bigg\{
V_M^{H(0)}(t,z)I_{M'}^{H(0)}(t,z)\kappa_0(z)
\int_\DD r\cos\theta({\hashat}\D\Psi_M)\wedge\D\Psi_{M'}\non\\
&& +\bigg(V_M^{H(0)}(t,z)I_{M'}^{H(1)}(t,z) +
V_M^{H(1)}(t,z)I_{M'}^{H(0)}(t,z)\bigg)\non\\
&&\qquad\qquad\times
\int_\DD({\hashat}\D\Psi_M)\wedge\D\Psi_{M'}\bigg\}.\non\\
\label{eqn:pow-1st}
\eeqa
With the aid of the symmetries,
\fr{eqn:symmetry-dphi-dpsi} and \fr{eqn:symmetry-e-h}, and the fact
that
$$
\int_\DD\D\Phi_N\wedge{\hashat}\D\Phi_{N'}
\propto\delta_{n',-n},\qquad
\int_\DD({\hashat}\D\Psi_M)\wedge\D\Psi_{M'} \propto\delta_{m',-m},
$$
one obtains finally
\beq
w^{(0)}=\Re\bigg\{\sum_N \beta_N^2\NN{N}^2
V_N^{E(0)}\overline{I_N^{E(0)}} -\sum_M \alpha_M^2\MM{M}^2
V_M^{H(0)}\overline{I_M^{H(0)}} \bigg\},
\label{eqn:pow-0th-result} \eeq
where $\Re$ represents the real part of 
its argument.

Similarly the first order correction to the power is
\beqa
w^{(1)}&=&\Re\bigg\{\sum_N
\beta_N^2\NN{N}^2 (V_N^{E(0)}\overline{I_N^{E(1)}}
+V_N^{E(1)}\overline{I_N^{E(0)}})\non\\
&&-\sum_M \alpha_M^2\MM{M}^2 (V_M^{H(0)}\overline{I_M^{H(1)}}
+V_M^{H(1)}\overline{I_M^{H(0)}})\non\\
&&+\kappa_0(z) \sum_N\sum_{N'}F^{\Phi}_{N,N'}
V_N^{E(0)} \overline{I_{N'}^{E(0)}}\non\\
&&-\kappa_0(z)\sum_M\sum_{M'} F^\Psi_{M,M'} V_M^{H(0)}
\overline{I_{M'}^{H(0)}}\bigg\}.
\label{eqn:pow-1st-result}
\eeqa

The analysis above is general and accommodates arbitrary smooth
continuous conserved source currents. It should be stressed that in
general one must specify how the prescribed sources should depend on
both space and time variables so that their perturbative expansions
in $\kappa$ can be determined. In the following sections we consider
particular sources of relevance to the issues mentioned in the
introduction.


\section{Smooth Longitudinal Convective Currents}
\label{sec:traveling-charge-source1}

In this section a particular current source is considered. It is generated by an arbitrary smooth {\it convective} charge density $\rho$ that is taken independent of $\kappa$. It has components
 $J_r=J_\theta=0,\, J_0(z-vt,r,\theta)=v\,\rho(z-vt,r,\theta)$ with $v$ less than the speed of light.
 In this source  model   $J_0^{(0)}=J_0$ and all higher
orders are taken zero.
 This is a contrived current source but serves as a comparison with the localised sources that will be used to model accelerated bunches of charge in subsequent sections.

\subsection{Lowest Order System with Smooth Longitudinal Currents}
 The equations for
$\gamma_N^{H(0)}$ and $\gamma_N^{E(0)}$ follow for these sources
from above:
 \beq
\ddot{\gamma}_N^{H(0)}-c^2\gamma_N^{H(0)\prime\prime}
+c^2\alpha_N^2\gamma_N^{H(0)}=0.
\label{eqn:Sol-gammaH:0}
\eeq
\beq
\ddot{\gamma}_N^{E(0)}-c^2\gamma_N^{E(0)\prime\prime}
+c^2\beta_N^2\gamma_N^{E(0)}=-\frac{c^2\mu}{\NN{N}^2}(c^2 - v^2)
\overline{\rho^{(0)\prime}_N}.
\label{eqn:Sol-gammaE:0}
\eeq
where the  conservation relation
$$
\frac{\partial}{\partial t}{\rho}^{(0)}(z-vt)=
-v\frac{\partial}{\partial z}{\rho}^{(0)}(z-vt),
$$
has been used and
for $k\in\mathbb{Z}_{\geq 0}$
we define
\beq
\rho^{(k)}_N:=\int_\DD\rho^{(k)}\Phi_N{\hashat}1.
\label{eqn:def-rho-j-N}
\eeq

In terms of $\gamma_N^{H(0)}$ and $\gamma_N^{E(0)}$ and the
projected convective longitudinal sources the equations
\fr{eqn:E4:0}, \fr{eqn:H2:0}, \fr{eqn:E3:0} and \fr{eqn:H4:0} yield
\beqa V_N^{E(0)}&=&\frac{1}{\beta_N^2}\bigg(
\gamma_N^{E(0)\prime}-\frac{1}{\NN{N}^2\EPS}\overline{\rho^{(0)}_N}
\bigg),
\label{eqn:Sol-VE:0}\\
V_N^{H(0)}&=&\frac{\mu}{\alpha_N^2}\dot{\gamma}_N^{H(0)},
\label{eqn:Sol-VH:0}\\
I_N^{E(0)}&=&-\frac{1}{\beta_N^2}\bigg( \EPS\dot{\gamma}_N^{E(0)}
+\frac{v}{\NN{N}^2} \overline{\rho^{(0)}_N}\bigg),
\label{eqn:Sol-IE:0}\\
I_N^{H(0)}&=&\frac{1}{\alpha_N^2}\gamma_N^{H(0)\prime},
\label{eqn:Sol-IH:0}
\eeqa
and from
\fr{eqn:pow-0th-result}, \fr{eqn:Sol-VE:0},
 \fr{eqn:Sol-VH:0}, \fr{eqn:Sol-IE:0}, and
\fr{eqn:Sol-IH:0} the lowest order contribution to the power flux
becomes
\beqa
w^{(0)}(t,z)&=&\Re\bigg\{\sum_N\bigg( \frac{v
|\rho^{(0)}_N|^2}{\NN{N}^2\beta_N^2\EPS}
+\frac{\overline{\dot{\gamma}_N^{E(0)}}}{\beta_N^2}
\overline{\rho^{(0)}_N} -\frac{v\gamma_N^{E(0)\prime}}{\beta_N^2}
\rho^{(0)}_N\non\\
&&-\frac{\EPS}{\beta_N^2}\NN{N}^2
\gamma_N^{E(0)\prime}\overline{\dot{\gamma}_N^{E(0)}} \bigg)
-\mu\sum_M\frac{\MM{M}^2}{\alpha_M^2}\dot{\gamma}_M^{H(0)}
\overline{\gamma_M^{H(0)\prime}}\bigg\}.
\label{eqn:pow-0th-tra-charge}
\eeqa

\subsection{First Order System with Smooth Longitudinal Currents}
\label{sec:traveling-charge-source}
The equations for
 $\gamma_M^{H(1)}$ and
$\gamma_M^{E(1)}$  follow similarly. From \fr{eqn:H2:1},
\fr{eqn:H4:1}, \fr{eqn:H3:1} one has
 \beqa
&&\ddot{\gamma}_M^{H(1)}-c^2\gamma_M^{H(1)\prime\prime}
+c^2\alpha_M^2\gamma_M^{H(1)} \non\\
&&= -\frac{c^2\kappa_0(z)}{\MM{M}^2}
{\sum_N}' (I_N^{E(0)\prime}-\EPS\dot{V}_N^{E(0)})
G_{M,N}^{\bar{\Psi},\Phi}
-\frac{c^2\kappa_0'(z)}{\MM{M}^2}{\sum_N}'I_N^{E(0)}G_{M,N}^{\bar{\Psi},\Phi}
\non\\
&&-\frac{c^2\kappa_0(z)}{\MM{M}^2} {\sum_N}'\bigg\{
(\EPS\dot{V}_N^{H(0)}+I_N^{H(0)\prime})
(C_{M,N}^{\Psi}-D_{M,N}^{\Psi}) \non\\
&&\qquad -
(\gamma_N^{H(0)\prime\prime}+\alpha_N^2I_N^{H(0)\prime})E_{M,N}^{\Psi}
\bigg\}
+\frac{c^2\kappa_0'(z)}{\MM{M}^2}
{\sum_N}'\bigg\{
(\gamma_N^{H(0)\prime}\non\\
&&\qquad
+\alpha_N^2I_N^{H(0)}) E_{M,N}^{\Psi}-I_N^{H(0)}(C_{M,N}^{\Psi}-D_{M,N}^{\Psi}) \bigg\}.
\label{eqn:Sol-gammaH:1:smooth}
\eeqa
and  \fr{eqn:E3:1}, \fr{eqn:E4:1}, \fr{eqn:E1:1} yield
\beqa
&&\ddot{\gamma}_M^{E(1)}-c^2\gamma_M^{E(1)\prime\prime}
+c^2\beta_M^2\gamma_M^{E(1)} \non\\
&&=-\frac{c^2\kappa_0(z)}{\NN{M}^2}
{\sum_N}' (V_N^{H(0)\prime}+\mu\dot{I}_N^{H(0)})
G_{M,N}^{\bar{\Phi},\Psi}
-\frac{c^2\kappa_0'(z)}{\NN{M}^2}{\sum_N}'V_N^{H(0)}G_{M,N}^{\bar{\Phi},\Psi}
\non\\
&&
-\frac{c^2\mu\kappa_0(z)}{\NN{M}^2}{\sum_N}'
\bigg\{
(\beta_N^2\dot{I}_N^{E(0)}+\EPS\ddot{\gamma}_N^{E(0)})
-\dot{I}_N^{E(0)}(C_{M,N}^{\Phi}-D_{M,N}^{\Phi})
\bigg\}
\non\\
&&+\frac{c^2\kappa_0(z)}{\NN{M}^2} {\sum_N}'\bigg\{ (\beta_N^2
V_N^{E(0)\prime}+\gamma_N^{E(0)\prime\prime})
E_{M,N}^{\Phi}-V_N^{E(0)\prime}(C_{M,N}^{\Phi}-D_{M,N}^{\Phi})\bigg\}\non\\
&&+\frac{\mu c^4}{\NN{M}^2}
\frac{\partial}{\partial z}
 \int_\DD ( {\rho}^{(0)}\kappa_0(z)
r\cos\theta) \overline{\Phi_M}{\hashat}1
+\frac{c^2\kappa_0'(z)}{\NN{M}^2}{\sum_N}'
\bigg\{
(\beta_N^2V_N^{E(0)}\non\\
&&\qquad
+\gamma_N^{E(0)\prime}) E_{M,N}^{\Phi}
-V_N^{E(0)}(C_{M,N}^{\Phi}-D_{M,N}^{\Phi})
\bigg\}.
\label{eqn:Sol-gammaE:1:smooth}
\eeqa

The right hand sides of these expressions for $\gamma_M^{H(1)}$ and
$\gamma_M^{E(1)}$ can be rewritten in terms of the lower order
fields  $\gamma_N^{E(0)}, \gamma_N^{H(0)}$ and the sources. Thus the
following terms in \fr{eqn:Sol-gammaH:1:smooth}
can be expressed as
\beqa
&& I_N^{E(0)\prime}-\EPS\dot{V}_N^{E(0)}= -2\bigg(
\frac{\EPS}{\beta_N^2}\dot{\gamma}_N^{E(0)\prime}
+\frac{v}{\NN{N}^2\beta_N^2} \overline{\rho^{(0)\prime}_N}\bigg),
\non\\
&&\EPS\dot{V}_N^{H(0)}+I_N^{H(0)\prime} =\frac{1}{\alpha_N^2}\bigg(
\frac{1}{c^2}\ddot{\gamma}_N^{H(0)}+\gamma_N^{H(0)\prime\prime}
\bigg),\non\\
&&\gamma_N^{H(0)\prime}+\alpha_N^2I_N^{H(0)}
=2\gamma_N^{H(0)\prime}.\non
\eeqa
Similarly the following terms in
\fr{eqn:Sol-gammaE:1:smooth}
can be expressed as
\beqa
&&V_N^{H(0)\prime}+\mu\dot{I}_N^{H(0)}
=2\frac{\mu}{\alpha_N^2}\dot{\gamma}_N^{H(0)\prime},\non\\
&&
\beta_N^2\dot{I}_N^{E(0)}+\EPS\ddot{\gamma}_N^{E(0)}=
\frac{v^2}{\NN{N}^2}\overline{\rho_N^{(0)\prime}}
,\non\\
&&
\dot{I}_N^{E(0)}=-\frac{1}{\beta_N^2}
\bigg(
\EPS\ddot{\gamma}_N^{E(0)}-\frac{v^2}{\NN{N}^2}\overline{\rho_N^{(0)\prime}}
\bigg),\non\\
&&\beta_N^2V_N^{E(0)}+\gamma_N^{E(0)\prime}
=2\gamma_N^{E(0)\prime}-\frac{1}{\NN{N}^2\EPS}
\overline{\rho^{(0)}_N},\non\\
&&V_N^{E(0)}=\frac{1}{\beta_N^2} \bigg(
\gamma_N^{E(0)\prime}-\frac{1}{\NN{N}^2\EPS}
\overline{\rho^{(0)}_N}\bigg).\non
\eeqa
Finally once
$\gamma_M^{H(1)}$ and $\gamma_M^{E(1)}$ have been determined from
these decoupled equations it follows from \fr{eqn:E4:1},
\fr{eqn:H2:1}, \fr{eqn:E3:1}, \fr{eqn:H4:1} that the remaining
fields can be readily determined
as:
\beqa
V_M^{E(1)}&=&\frac{1}{\beta_M^2}\bigg[
\gamma_M^{E(1)\prime}+\frac{1}{\EPS\NN{M}^2}
\int_{\DD}({\rho}^{(0)}\kappa_0(z) r\cos\theta)
\overline{\Phi_M}{\hashat}1\non\\
&&+\frac{\kappa_0(z)}{\NN{M}^2}{\sum_N}' \bigg\{
(V_N^{E(0)}\beta_N^2+\gamma_N^{E(0)\prime})E_{M,N}^{\Phi}
-V_N^{E(0)}(C_{M,N}^{\Phi}-D_{M,N}^{\Phi}) \bigg\}
\bigg],\non\\
V_M^{H(1)}&=&\frac{1}{\alpha_M^2}\bigg\{
\mu\dot{\gamma}_M^{H(1)}+\frac{\kappa_0(z)}{\MM{M}^2} {\sum_N}'
V_N^{H(0)}(C_{M,N}^{\Psi}-D_{M,N}^{\Psi})
\bigg\},\non\\
I_M^{E(1)}&=&-\frac{1}{\beta_M^2}\bigg[
\EPS\dot{\gamma}_M^{E(1)}+\frac{\kappa_0(z)}{\NN{M}^2}{\sum_N}' \bigg\{
(\beta_N^2I_N^{E(0)}+\EPS\dot{\gamma}_N^{E(0)})E_{M,N}^{\Phi}\non\\
&&-I_N^{E(0)}(C_{M,N}^{\Phi}-D_{M,N}^{\Phi}) \bigg\}
\bigg],\non\\
I_M^{H(1)}&=&\frac{1}{\alpha_M^2}\bigg[
\gamma_M^{H(1)\prime}+\frac{\kappa_0(z)}{\MM{M}^2}{\sum_N}' \bigg\{
(\gamma_N^{H(0)\prime}+\alpha_N^2 I_N^{H(0)})E_{M,N}^{\Psi}\non\\
&&-I_N^{H(0)}(C_{M,N}^{\Psi}-D_{M,N}^{\Psi}) \bigg\} \bigg].\non
\eeqa

\subsubsection{Radiated  Power Dependence on Local Curvature}

A straightforward but tedious calculation leads to
the dependence on
local curvature of the radiated power  \fr{eqn:pow-1st-result} for
the convective source model above, in terms of
$\gamma_N^{H(0)}$,$\gamma_N^{E(0)}$,
$\gamma_N^{H(1)}$,$\gamma_N^{E(1)}$, various overlap coefficients
and projections of $\rho$.

Each term in \fr{eqn:pow-1st-result} can be reduced as follows:
\beqa
&&\sum_N\beta_N^2\NN{N}^2V_N^{E(0)}\overline{I_N^{E(1)}}=
-\sum_N\frac{\NN{N}^2}{\beta_N^2}\bigg(
\gamma_N^{E(0)\prime}-\frac{1}{\NN{N}^2\EPS} \overline{\rho^{(0)}_N}
\bigg)\non\\
&&\times\bigg[ \EPS\overline{\dot{\gamma}_N^{E(1)}}
+\frac{\kappa_0(z)}{\NN{N}^2} {\sum_M}'\bigg\{
-\frac{E_{M,N}^{\Phi}v}{\NN{N}^2}
\rho^{(0)}_M\non\\
&&\quad +\frac{C_{N,M}^{\Phi}-D_{N,M}^{\Phi}}{\beta_M^2}
\bigg(\EPS\overline{\dot{\gamma}_M^{E(0)}}+\frac{v}{\NN{M}^2}
\rho^{(0)}_M \bigg) \bigg\}  \bigg],
\label{eqn:pow-1st-explicit-1}
\eeqa
\beqa
&&\sum_N\beta_N^2\NN{N}^2V_N^{E(1)}\overline{I_N^{E(0)}}\non\\
&&= -\sum_N\frac{\NN{N}^2}{\beta_N^2}\bigg[ \gamma_N^{E(1)\prime}
+\frac{1}{\EPS\NN{N}^2} \int_\DD
{\rho}^{(0)}\kappa_0(z) r\cos\theta
\overline{\Phi_N}{\hashat}1\non\\
&&\quad +\frac{\kappa_0(z)}{\NN{N}^2}{\sum_M}'\bigg\{
\bigg(2E_{N,M}^{\Phi}-\frac{C_{M,N}^{\Phi}-D_{M,N}^{\Phi}}{\beta_M^2}\bigg)
\gamma_M^{E(0)\prime}\non\\
&&\quad +\bigg(
-E_{N,M}^{\Phi}+\frac{C_{N,M}^{\Phi}-D_{N,M}^{\Phi}}{\beta_M^2}\bigg)
\frac{\overline{\rho^{(0)}_M}}{\NN{M}^2} \bigg\}\bigg]\bigg(
\overline{\dot{\gamma}_N^{E(0)}}+\frac{v {\rho}^{(0)}_N}{\NN{N}^2
\EPS} \bigg), \label{eqn:pow-1st-explicit-2}
\eeqa
\beqa &&\sum_M\alpha_M^2\MM{M}^2V_M^{H(0)}\overline{I_M^{H(1)}}=
\mu\sum_M\frac{\MM{M}^2}{\alpha_M^2}\dot{\gamma}_M^{H(0)} \bigg\{
\overline{\gamma_M^{H(1)\prime}}\non\\
&&\qquad\qquad +\frac{\kappa_0(z)}{\MM{M}^2}{\sum_N}'
\bigg(2E_{M,N}^{\Psi}-\frac{C_{M,N}^{\Psi}-D_{M,N}^{\Psi}}{\alpha_N^2}
\bigg)\overline{\gamma_N^{H(0)\prime}} \bigg\},
\label{eqn:pow-1st-explicit-3}
\eeqa
\beqa
&&\sum_M\alpha_M^2\MM{M}^2V_M^{H(1)}\overline{I_M^{H(0)}}\non\\
&&=\mu\sum_M\frac{\MM{M}^2}{\alpha_M^2} \bigg(
\dot{\gamma}_M^{H(1)}+\frac{\kappa_0(z)}{\MM{M}^2}{\sum_N}'
\frac{C_{M,N}^{\Psi}-D_{M,N}^{\Psi}}{\alpha_N^2}
\dot{\gamma}_{N}^{H(0)}
\bigg)\overline{\gamma_M^{H(0)\prime}},
\label{eqn:pow-1st-explicit-4}
\eeqa
\beqa
&&\sum_N\sum_{N'}F_{N,N'}^{\Phi}V_N^{E(0)}
\overline{I_{N'}^{E(0)}}\non\\
&&=-\sum_N\sum_{N'} \frac{F_{N,N'}^{\Phi}}{\beta_N^2\beta_{N'}^2}
\bigg(\gamma_N^{E(0)\prime}-\frac{\overline{\rho^{(0)}_N}}{\NN{N}^2\EPS}
\bigg)\bigg( \EPS\overline{\dot{\gamma}_{N'}^{E(0)}}
+\frac{v\rho^{(0)}_{N'}}{\NN{N'}^2} \bigg).
\label{eqn:pow-1st-explicit-5}
\eeqa
Finally,
\beq
\sum_M\sum_{M'}F_{M,M'}^{\Psi}V_{M}^{H(0)}
\overline{I_{M'}^{H(0)}} =\mu\sum_M\sum_{M'}F_{M,M'}^{\Psi}
\frac{\dot{\gamma}_M^{H(0)}\overline{\gamma_{M'}^{H(0)\prime}}}{\alpha_M^2\alpha_{M'}^2}.
\label{eqn:pow-1st-explicit-6}
\eeq

\section{Moving Point Charge Source}
\label{sec:Traveling-point-charge-source-infinite}

In this section the above formalism is applied to the determination
of the fields produced by the motion of a single point charge
source moving in a curved beam pipe.
 A full account would involve generalising the formalism to
accommodate  fields from point distributional sources. Following established
custom  the  formalism is extended here  by  modelling a point
source  as a  moving Dirac distribution on $\cal U$.  One can then
explicitly remove some of the integrations that arise in the smooth
continuous charge source model. However a point charge implies
singularities in the electromagnetic fields and these should not be
ignored.   The distributional charge density must model a
charged particle with constant electric charge $Q_{tot}$ so we
demand that
\beq
\int_{{\cal U}}{\rho}\hash 1 = Q_{tot}.
\label{eqn:const-e-charge}
\eeq
Furthermore  the motion of the charge is maintained  (e.g. by
externally applied magnetostatic fields) on a curved path parallel
to the {\it design-orbit} with curvature $\kappa(z)$ and constant speed
$v$. These conditions are satisfied if
$$
{\rho}(\epsilon,t,z,x_1,x_2)= Q(\epsilon,t)
\delta(x_1-x_{1,0})\delta(x_2-x_{2,0}) \delta(z-vt),
$$
with \BE{ Q(\epsilon,t):=\frac{Q_{tot}}{1-\epsilon\kappa_0(vt)
x_{1,0}} =Q_{tot}+\epsilon\kappa_0(vt) x_{1,0}Q_{tot}+{\cal
O}(\epsilon^2),\label{QQ} }
 in terms of the symbolic representation of the
Cartesian three-dimensional Dirac distribution with moving point
support at
\newline
$
(x_{1,0},x_{2,0}, vt)=(r_0\cos\theta_0,r_0\sin\theta_0, vt),
$
determining the location of the point charge in $\U$ at time $t$.

Thus it follows from (\ref{QQ}) that
\beqa
{\rho}&=&{\rho}^{(0)}+\epsilon{\rho}^{(1)}
+{\cal O}(\epsilon^2),\non\\
{\rho}^{(0)}(z-vt,x_1,x_2)&=&Q_{tot}
\delta(x_1-x_{1,0})\delta(x_2-x_{2,0})
\delta(z-vt),\non\\
{\rho}^{(1)}(t,z,x_1,x_2)&=&\kappa_0(vt) x_{1,0} Q_{tot}
\delta(x_1-x_{1,0})\delta(x_2-x_{2,0}) \delta(z-vt).\non
\eeqa
In adapted coordinates
$$
({\rho}{\hashat}1)(\epsilon,t,z,r,\theta)=
Q(\epsilon,t)\frac{\delta(r-r_0)}{r}\delta(\theta-\theta_0) \delta
(z-vt) r\D r\wedge\D\theta.
$$
The associated electric current components are $J_r=J_\theta=0$ and
 \beqa
&&J_0^{(0)}=v
{\rho}^{(0)}=vQ_{tot}
\delta(x_1-x_{1,0})\delta(x_2-x_{2,0})\delta(z-vt),
\label{eqn:TPS-J0}\\
&&J_0^{(1)}=v {\rho}^{(1)}=v\kappa_0(vt) x_{1,0} Q_{tot}\non\\
&&\qquad\qquad\qquad \times\delta(x_1-x_{1,0})\delta(x_2-x_{2,0})
\delta(z-vt). \label{eqn:TPS-J1}
\eeqa
Note that in this source
model both  $J_0^{(0)}, J_0^{(1)}$ and $\rho^{(0)}, \rho^{(1)}$ are
non-zero with all higher orders zero.

\subsection{Lowest Order Fields from a Moving Point Charge}
Using this distributional model the projected sources in
\fr{eqn:Sol-gammaE:0} can be evaluated  as
\beq
\int_{\DD}
\frac{\partial}{\partial z}{\rho}^{(0)}(z-vt,r,\theta)
\overline{\Phi_N}{\hashat}1 =
Q_{tot}J_n\left(x_{q(n)}\frac{r_0}{a}\right)
e^{-in\theta_0} \delta'(z-vt).
\label{eqn:int-rho-diff-phi}
\eeq
so the solutions for
$\gamma_N^{H(0)}$ and $\gamma_N^{E(0)}$ following from
\fr{eqn:Pinsky} are
\beqa
&&\gamma_N^{H(0)}(t,z)={\cal H}_{\alpha_N}
 [\gamma_N^{H(0)init}](t,z),\non\\
&&\gamma_N^{E(0)}(t,z)={\cal H}_{\beta_N}
 [\gamma_N^{E(0)init}](t,z)\non\\
 &&\qquad\qquad
 -\frac{c^2\mu}{\NN{N}^2}(c^2-v^2)Q_{tot}J_n\left(x_{q(n)}\frac{r_0}{a}\right)
e^{-in\theta_0}\Upsilon_N(t,z),\non
\eeqa
where
$$
\Upsilon_N(t,z):=\int_0^{t}dt'\int_{z-c(t-t')}^{z+c(t-t')}d\zeta
\delta'(\zeta-vt') J_0(\beta_N\sqrt{c^2(t-t')^2-(z-\zeta)^2}~)
$$
and ${\cal H}_{\alpha_N}$ is given by \fr{eqn:def-H}.
For $v>0$ (See Fig.\ref{fig:range-Upsilon}) it is shown in Appendix B that:
\begin{itemize}
\item{for $z \leq -ct$, (Outside  the Regions $R_1$ and $R_2$)}
\beq \Upsilon_N(t,z)=0,
\label{eqn:upsilon-before}
\eeq
\item{for $-ct<z\leq vt$, (Region $R_2$)}
\beq
\Upsilon_N(t,z)=\beta_N\int_0^{t'_+(t,z)} dt'
\frac{(z-vt')J_1(\beta_N s(t,z|t'))}{s(t,z|t')},
\label{eqn:upsilon-before-vt}
\eeq
\item{for $vt\leq z< ct$, (Region $R_1$)}
\beq
\Upsilon_N(t,z)=\beta_N\int_0^{t'_-(t,z)} dt'
\frac{(z-vt')J_1(\beta_N s(t,z|t'))}{s(t,z|t')},
\label{eqn:upsilon-after-vt}
\eeq
\item{for $z\geq ct$, (Outside the regions $R_1$ and $R_2$)}
\beq
\Upsilon_N(t,z)=0,
\label{eqn:upsilon-after}
\eeq
\end{itemize}
where
\BAE{
t'_+(t,z):=\frac{t+z/c}{1+v/c},\qquad
t'_-(t,z):=\frac{t-z/c}{1-v/c},\\
\mbox{and}\qquad s(t,z|t'):=\sqrt{c^2(t-t')^2-(z-vt')^2}.
\label{eqn:abb-int}
}
The field components
$V_N^{E(0)},V_N^{H(0)},I_N^{E(0)},I_N^{H(0)}$ follow from
\fr{eqn:GEN-Sol-VE:0}, \fr{eqn:GEN-Sol-VH:0}, \fr{eqn:GEN-Sol-IE:0},
\fr{eqn:GEN-Sol-IH:0} in terms of the solutions for
$\gamma_N^{H(0)},\gamma_N^{E(0)}$.
\begin{figure}[h]
\unitlength 1mm
\begin{picture}(160,75)
\put(10,0){\includegraphics[width=12.0cm]{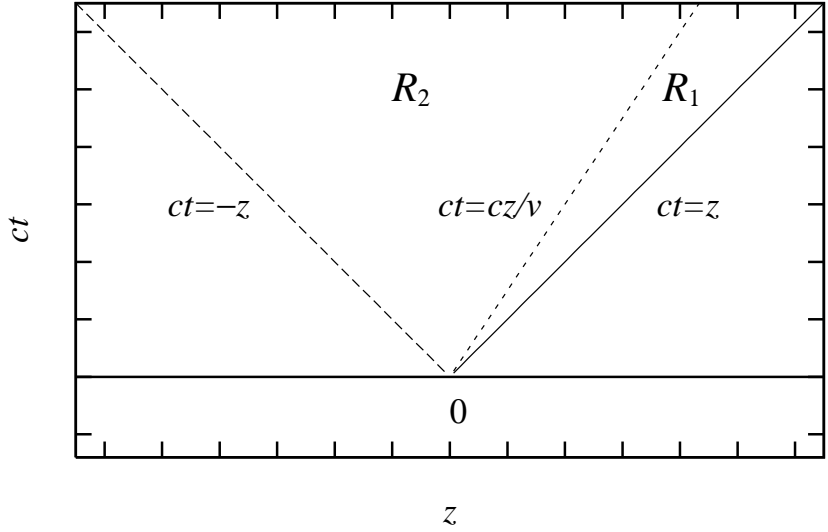}}
\end{picture}
\caption{
Domains for $\Upsilon_N(t,z)$  and $A_\sigma(t,z)$, $B_\sigma(t,z)$ in \S\ref{sec:1st-point}. $R_1$ is the triangle in the upper
  right-hand corner $(vt\leq z<ct)$.
 $R_2$ is the adjacent large triangle
$(-ct<z\leq vt)$ where the source world-line has $z=vt$.
}
\label{fig:range-Upsilon}
\end{figure}

\subsubsection{Lowest Order Contribution to the Instantaneous  Power
  from a Moving Point Charge}

We note from \fr{eqn:pow-0th-tra-charge} that the lowest order
contribution $w^{(0)}$  to the instantaneous power contains terms
derived  from  the projections
 \beq
\int_{\DD}{\rho}^{(0)}(z-vt,r,\theta)\overline{\Phi_N}{\hashat}1
=Q_{tot}J_n\left(x_{q(n)}\frac{r_0}{a}\right)e^{-in\theta_0}\delta(z-v t).
\label{eqn:integral-trave-point-charg}
\eeq
Such distributional
contributions  are absent in pipe cross-sections that do not contain
the point source.  Thus for sections with $z\neq vt$ the
instantaneous power is
given to lowest order as
\beq
\breve{w}^{(0)}
=-\Re\bigg\{\EPS\sum_N\frac{\NN{N}^2}{\beta_N^2}
\gamma_N^{E(0)\prime}\overline{\dot{\gamma}_N^{E(0)}}
+\mu\sum_M\frac{\MM{M}^2}{\alpha_M^2}\dot{\gamma}_M^{H(0)}
\overline{\gamma_M^{H(0)\prime}}\bigg\},
\non
\eeq
explicitly in terms of the above solutions. This is
independent of the curvature $\kappa$.

\subsection{First Order Fields from a Moving Point Charge}
\label{sec:1st-point}
The explicit computation of the contribution of the radiated power
to the next order, involving $\kappa$, is somewhat  more complicated
since it requires the solutions to the zeroth-order fields as
sources for the first-order fields as well as evaluation of
projected sources that now involve $\rho^{(1)}$. The  equations for
$\gamma_M^{H(1)}$ and $\gamma_M^{E(1)}$ also contain additional
source terms from these first order contributions to the sources.
From \fr{eqn:H2:1}, \fr{eqn:H4:1}, \fr{eqn:H3:1} one has
\beqa
&&\ddot{\gamma}_M^{H(1)}-c^2\gamma_M^{H(1)\prime\prime}
+c^2\alpha_M^2\gamma_M^{H(1)} \non\\
&&= -\frac{c^2\kappa_0(z)}{\MM{M}^2}
{\sum_N}' (I_N^{E(0)\prime}-\EPS\dot{V}_N^{E(0)})
G_{M,N}^{\bar{\Psi},\Phi}
-\frac{c^2\kappa_0'(z)}{\MM{M}^2}
{\sum_N}'I_N^{E(0)}G_{M,N}^{\bar{\Psi},\Phi}
\non\\
&&-\frac{c^2\kappa_0(z)}{\MM{M}^2} {\sum_N}'\bigg\{
(\EPS\dot{V}_N^{H(0)}+I_N^{H(0)\prime})
(C_{M,N}^{\Psi}-D_{M,N}^{\Psi}) \non\\
&&\qquad
-(\gamma_N^{H(0)\prime\prime}+\alpha_N^2I_N^{H(0)\prime})E_{M,N}^{\Psi}
\bigg\}+\frac{c^2\kappa_0'(z)}{\MM{M}^2}{\sum_N}'\bigg\{
(\gamma_N^{H(0)\prime}\non\\
&&\qquad
+\alpha_N^2 I_N^{H(0)})E_{M,N}^{\Psi}
-I_N^{H(0)}(C_{M,N}^{\Psi}-D_{M,N}^{\Psi})
\bigg\},
\label{eqn:Sol-gammaH:1}
\eeqa
and  \fr{eqn:E3:1}, \fr{eqn:E4:1}, \fr{eqn:E1:1}
yield
\beqa
&&\ddot{\gamma}_M^{E(1)}-c^2\gamma_M^{E(1)\prime\prime}
+c^2\beta_M^2\gamma_M^{E(1)} \non\\
&&=-\frac{c^2\kappa_0(z)}{\NN{M}^2}
{\sum_N}' (V_N^{H(0)\prime}+\mu\dot{I}_N^{H(0)})
G_{M,N}^{\bar{\Phi},\Psi}
-\frac{c^2\kappa_0'(z)}{\NN{M}^2}{\sum_N}'V_N^{H(0)}G_{M,N}^{\bar{\Phi},\Psi}
\non\\
&&
-\frac{\mu c^2\kappa_0(z)}{\NN{M}^2} {\sum_N}'
\bigg\{
(\beta_N^2\dot{I}_N^{E(0)}+\EPS\ddot{\gamma}_N^{E(0)}) E_{M,N}^{\Phi}
-\dot{I}_N^{E(0)} (C_{M,N}^{\Phi}-D_{M,N}^{\Phi})
\bigg\}
\non\\
&&+\frac{c^2\kappa_0(z)}{\NN{M}^2} {\sum_N}'\bigg\{ (\beta_N^2
V_N^{E(0)\prime}+\gamma_N^{E(0)\prime\prime})
E_{M,N}^{\Phi}-V_N^{E(0)\prime}(C_{M,N}^{\Phi}-D_{M,N}^{\Phi})\bigg\}\non\\
&&-\frac{c^2}{\EPS\NN{M}^2}
\int_\DD \{ {\rho}^{(1)\prime}
-(\rho^{(0)}\kappa_0'(z)+{\rho}^{(0)\prime}\kappa_0(z)) r\cos\theta\}
\overline{\Phi_M}{\hashat}1\non\\
&&\quad -\frac{\mu c^2 v}{\NN{M}^2}\overline{\dot{\rho}^{(1)}_M}
+\frac{c^2\kappa_0'(z)}{\NN{M}^2}{\sum_N}'\bigg\{
(\beta_N^2V_N^{E(0)}+\gamma_N^{E(0)\prime})E_{M,N}^{\Phi}\non\\
&&\qquad -V_N^{E(0)}(C_{M,N}^{\Phi}-D_{M,N}^{\Phi})
\bigg\}.
\label{eqn:Sol-gammaE:1}
\eeqa
The right hand sides of these
expressions for $\gamma_M^{H(1)}$ and $\gamma_M^{E(1)}$ can be
rewritten in terms of the lower order fields  $\gamma_N^{E(0)},
\gamma_N^{H(0)}$ and the sources. Thus the following terms in
\fr{eqn:Sol-gammaH:1} can be expressed as
\beqa
&&
I_N^{E(0)\prime}-\EPS\dot{V}_N^{E(0)}= -2\bigg(
\frac{\EPS}{\beta_N^2}\dot{\gamma}_N^{E(0)\prime}
+\frac{v}{\NN{N}^2\beta_N^2} \overline{\rho^{(0)\prime}_N}\bigg),
\non\\
&&\EPS\dot{V}_N^{H(0)}+I_N^{H(0)\prime} =\frac{1}{\alpha_N^2}\bigg(
\frac{1}{c^2}\ddot{\gamma}_N^{H(0)}+\gamma_N^{H(0)\prime\prime}
\bigg),\non\\
&&\gamma_N^{H(0)\prime}+\alpha_N^2I_N^{H(0)}
=2\gamma_N^{H(0)\prime}.\non
\eeqa
Similarly the following
terms in \fr{eqn:Sol-gammaE:1} can be expressed as
\beqa
&&V_N^{H(0)\prime}+\mu\dot{I}_N^{H(0)}
=2\frac{\mu}{\alpha_N^2}\dot{\gamma}_N^{H(0)\prime},\non\\
&&
\beta_N^2\dot{I}_N^{E(0)}+\EPS\ddot{\gamma}_N^{E(0)}=
\frac{v^2}{\NN{N}^2}\overline{\rho_N^{(0)\prime}}
,\non\\
&&
\dot{I}_N^{E(0)}=-\frac{1}{\beta_N^2}
\bigg(
\EPS\ddot{\gamma}_N^{E(0)}-\frac{v^2}{\NN{N}^2}\overline{\rho_N^{(0)\prime}}
\bigg)
,\non\\
&&\beta_N^2V_N^{E(0)\prime}+\gamma_N^{E(0)\prime\prime}
=2\gamma_N^{E(0)\prime\prime}-\frac{1}{\NN{N}^2\EPS}
\overline{\rho^{(0)\prime}_N},\non\\
&&V_N^{E(0)\prime}=\frac{1}{\beta_N^2} \bigg(
\gamma_N^{E(0)\prime\prime}-\frac{1}{\EPS}
\overline{\rho^{(0)\prime}_N}\bigg).\non
\eeqa
Finally once
$\gamma_M^{H(1)}$ and $\gamma_M^{E(1)}$ have been determined from
these decoupled equations it follows from \fr{eqn:E4:1},
\fr{eqn:H2:1}, \fr{eqn:E3:1}, \fr{eqn:H4:1} that the remaining
fields can be readily determined:
 \beqa
V_M^{E(1)}&=&\frac{1}{\beta_M^2}\bigg[
\gamma_M^{E(1)\prime}-\frac{1}{\EPS\NN{M}^2}
\int_{\DD}({\rho}^{(1)}-{\rho}^{(0)}\kappa_0(z) r\cos\theta)
\overline{\Phi_M}{\hashat}1\non\\
&&+\frac{\kappa_0(z)}{\NN{M}^2}{\sum_N}' \bigg\{
(V_N^{E(0)}\beta_N^2+\gamma_N^{E(0)\prime})E_{M,N}^{\Phi}
-V_N^{E(0)}(C_{M,N}^{\Phi}-D_{M,N}^{\Phi} )\bigg\}
\bigg],\non\\
V_M^{H(1)}&=&\frac{1}{\alpha_M^2}\bigg\{
\mu\dot{\gamma}_M^{H(1)}+\frac{\kappa_0(z)}{\MM{M}^2} {\sum_N}'
V_N^{H(0)}(C_{M,N}^{\Psi}-D_{M,N}^{\Psi})
\bigg\},\non\\
I_M^{E(1)}&=&-\frac{1}{\beta_M^2}\bigg[
\EPS\dot{\gamma}_M^{E(1)}+\frac{\kappa_0(z)}{\NN{M}^2}{\sum_N}' \bigg\{
(\beta_N^2I_N^{E(0)}+\EPS\dot{\gamma}_N^{E(0)})E_{M,N}^{\Phi}\non\\
&&-I_N^{E(0)}(C_{M,N}^{\Phi}-D_{M,N}^{\Phi})
\bigg\}+\frac{v}{\NN{M}^2}\overline{\rho^{(1)}_M}
\bigg],\non\\
I_M^{H(1)}&=&\frac{1}{\alpha_M^2}\bigg[
\gamma_M^{H(1)\prime}+\frac{\kappa_0(z)}{\MM{M}^2}{\sum_N}' \bigg\{
(\gamma_N^{H(0)\prime}+\alpha_N^2 I_N^{H(0)})E_{M,N}^{\Psi}\non\\
&&-I_N^{H(0)}(C_{M,N}^{\Psi}-D_{M,N}^{\Psi}) \bigg\} \bigg].\non
\eeqa
The  distributional nature of the source permits evaluation of
the integrals in \fr{eqn:Sol-gammaH:1} and
\fr{eqn:Sol-gammaE:1} yielding
\beqa
&&\int_\DD
\bigg\{{\rho}^{(1)\prime}
-\bigg(
\rho^{(0)}\kappa_0'(z)+{\rho}^{(0)\prime}\kappa_0(z)
\bigg)
r\cos\theta\bigg\}
\overline{\Phi_M}{\hashat}1\non\\
&&=
r_0\cos\theta_0 Q_{tot}J_m\left(x_{p(m)}\frac{r_0}{a}\right)
e^{-im\theta_0}\non\\
&&\times
\bigg\{
\delta'(z-vt)\bigg(\kappa_0(vt)-\kappa_0(z)\bigg)
-\delta(z-vt)\kappa_0'(z)\bigg\},
\non
\eeqa
and
\beqa
v\int_\DD\dot{\rho}^{(1)}\overline{\Phi_M}{\hashat}1
&=&v^2 x_{1,0} Q_{tot}J_m\left(x_{p(m)}\frac{r_0}{a}\right)
e^{-im\theta_0} \non\\
&&\times\bigg(-\kappa_0(vt)\delta'(z-vt)+\kappa'_0(vt)\delta(z-vt)\bigg)
\non
\eeqa
where
$\kappa_0'(vt)=d\kappa_0(z)/dz|_{z=vt}$.

The solutions for $\gamma_M^{H(1)}$ and $\gamma_M^{E(1)}$ follow
similarly from \fr{eqn:Pinsky}. The source term in the equation for
$\gamma_M^{H(1)}$ is
$$
g_M^{H(1)}(t,z)=K_M^{H(1)}(t,z)
+L_M^{H(1)}(t,z),
$$
where, for $\kappa_0$ varying with $z$, \footnote{Note that $L_M^{H(1)}$
 and $S_M^{H(1)}$ come from
$\overline{\rho_N^{(0)\prime}}$ and $\overline{\rho_N^{(0)}}$, respectively.}
\beqa
&&
K_M^{H(1)}(t,z):=\frac{c^2\kappa_0(z)}{\MM{M}^2}{\sum_N}'
\bigg\{2G_{M,N}^{\bar{\Psi},\Phi}
\frac{\EPS}{\beta_N^2}\dot{\gamma}_N^{E(0)\prime}\non\\
&&\qquad\qquad
-\frac{C_{M,N}^{\Psi}-D_{M,N}^{\Psi}}{\alpha_N^2}\bigg(
\frac{\ddot{\gamma}_N^{H(0)}}{c^2}+\gamma_N^{H(0)\prime\prime}
\bigg)+2E_{M,N}^{\Psi}\gamma_N^{H(0)\prime\prime} \bigg\},\non\\
&&
\qquad\qquad +\frac{c^2\kappa_0'(z)}{\MM{M}^2}{\sum_N}'
\bigg\{
\frac{\EPS}{\beta_N^2}G_{M,N}^{\bar{\Psi},\Phi}\dot{\gamma}_N^{E(0)}
\non\\
&&
\qquad\qquad +\bigg(
2E_{M,N}^{\Psi}-\frac{C_{M,N}^{\Psi}-D_{M,N}^{\Psi}}{\alpha_N^2}
\bigg)\gamma_N^{H(0)\prime}
\bigg\},
\label{eqn:K-M-H:1}\\
&&
L_M^{H(1)}(t,z):=\frac{2c^2v Q_{tot}}{\MM{M}^2}
\kappa_0(z)\delta'(z-vt) {\sum_N}'
\frac{G_{M,N}^{\bar{\Psi},\Phi}}{\NN{N}^2\beta_N^2}
J_n\left(x_{q(n)}\frac{r_0}{a}\right)e^{-in\theta_0},\non \\
&&
S_M^{H(1)}(t,z):=\frac{c^2 vQ_{tot}}{\MM{M}^2}\kappa_0'(vt)\delta(z-vt)
{\sum_N}'\frac{G_{M,N}^{\bar{\Psi},\Phi}}{\NN{N}^2\beta_N^2}
J_n\left(x_{q(n)}\frac{r_0}{a}\right)e^{-in\theta_0}.
\non
\eeqa
The source
term in the equation for $\gamma_M^{E(1)}$ is
$$
g_M^{E(1)}(t,z)=K_M^{E(1)}(t,z)+
L_M^{E(1)}(t,z)+P_M^{E(1)}(t,z)+R_M^{E(1)}(t,z)
+S_M^{E(1)}(t,z),
$$
where \footnote{
The terms proportional to
$L_M^{E(1)}$,
$P_M^{E(1)}$, $R_M^{E(1)}$ and
$S_M^{E(1)}$ arise from
$\overline{\rho_N^{(0)\prime}}$,
$\overline{\dot{\rho}_M^{(1)}}$,
$\{\rho^{(1)\prime}-(\rho^{(0)}\kappa_0'
+\rho^{(0)\prime}\kappa_0) r\cos\theta\}$ and
$\overline{\rho_N^{(0)}}$,
 respectively.}

\beqa
&&K_M^{E(1)}(t,z):=\frac{c^2 \kappa_0(z)}{\NN{M}^2}
{\sum_N}'\bigg\{ \frac{-2\mu }{\alpha_N^2}
G_{M,N}^{\bar{\Phi},\Psi}{\dot{\gamma}_N^{H(0)\prime}}
-\frac{C_{M,N}^{\Phi}-D_{M,N}^{\Phi}}{c^2\beta_N^2}
\ddot{\gamma}^{E(0)}_N
\non\\
&&\qquad\qquad +
\bigg(2E_{M,N}^{\Phi}
-\frac{C_{M,N}^{\Phi}-D_{M,N}^{\Phi}}{\beta_N^2}\bigg)
\gamma_N^{E(0)\prime\prime}\bigg\}\non\\
&&\qquad
+\frac{c^2\kappa_0'(z)}{\NN{M}^2}{\sum_N}'
\bigg\{
\frac{-\mu}{\alpha_N^2}G_{M,N}^{\bar{\Phi},\Psi}\dot{\gamma}_N^{H(0)}
\non\\
&&\qquad +\bigg(2E_{M,N}^{\Phi}
-\frac{C_{M,N}^{\Phi}-D_{M,N}^{\Phi}}{\beta_N^2}
\bigg)\gamma_N^{E(0)\prime}
\bigg\}
\label{eqn:K-M-E-1},\\
&&
L_M^{E(1)}(t,z):=
\frac{\mu c^2 (c^2+v^2) Q_{tot}}{\NN{M}^2}
\kappa_0(z)\delta'(z-vt)\non\\
&&\qquad\qquad \times{\sum_N}' \frac{1}{\NN{N}^2}
\bigg(
-E_{M,N}^{\Phi}
+\frac{C_{M,N}^{\Phi}-D_{M,N}^{\Phi}}{\beta_N^2} \bigg)
J_n\left(x_{q(n)}\frac{r_0}{a}\right)e^{-in\theta_0},\non\\
&&
P_M^{E(1)}(t,z):= -\frac{\mu c^2 v^2
Q_{tot}}{\NN{M}^2}r_0\cos\theta_0
J_m\left(x_{p(m)}\frac{r_0}{a}\right)e^{-im\theta_0}\non\\
&& \qquad\qquad \times
\bigg\{\kappa_0'(vt)\delta(z-vt)-\kappa_0(vt)\delta'(z-vt)
\bigg\},\non \\
&&
R_M^{E(1)}(t,z):=-\frac{\mu c^4 Q_{tot}}{\NN{M}^2}
r_0\cos\theta_0 J_m\left(x_{p(m)}\frac{r_0}{a}\right)e^{-im\theta_0} \non\\
&&\qquad\qquad
\times
\bigg(\delta'(z-vt)\{\kappa_0(vt)-\kappa_0(z)\}
-\delta(z-vt)\kappa_0'(z)\bigg),
\non\\
&&
S_M^{E(1)}(t,z):=
\frac{\mu c^4 Q_{tot}}{\NN{M}^2}\kappa_0'(z)\delta(z-vt)\non\\
&&\qquad\qquad
\times{\sum_N}'\frac{1}{\NN{N}^2}\bigg(
-E_{M,N}^{\Phi}+\frac{C_{M,N}^{\Phi}-D_{M,N}^{\Phi}}{\beta_N^2}
\bigg)J_n\left(x_{q(n)}\frac{r_0}{a}\right)e^{-in\theta_0}
.\non
\eeqa


In the next subsection, \S\ref{sec:wake},
the fields associated with an {\it ultra-relativistic} point
source will be of interest.
Then the contributions to the solution $\gamma_N^{E(0)}$
that depend on $Q_{tot}$  tend to zero and
the solution $\gamma_N^{H(0)}$ depends only on external and
static magnetic fields since there are no magnetic charges
in existence. In this limit the source contributions  $K_M^{H(1)}$ and
$K_M^{E(1)}$ to $g_M^{H(1)}$  and $g_M^{E(1)}$  drop out.

In general,  an analytic form for $\gamma_M^{H(1)}$ follows
from \fr{eqn:def-H} and \fr{eqn:def-I},
by applying the integral operators ${\cal H}_{\alpha_M}$
and ${\cal I}_{\alpha_M}$ to  the source functions $K_M^{E(1)}$,
$L_M^{E(1)},P_M^{E(1)},R_M^{E(1)},S_M^{E(1)}$.
These source functions are  simple functions of $z,t$
multiplied by (complex) numerical coefficients. Thus with
\beqa
f_1(t,z)&:=&\kappa_0(z)\delta'(z-vt),
\label{eqn:def-f1}\\
f_2(t,z)&:=&\kappa_0'(vt)\delta(z-vt),
\label{eqn:def-f2}\\
f_3(t,z)&:=&\kappa_0(vt)\delta'(z-vt),
\label{eqn:def-f3}
\eeqa
we write the application of ${\cal I}_{\sigma}$  on them as
\beqa
{\cal I}_\sigma[f_1](t,z)&=&
-{\cal A}_\sigma(t,z)+{\cal B}_\sigma(t,z),\non\\
{\cal I}_\sigma[f_2](t,z)&=&
{\cal A}_\sigma(t,z),\non\\
{\cal I}_\sigma[f_3](t,z)&=&
{\cal B}_\sigma(t,z),\non\
\eeqa
where ${\cal A}_\sigma$
and ${\cal B}_\sigma$, are calculated to be
(See Fig.\ref{fig:range-Upsilon}):
\begin{itemize}
\item{for $z\leq -ct$, (Outside the regions $R_1$ and $R_2$)}
\beq
{\cal A}_{\sigma}(t,z):=0,\qquad\qquad
{\cal B}_{\sigma}(t,z):=0,
\label{eqn:AB-before}
\eeq
\item{for $-ct<z\leq vt$, (Region $R_2$)}
\BAE{
{\cal A}_{\sigma}(t,z):=\frac{1}{2c}\int_0^{t'_+(t,z)}dt'\kappa_0'(vt')
J_0(\sigma s(t,z|t')),\\
{\cal B}_{\sigma}(t,z):=\frac{\sigma}{2c}\int_0^{t'_+(t,z)}dt'
\frac{\kappa_0(vt')(z-vt')}{s(t,z|t')}J_1(\sigma s(t,z|t')),
\label{eqn:AB-before-vt}
}
\item{for $vt\leq z <ct$, (Region $R_1$)}
\BAE{
{\cal A}_{\sigma}(t,z):=\frac{1}{2c}\int_0^{t'_-(t,z)}dt'\kappa_0'(vt')
J_0(\sigma s(t,z|t')),\\
{\cal B}_{\sigma}(t,z):=\frac{\sigma}{2c}\int_0^{t'_-(t,z)}dt'
\frac{\kappa_0(vt')(z-vt')}{s(t,z|t')}J_1(\sigma s(t,z|t')),
\label{eqn:AB-after-vt}
}
\item{for $z\geq ct$, (Outside the regions $R_1$ and $R_2$)}
\beq
{\cal A}_{\sigma}(t,z):=0,\qquad\qquad
{\cal B}_{\sigma}(t,z):=0,
\label{eqn:AB-after}
\eeq
\end{itemize}
with $t_+'(t,z), t_-'(t,z)$ and $s(t,z|t')$ defined
in \fr{eqn:abb-int}.
Thus one sees that to ${\cal O}(\epsilon)$, ${\cal A}_{\sigma}$
containes $\kappa_0'$ while ${\cal B}_\sigma$ containes $\kappa_0$.

The causal solution for $\gamma_M^{H(1)}$ generated by  the source
term $g_M^{H(1)}$  above  then follows from \fr{eqn:def-I}
and can be written:
\beqa
\gamma_M^{H(1)}(t,z)&=&{\cal I}_{\alpha_M}[g_M^{H(1)}](t,z) \non\\
&=&{\cal I}_{\alpha_M}[K_M^{H(1)}](t,z)
+{\cal I}_{\alpha_M}[L_M^{H(1)}](t,z)
+{\cal I}_{\alpha_M}[S_M^{H(1)}](t,z).
\non
\eeqa
The first term on the right above depends on initial data but
 explicit expressions for the last two terms  are:
\beqa
{\cal I}_{\alpha_M}[L_M^{H(1)}](t,z)&=&
\frac{2c^2v Q_{tot}}{\MM{M}^2}
\bigg(-{\cal A}_{\alpha_M}(t,z)+{\cal B}_{\alpha_M}(t,z)\bigg) \non\\
&&\times{\sum_N}'
\frac{G_{M,N}^{\bar{\Psi},\Phi}}{\NN{N}^2\beta_N^2}
J_n\left(x_{q(n)}\frac{r_0}{a}\right)e^{-in\theta_0},
\label{eqn:I-alphaM-k-delta-prime}
\eeqa
and
\beqa
{\cal I}_{\alpha_M}[S_M^{H(1)}](t,z)&=&
\frac{c^2 vQ_{tot}}{\MM{M}^2}{\cal A}_{\alpha_M}(t,z)\non\\
&&\qquad\times
{\sum_N}'\frac{G_{M,N}^{\bar{\Psi},\Phi}}{\NN{N}^2\beta_N^2}
J_n\left(x_{q(n)}\frac{r_0}{a}\right)e^{-in\theta_0}.
\eeqa

Similarly the causal solution for $\gamma_M^{E(1)}$ generated by the source
term $g_M^{E(1)}$ can be written as:
\beqa
\gamma_M^{E(1)}(t,z)&=&{\cal I}_{\beta_M}[g_M^{E(1)}](t,z)\non\\
&=&{\cal I}_{\beta_M}[K_M^{E(1)}](t,z) +{\cal I}_{\beta_M}
[L_M^{E(1)}](t,z)+{\cal I}_{\beta_M}[P_M^{E(1)}](t,z)\non\\
&&
+{\cal I}_{\beta_M}[R_M^{E(1)}](t,z)
+{\cal I}_{\beta_M}[S_M^{E(1)}],
\non
\eeqa
with
\beqa
&&
{\cal I}_{\beta_M}[L_M^{E(1)}](t,z)=
\frac{\mu c^2(c^2+v^2) Q_{tot}}{\NN{M}^2}
\bigg(-{\cal A}_{\beta_M}(t,z)+{\cal B}_{\beta_M}(t,z)\bigg)\non\\
&&\qquad\qquad \times{\sum_N}' \frac{1}{\NN{N}^2}
\bigg(
-E_{M,N}^{\Phi}
+\frac{C_{M,N}^{\Phi}-D_{M,N}^{\Phi}}{\beta_N^2} \bigg)
J_n\left(x_{q(n)}\frac{r_0}{a}\right)e^{-in\theta_0},\non
\eeqa
\beqa
{\cal I}_{\beta_M}[P_M^{E(1)}](t,z)&=&
-\frac{\mu c^2 v^2 Q_{tot}}{\NN{M}^2}\bigg(
{\cal A}_{\beta_M}(t,z)-{\cal B}_{\beta_M}(t,z)\bigg)\non\\
&&\times
r_0\cos\theta_0
J_m\left(x_{p(m)}\frac{r_0}{a}\right)e^{-im\theta_0}
,\non
\eeqa
$$
{\cal I}_{\beta_M}[R_M^{E(1)}](t,z)= 0,
$$

\beqa
&&{\cal I}_{\beta_M}[S_M^{E(1)}](t,z)=
\frac{\mu c^4 Q_{tot}}{\NN{M}^2}{\cal A}_{\beta_M}(t,z)\non\\
&&\qquad \times{\sum_N}'\frac{1}{\NN{N}^2}\bigg(
-E_{M,N}^{\Phi}+\frac{C_{M,N}^{\Phi}-D_{M,N}^{\Phi}}{\beta_N^2}
\bigg)J_n\left(x_{q(n)}\frac{r_0}{a}\right)e^{-in\theta_0}.\non
\eeqa

The  contribution  ${\cal I}_{\beta_M}[K_M^{E(1)}](t,z)$
depends on initial data and electric currents
that vanish in the ultra-relativistic limit.

\subsubsection{First Order Contribution to the Instantaneous  Power
  from a Moving Point Charge}

In terms of the lowest order modal solutions  the instantaneous power
$\breve{w}^{(1)}(t,z)$ for $z\ne vt, t>0$ can now be computed to
${\cal O}(\epsilon^1)$ from \fr{eqn:pow-1st-result}:
 \beqa
&&\breve{w}^{(1)}=-\Re\bigg[
\EPS\sum_N\frac{\NN{N}^2}{\beta_N^2}\gamma_N^{E(0)\prime} \bigg(
\overline{\dot{\gamma}_N^{E(1)}}+\frac{\kappa_0(z)}{\NN{N}^2}
{\sum_M}'\frac{C_{N,M}^{\Phi}-D_{N,M}^{\Phi}}{\beta_M^2}
\overline{\dot{\gamma}_M^{E(0)}}
\bigg)\non\\
&&+\EPS\sum_N\frac{\NN{N}^2}{\beta_N^2} \bigg\{
\gamma_N^{E(1)\prime} +\frac{\kappa_0(z)}{\NN{N}^2}
{\sum_M}'\bigg(2E_{N,M}^{\Phi}-\frac{C_{M,N}^{\Phi}-D_{M,N}^{\Phi}}{\beta_M^2}
\bigg)\gamma_M^{E(0)\prime}
\bigg\}\overline{\dot{\gamma}_N^{E(0)}}\non\\
&&+\mu\sum_M\frac{\MM{M}^2}{\alpha_M^2}\dot{\gamma}_M^{H(0)} \bigg\{
\overline{\gamma_M^{H(1)\prime}}+\frac{\kappa_0(z)}{\MM{M}^2}{\sum_N}'
\bigg(2E_{M,N}^{\Psi}-\frac{C_{M,N}^{\Psi}-D_{M,N}^{\Psi}}{\alpha_N^2}
\bigg)\overline{\gamma_N^{H(0)\prime}}
\bigg\}\non\\
&&+\mu\sum_M\frac{\MM{M}^2}{\alpha_M^2} \bigg(
\dot{\gamma}_M^{H(1)}+\frac{\kappa_0(z)}{\MM{M}^2}{\sum_N}'
\frac{C_{M,N}^{\Psi}-D_{M,N}^{\Psi}}{\alpha_N^2}
\dot{\gamma}_{N}^{H(0)}
\bigg)\overline{\gamma_M^{H(0)\prime}}\non\\
&&+\kappa_0(z)\EPS\sum_N\sum_{N'}F_{N,N'}^{\Phi}
\frac{\gamma_N^{E(0)\prime}\overline{\dot{\gamma}_{N'}^{E(0)}}}{\beta_N^2\beta_{N'}^2}+\kappa_0(z)\mu\sum_M\sum_{M'}F_{M,M'}^{\Psi}
\frac{\dot{\gamma}_M^{H(0)}\overline{\gamma_{M'}^{H(0)\prime}}}{\alpha_M^2\alpha_{M'}^2}\bigg].
\non
\eeqa
%

\subsection{Ultra-relativistic Longitudinal Wake Potentials}
\label{sec:wake}

The wakefield formalism is designed to exploit the simplifications
that arise by considering the unphysical (ultra-relativistic) limit
obtained from charged sources moving at the speed of light. The
resulting electromagnetic fields give rise to various wake-potentials
from which wake-impedances may be computed for ultra-relativistic
charged bunches with prescribed charged distributions. The formalism
is based on calculating the emf induced on a spectator (test)
ultra-relativistic point particle moving behind a leading
ultra-relativistic charged particle with the same velocity but in
general on a different orbit. Since the section above provides the
electromagnetic fields for a point particle moving with arbitrary
speed on an orbit (in general) off the tube axis (with transverse
coordinates $(r_0,\theta_0)$) one may readily calculate the general
longitudinal wake potential to the same order as the fields, by having
the spectator charge, with transverse coordinates $(r,\theta)$,  at a
fixed longitudinal separation $\ts>0$ behind a right moving  source particle.

The definition  \cite{wakedef} of the ultra-relativistic longitudinal wake
potential is taken as
\beq
{\cal W}_{\parallel}^{(r_0, \theta_0)}(\epsilon, r, \theta, \ts ):=
-\frac{1}{Q_{tot}}\int_{-\ts/2}^{\infty}dz\,
{\cal E}_z^{(r_0,\theta_0)}\left(\epsilon, \frac{z+\ts}{c},z, r ,\theta\right),
\label{eqn:def-wake-dz}
\eeq
where ${\cal E}_z^{(r_0,\theta_0)}(\epsilon,t,z,r,\theta)$
is the $z$-component of the
electric field generated by the point source with speed $v=c$
and charge $Q_{tot}$.\footnote{
On the test spectator particle worldline, $z=ct-\ts$, so
one may use this relation to express the longitudinal wake potential as an integral over the worldline parameter $t$ rather than $z$.}
Since the $z$-component of the {\it total} electric field is
\beqa
i_{\partial_z}\bfe(\epsilon,t,z,r,\theta)
&=&
\sum_N\gamma_N^{E}(\epsilon,t,z)\Phi_N(r,\theta)\non\\
&=&
\epsilon\sum_N\gamma_N^{E(1)}(t,z)\Phi_N(r,\theta)+{\cal O}(\epsilon^2),\non
\eeqa
one has
$$
{\cal E}_z^{(r_0,\theta_0)}(\epsilon,t,z,r,\theta)=
\epsilon\sum_M\breve{\gamma}_M^{E(1)}(t,z)\Phi_M(r,\theta)+{\cal O}(\epsilon^2),
$$
with
$$
\breve{\gamma}_M^{E(1)}(t,z):=
\gamma_M^{E(1)}(t,z)-{\cal I}_{\beta_M}[K_M^{E(1)}](t,z).
$$
Thus,
\beq
{\cal W}_{\parallel}^{(r_0, \theta_0)}(\epsilon, r, \theta, \ts )=
-\frac{\epsilon}{Q_{tot}}\sum_M\int_{-\ts/2}^{\infty}dz\,
\breve{\gamma}_M^{E(1)}\left(\frac{z+\ts}{c},z\right)\Phi_M(r,\theta)
+{\cal O}(\epsilon^2).
\label{eqn:wake-dz-perturb}
\eeq
From the orthogonality relation \fr{eqn:Dirichelet-orthogonal},
one calculates the projected longitudinal wake potentials:
\beq
{\cal W}_{\parallel\quad M}^{(r_0, \theta_0)}(\epsilon, \ts)
:=
\int_{\DD}
{\cal W}_{\parallel}^{(r_0, \theta_0)}\Phi_M\hashat 1.
\label{eqn:def-integrated-waks-dz}
\eeq
Hence
\beq
{\cal W}_{\parallel\quad M}^{(r_0, \theta_0)}(\epsilon, \ts)=
-\epsilon\frac{\NN{M}^2}{Q_{tot}}\int_{-\ts/2}^{\infty}dz\,
\overline{\breve{\gamma}_M^{E(1)}}\left(\frac{z+\ts}{c},z\right)
+{\cal O}(\epsilon^2).
\label{eqn:integrated-waks-dz}
\eeq
From \fr{eqn:def-wake-dz}, the definition of
the ultra-relativistic longitudinal impedance is taken as
$$
Z_{\parallel}^{(r_0,\theta_0)}
(\epsilon, r, \theta,\omega ):=
\frac{1}{c}\int_{0}^{\infty}d\ts\,e^{i\omega \ts/c}
{\cal W}_{\parallel}^{(r_0, \theta_0)}
(\epsilon, r, \theta, \ts),
$$
and from the orthogonality relation \fr{eqn:Dirichelet-orthogonal},
one calculates the projected longitudinal impedances:
$$
\langle{Z_{\parallel}^{(r_0, \theta_0)}\rangle_M}(\epsilon,\omega)
:=
\int_{\DD}
Z_{\parallel}^{(r_0, \theta_0)}
(\epsilon,r,\theta,\omega)\,
\overline{\Phi_M}(r,\theta)\hashat 1.
$$

To calculate \fr{eqn:integrated-waks-dz}, one needs
$\breve{\gamma}_M^{E(1)}(t,z)$.
In the ultra-relativistic limit the expressions
\fr{eqn:def-f1},\fr{eqn:def-f2},\fr{eqn:def-f3} become
\beqa
\check{f}_1(t,z)&:=&\kappa_0(z)\delta'(z-ct),
\label{eqn:def-chcek-f1}\\
\check{f}_2(t,z)&:=&\kappa_0'(ct)\delta(z-ct),
\label{eqn:def-chcek-f2}\\
\check{f}_3(t,z)&:=&\kappa_0(ct)\delta'(z-ct).
\label{eqn:def-chcek-f3}
\eeqa
One may write the  application of the  integral operator
${\cal I}_\sigma$ on
these as
\beqa
{\cal I}_\sigma[\check{f}_1](t,z)&=&
-{\cal\check{A}}_\sigma(t,z)+{\cal\check{B}}_\sigma(t,z),\non\\
{\cal I}_\sigma[\check{f}_2](t,z)&=&
{\cal\check{A}}_\sigma(t,z),\non\\
{\cal I}_\sigma[\check{f}_3](t,z)&=&
{\cal\check{B}}_\sigma(t,z).\non
\eeqa
where  ${\cal \check{A}}_\sigma$ and ${\cal \check{B}}_\sigma$
are given in the following domains:
\begin{itemize}
\item{for $z\leq -ct$,}
\beq
{\cal \check{A}}_{\sigma}(t,z):=0,\qquad\qquad
{\cal \check{B}}_{\sigma}(t,z):=0,
\label{eqn:cAB-before}
\eeq
\item{for $-ct<z<ct$,}
\BAE{
{\cal\check{A}}_{\sigma}(t,z):=\frac{1}{2c}\int_0^{t'_c(t,z)}dt'
\kappa_0'(ct')
J_0(\sigma s_c(t,z|t')),\\
{\cal \check{B}}_{\sigma}(t,z):=\frac{\sigma}{2c}\int_0^{t'_c(t,z)}dt'
\frac{\kappa_0(ct')(z-ct')}{s_c(t,z|t')}J_1(\sigma s_c(t,z|t')),
\label{eqn:cAB-before-ct}
}
\item{for $z=ct$,}
\BAE{
{\cal \check{A}}_{\sigma}(t,z):=\frac{1}{4c}\int_0^{t}dt' \kappa_0'(ct')
J_0(\sigma s_c(t,z|t')),\\
{\cal \check{B}}_{\sigma}(t,z):=\frac{\sigma}{4c}\int_0^{t}dt'
\frac{\kappa_0(ct')(z-ct')}{s_c(t,z|t')}J_1(\sigma s_c(t,z|t')),
\label{eqn:cAB-after-ct}
}
\item{for $z\geq ct$,}
\beq
{\cal \check{A}}_{\sigma}(t,z):=0,\qquad\qquad
{\cal \check{B}}_{\sigma}(t,z):=0,
\label{eqn:cAB-after}
\eeq
\end{itemize}
where
\beq
s_c(t,z|t'):=\sqrt{c^2(t-t')^2-(z-ct')^2},
\label{eqn:def-sc}
\eeq
and
\beq
t_c'(t,z):=\frac{1}{2}\bigg(t+\frac{z}{c}\bigg).
\label{eqn:def-tc}
\eeq
It immediately follows that
$$
s_c(t,ct|t')=0.
$$
The explicit form  of ${\cal\check{A}}_\sigma(t,z)$ and
${\cal\check{B}}_\sigma(t,z)$ follows from a calculation  similar to that outlined  for
${\cal A}_\sigma(t,z)$ and ${\cal B}_\sigma(t,z)$ respectively
 in Appendix A.
Thus ${\cal \check{A}}_\sigma$ and ${\cal \check{B}}_\sigma$
correspond to ${\cal A}_\sigma$ and ${\cal B}_\sigma$  respectively in the case when
$0<v<c$.

Using the results in \S\ref{sec:1st-point} and
\fr{eqn:cAB-before}, \fr{eqn:cAB-after-ct}, \fr{eqn:cAB-after-ct},
\fr{eqn:cAB-after} one finds
\beqa
&&\breve{\gamma}_M^{E(1)}(t,z)=
{\cal I}_{\beta_M}[L_M^{E(1)}+P_M^{E(1)}+R_M^{E(1)}+S_M^{E(1)}](t,z),\non\\
&& =
l_M^{(r_0,\theta_0)}(-{\cal \check{A}}_{\beta_M}+{\cal \check{B}}_{\beta_M})
+p_M^{(r_0,\theta_0)}({\cal \check{A}}_{\beta_M}-{\cal \check{B}}_{\beta_M})
+s_M^{(r_0,\theta_0)}{\cal \check{A}}_{\beta_M}\non\\
&&
=(-l_M^{(r_0,\theta_0)}+p_M^{(r_0,\theta_0)}+s_M^{(r_0,\theta_0)}){\cal \check{A}}_{\beta_M}
+(l_M^{(r_0,\theta_0)}-p_M^{(r_0,\theta_0)})
{\cal\check{B}}_{\beta_M},
\label{eqn:breve-gamma-E-first}
\eeqa
where
\beqa
l_M^{(r_0,\theta_0)}&:=&
\frac{2\mu c^4}{\NN{M}^2}Q_{tot}{\sum_N}'
\frac{1}{\NN{N}^2}
\bigg(-E^{\Phi}_{M,N}
+\frac{C^{\Phi}_{M,N}-D^{\Phi}_{M,N}}{\beta_N^2}
\bigg)\non\\
&&\qquad\qquad \times
J_n\left(x_{q(n)}\frac{r_0}{a}\right)e^{-in\theta_0},
\label{eqn:lM}\\
p_M^{(r_0,\theta_0)}&:=&\frac{\mu c^4}{\NN{M}^2}Q_{tot}r_0\cos\theta_0
J_m\left(x_{p(m)}\frac{r_0}{a}\right)e^{-im\theta_0},
\label{eqn:pM}\\
s_M^{(r_0,\theta_0)}&:=&
\frac{\mu c^4}{\NN{M}^2}Q_{tot}{\sum_N}'
\frac{1}{\NN{N}^2}
\bigg(-E^{\Phi}_{M,N}
+\frac{C^{\Phi}_{M,N}-D^{\Phi}_{M,N}}{\beta_N^2}
\bigg)\non\\
&&\qquad\qquad
\times J_n\left(x_{q(n)}\frac{r_0}{a}\right)e^{-in\theta_0}.
\label{eqn:sM}
\eeqa
From these expressions one  calculates
\fr{eqn:integrated-waks-dz} and  \fr{eqn:def-integrated-waks-dz}.
 Thus  $\breve{\gamma}_M^{E(1)}$ and hence
\fr{eqn:integrated-waks-dz} can be
expressed
in terms of  separate contributions from $\kappa_0(z)$ and
 $\kappa_0'(z)$.


\subsubsection{Longitudinal Wake Potential for a Pipe with Piecewise  Constant Curvature}
\begin{figure}[h]
\unitlength 1mm
\begin{picture}(160,25)
\put(0,0){\includegraphics[width=11.0cm]{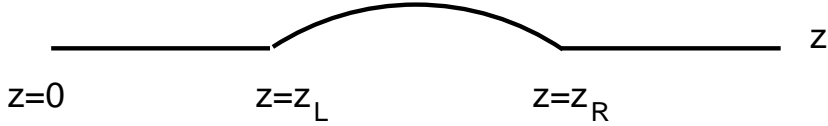}}
\end{picture}
\caption{Profile of beam pipe with a segment of constant curvature.}
\label{fig:piecewise}
\end{figure}

In the last section explicit formulae are given for the computation to
leading order of the longitudinal wake potential in a pipe with
arbitrary smooth curvature and $\vert\kappa(z)a\vert < 1$. From such
potentials one may calculate the longitudinal impedance to the
same order. These expressions involve integrals of the curvature with
Bessel functions and such integrals in general require numerical
analysis. However in  cases where   segments of the beam pipe are
connected by planar segments of arcs with constant radius of curvature
(See Fig.\ref{fig:piecewise}) one may perform these integrals
analytically and hence generate analytic expressions
for the corresponding wake impedances.
In principle there is an element of further approximation
involved if one assumes that the tangent to the axial space-curve
is discontinuous where the straight segment joins the curved segment.
However bearing this in mind consider the case of an infinitely
long planar pipe with axial curvature  given by
$$
\kappa_0(z)=\left(\Theta(z-z_L)-\Theta(z-z_R)\right)\check{\kappa}_0,
$$
where $z_L,z_R,(0<z_L<z_R)$, $\check{\kappa}_0(\neq 0)$ are constants
and $\Theta(z)$ is the Heaviside function
\beqa
\Theta(z)=
\left\{
\begin{array}{ll}
1,&\qquad\mbox{for }z\geq 0,\\
0,&\qquad\mbox{otherwise}.\\
\end{array}\right.
\eeqa
In this case one can  calculate
${\cal\check{A}}_\sigma, {\cal\check{B}}_\sigma$ and write
\fr{eqn:integrated-waks-dz}  in terms of known functions.

One finds that the terms in the wake potential proportional to ${\cal\check{A}}_{\beta_M}$
express the contributions to $\breve{\gamma}_M^{E(1)}$
from the transitions at $z=z_L$ and $z=z_R$
and the terms proportional
to ${\cal\check{B}}_{\beta_M}$ express the contributions coming from the region
$z_L<z<z_R$ where the curvature  is the constant $\check{\kappa}_0$.

\begin{figure}[h]
\unitlength 1mm
\begin{picture}(160,75)
\put(10,0){\includegraphics[width=12.0cm]{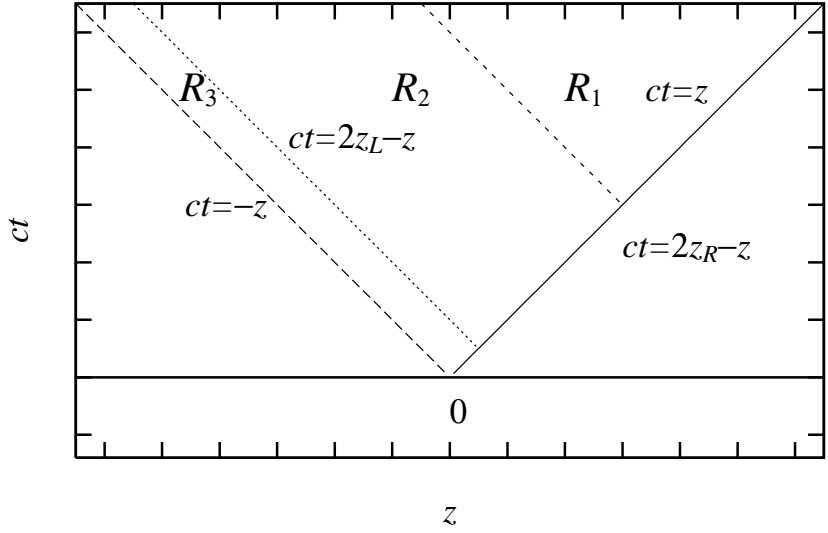}}
\end{picture}
\caption{Integration ranges for $\check{A}_\sigma(t,z)$ and
  $\check{B}_\sigma(t,z)$. $R_1$ is the triangle in the upper
  right-hand corner i.e.,$(-ct<z<ct)\cap (z_R< c t_c'(t,z))$.
 $R_2$ is the adjacent large rectangle:
$(-ct<z<ct)\cap (z_R\geq ct_c'(t,z))\cap (z_L< ct_c'(t,z))$, and
 $R_3$ is the adjacent smaller rectangle:
$(-ct<z<ct)\cap (z_L\geq ct_c'(t,z))$. }
\label{fig:wake-int-AB}
\end{figure}

The only non-zero contribution to ${\cal\check{A}}_\sigma(t,z)$
arises from the region
$-ct<z\leq ct$.
The term involving $\kappa_0'(ct')$ in the integrand now
follows from the relations
$$
\frac{\partial}{\partial z}\bigg(\Theta(z-z_L)-\Theta(z-z_R)\bigg)
=\delta(z-z_L)-\delta(z-z_R),
$$
and $\delta(z/z_0)=|z_0|\delta(z)$ yielding
$$
{\cal \check{A}}_\sigma(t,z)=\frac{\check{\kappa}_0}{2c^2}
\int_0^{t_c'(t,z)}dt'\, \bigg(
\delta\left(t'-\frac{z_L}{c}\right)-\delta\left(t'-\frac{z_R}{c}\right)\bigg)J_0(\sigma s_c(t,z|t')).
$$
Hence, see Fig.\ref{fig:wake-int-AB},
\begin{itemize}
\item{For $(-ct<z<ct)\cap (z_R< c t_c'(t,z))$, (Region $R_1$)}
$$
{\cal \check{A}}_\sigma(t,z)=\frac{\check{\kappa}_0}{2c^2}
\bigg\{J_0\left(\sigma s_c\left(t,z\bigg|\,\frac{z_L}{c}\right)\right)
-J_0\left(\sigma s_c\left(t,z\bigg|\frac{z_R}{c}\right)\right)
\bigg\},
$$
\item{for $(-ct<z<ct)\cap (z_R\geq ct_c'(t,z))\cap (z_L< ct_c'(t,z))$,
(Region $R_2$)}
$$
{\cal \check{A}}_\sigma(t,z)=\frac{\check{\kappa}_0}{2c^2}
J_0\left(\sigma s_c\left(t,z\bigg|\frac{z_L}{c}\right)\right),
$$
\item{for $(-ct<z<ct)\cap (z_L\geq ct_c'(t,z))$, (Region $R_3$)}
$$
{\cal \check{A}}_\sigma(t,z)=0.
$$
\end{itemize}
Since at
$z_R=ct_c'$ and $z_L=ct_c'$,
$$
s_c\left(t,2z_R-ct\,\bigg|\,\frac{z_R}{c}\right)=0,
\qquad \mbox{and} \qquad
s_c\left(t,2z_L-ct\,\bigg|\,\frac{z_L}{c}\right)=0.
$$
A simple calculation from\fr{eqn:cAB-after-ct},
 shows that with $z=ct$,
%
\begin{itemize}
\item{for $z_R<z$,}
$$
{\cal \check{A}}_\sigma(t,z)=0,
$$
\item{for $z_R=z$,}
$$
{\cal \check{A}}_\sigma(t,z)=\frac{\check{\kappa}_0}{4c^2},
$$
\item{for $z_L<z<z_R$,}
$$
{\cal \check{A}}_\sigma(t,z)=\frac{\check{\kappa}_0}{2c^2},
$$
\item{for $z=z_L$,}
$$
{\cal \check{A}}_\sigma(t,z)=\frac{\check{\kappa}_0}{4c^2},
$$
\item{for $z<z_L$,}
$$
{\cal \check{A}}_\sigma(t,z)=0.
$$
\end{itemize}


Similarly  since
$$
{\cal \check{B}}_\sigma(t,z)=0,
$$
for $-ct<z<ct$, and  $( z<z_L ) \cup (z\geq z_R)$,
we must evaluate
$$
{\cal \check{B}}_\sigma(t,z)=
\frac{\sigma\check{\kappa}_0}{2c}\int_0^{t_c'(t,z)}{dt'}
\frac{z-ct'}{s_c(t,z|t')}J_1(\sigma s_c(t,z|t')),
$$
with $-ct<z<ct$ and $z_L\leq z< z_R$ where $\kappa_0(z)\neq 0$.

Write the integral in ${\cal\check{B}_\sigma}$ as
\beq
\Lambda_\sigma(t,z):=
\int_0^{t_c'(t,z)}{dt'}
\frac{z-ct'}{s_c(t,z|t')}J_1(\sigma s_c(t,z|t')),
\label{eqn:typical-int}
\eeq
and change  variables, $t'\mapsto s'_c(t')$ with fixed $t$ and
$z$,
\beq
s'_c(t'):=\sqrt{c^2(t-t')^2-(z-ct')^2}.
\label{eqn:def-sc-prime}
\eeq
Then with
$$
dt'=ds'_c\frac{s'_c}{c(z-ct)},\qquad
z-ct'=\frac{(z-ct)^2-s_c^{\prime 2}}{2(z-ct)},
$$
Eq.\fr{eqn:typical-int} can be expressed
\beq
\Lambda_\sigma=\int_{\sqrt{(ct)^2-z^2}}^0 ds'_c\,
\frac{(z-ct)^2-s_c^{\prime 2}}{2c(z-ct)^2}J_1(\sigma s'_c),
\label{eqn:typical-int-rewrite}
\eeq
since
$$
s_c(t,z|0)=\sqrt{(ct)^2-z^2},\qquad
s_c(t,z|t_c'(t,z))=0.
$$
This  integral can now be evaluated using
$$
\int dz J_1(\sigma z)=-\frac{1}{\sigma}J_0(\sigma z),\qquad
\int dz z^2 J_1(\sigma z)=\frac{z^2}{\sigma}J_2(\sigma z).
$$
Hence
\beqa
\Lambda_\sigma&=&\frac{1}{2c}\int_{\sqrt{(ct)^2-z^2}}^0 ds'_c\,
J_1(\sigma s'_c)
-\frac{1}{2c(z-ct)^2}\int_{\sqrt{(ct)^2-z^2}}^0 ds'_c\,
s_c^{\prime 2} J_1(\sigma s'_c)\non\\
&=&
\frac{1}{2c}\frac{(-1)}{\sigma}
\bigg[ J_0(\sigma s'_c)\bigg]_{\sqrt{(ct)^2-z^2}}^0
-\frac{1}{2c(z-ct)^2}\frac{1}{\sigma}
\bigg[ s_c^{\prime 2} J_2(\sigma s'_c)
\bigg]_{\sqrt{(ct)^2-z^2}}^0,
\non
\eeqa
or with $J_0(0)=1$ and
$J_2(0)=0$, 
\beq
\Lambda_\sigma(t,z)=\frac{1}{2c\sigma}
\bigg(
-1+J_0(\sigma\sqrt{(ct)^2-z^2}\,)
\bigg)+
\frac{(ct)^2-z^2}{2c(z-ct)^2\sigma}J_2(\sigma\sqrt{(ct)^2-z^2}\,),
\label{eqn:typical-int-rewrite}
\eeq
valid in the regime $-ct<z<ct$.

Finally for  $z=ct$,
$$
\Lambda_\sigma=0.
$$

Thus the function $\breve{\gamma}_M^{E(1)}(t,z)$ is given in the indicated
domains  (See Fig.\ref{fig:wake-int-AB} ) by:
\begin{itemize}
\item{for $(z\leq -ct)\cup(z\geq ct)$,
(Outside the regions $R_1,R_2$ and $R_3$) }
$$
\breve{\gamma}_M^{E(1)}(t,z)=0,
$$
\item{for $(-ct<z< ct)\cap(z_L\geq ct_c'(t,z))$, (Region $R_3$)}
$$
\breve{\gamma}_M^{E(1)}(t,z)
=(l_M^{(r_0,\theta_0)}-p_M^{(r_0,\theta_0)})
\frac{\beta_M\check{\kappa}_0}{2c}\Lambda_{\beta_M}(t,z),
$$
\item{for $(-ct<z< ct)\cap(z_L< ct_c'(t,z))\cap(z_R\geq ct_c'(t,z))$,
    (Region $R_2$)}
\beqa
\breve{\gamma}_M^{E(1)}(t,z)
&=&(-l_M^{(r_0,\theta_0)}+p_M^{(r_0,\theta_0)}+s_M^{(r_0,\theta_0)})
\frac{\check{\kappa}_0}{2c^2}J_0
\left(\beta_M s_c\left(t,z\bigg|\frac{z_L}{c}\right)\right)
\non\\
&&
+(l_M^{(r_0,\theta_0)}-p_M^{(r_0,\theta_0)})\frac{\beta_M\check{\kappa}_0}{2c}\Lambda_{\beta_M}(t,z),\non
\eeqa
\item{for $(-ct<z< ct)\cap(z_R< ct_c'(t,z))$, (Region $R_1$)}
\beqa
&&\breve{\gamma}_M^{E(1)}(t,z)
=(-l_M^{(r_0,\theta_0)}+p_M^{(r_0,\theta_0)}+s_M^{(r_0,\theta_0)})
\frac{\check{\kappa}_0}{2c^2}\non\\
&&\qquad\qquad \times
\left\{
J_0\left(\beta_M s_c\left(t,z\bigg|\frac{z_L}{c}\right)\right)
-J_0\left(\beta_M s_c\left(t,z\bigg|\frac{z_R}{c}\right)\right)
\right\}
\non\\
&&\qquad
+(l_M^{(r_0,\theta_0)}-p_M^{(r_0,\theta_0)})\frac{\beta_M\check{\kappa}_0}{2c}\Lambda_{\beta_M}(t,z).\non
\eeqa
\end{itemize}

The explicit longitudinal wake potential in this case
now follows from \fr{eqn:integrated-waks-dz}
and \fr{eqn:breve-gamma-E-first}:
\beq
{\cal W}_{\parallel\quad M}^{(r_0, \theta_0)}
(\epsilon,\ts\,)={\cal W}_{\parallel M, edges }^{(r_0, \theta_0)}
(\epsilon,\ts\,)\,+\,
{\cal W}_{\parallel M,\check{\kappa}_0 }^{(r_0, \theta_0)}
(\epsilon,\ts\,)+{\cal O}(\epsilon^2).
\label{eqn:decompo-wake-M}
\eeq
Here, ${\cal W}_{\parallel M, edges}^{(r_0,\theta_0)}$ expresses
the  contributions from the abrupt transitions
in curvature at $z=z_L$ and $z=z_R$,
while ${\cal W}_{\parallel M,\check{\kappa}_0 }^{(r_0,\theta_0)}$
denotes that from the region of constant curvature $z_L<z<z_R$.
Splitting the range of integration in \fr{eqn:integrated-waks-dz}
according the domains associated
with $\breve{\gamma}_M^{E(1)}$ (See Fig.\ref{fig:wake-int-range}) one has
\BAE{
\overline{{\cal W}_{\parallel M,edges}^{(r_0, \theta_0)}}(\epsilon,\ts\,)
=
\epsilon\frac{\NN{M}^2}{Q_{tot}}
(l_M^{(r_0,\theta_0)}-p_M^{(r_0,\theta_0)}-s_M^{(r_0,\theta_0)})\\
\times\left(
\int_{-\ts/2}^{z_L-\ts/2}+\int_{z_L-\ts/2}^{z_R+\ts/2}
+\int_{z_R+\ts/2}^{\infty}\right)
dz\,{\cal\check{A}}_{\beta_M}\left(\frac{z+\ts}{c},z\right)
+{\cal O}(\epsilon^2),
\label{eqn:ave-wake-M-edge}
}
and
\beq
\overline{{\cal W}_{\parallel M,\check{\kappa}_0}^{(r_0, \theta_0)}}(\epsilon,\ts\,)
=-\epsilon\frac{\NN{M}^2}{Q_{tot}}(l_M^{(r_0,\theta_0)}-p_M^{(r_0,\theta_0)})
\int_{z_L}^{z_R}dz\,{\cal\check{B}}_{\beta_M}
\left(\frac{z+\ts}{c},z\right)+{\cal O}
(\epsilon^2).
\label{eqn:ave-wake-M-kappa}
\eeq
\begin{figure}[h]
\unitlength 1mm
\begin{picture}(160,75)
\put(10,0){\includegraphics[width=12.0cm]{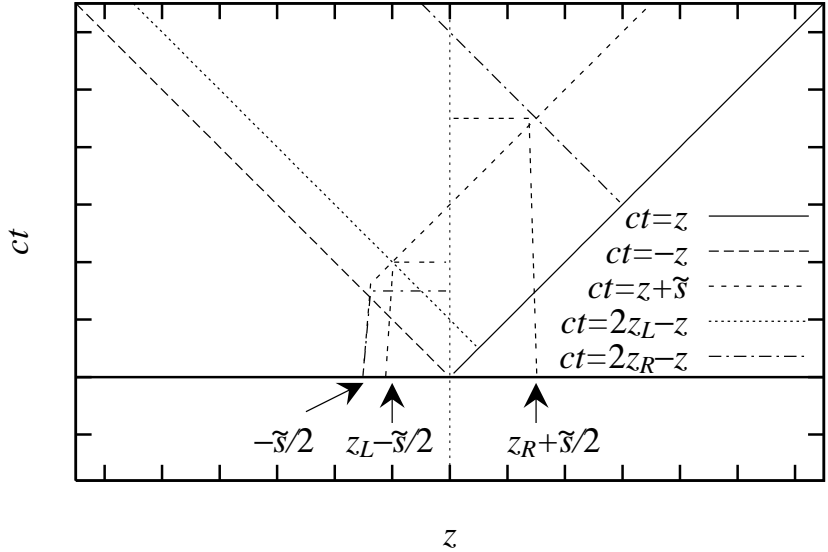}}
\end{picture}
\caption{Integration ranges for $\overline{{\cal W}_{\parallel M,edges}^{(r_0, \theta_0)}}$.}
\label{fig:wake-int-range}
\end{figure}

Again using the relation
\beq
\int dz\,z^{\nu/2} J_\nu(\sqrt{\sigma z})
=\frac{2}{\sqrt{\sigma}}z^{(1+\nu)/2}J_{\nu+1}(\sqrt{\sigma z}),
\label{eqn:bessel-int-formula}
\eeq
with $\nu=0$
the integrals in \fr{eqn:ave-wake-M-edge} are evaluated as
\beqa
&&\bigg(
\int_{-\ts/2}^{z_L-\ts/2}+\int_{z_L-\ts/2}^{z_R+\ts/2}
+\int_{z_R+\ts/2}^{\infty}
\bigg)
dz\,{\cal\check{A}}_{\beta_M}
\left(\frac{z+\ts}{c},z\right)\non\\
&&=0+
\frac{\check{\kappa}_0}{2c^2}\int_{z_L-\ts/2}^{z_R+\ts/2}dz
J_0\left(\beta_M s_c
\left(\frac{z+\ts}{c},z\bigg|\frac{z_L}{c}\right)\right)\non\\
&&+
\frac{\check{\kappa}_0}{2c^2}\int_{z_R+\ts/2}^{\infty}dz
\bigg\{
J_0\left(\beta_M s_c
\left(\frac{z+\ts}{c},z\bigg|\frac{z_L}{c}\right)\right)
-J_0\left(\beta_M s_c
\left(\frac{z+\ts}{c},z\bigg|\frac{z_R}{c}\right)\right)
\bigg\}\non\\
&&=
\frac{\check{\kappa}_0}{2c^2}\int_0^{\ts}
dz'J_0(\beta_M
\sqrt{2\ts z'})
=
\frac{\check{\kappa}_0}{\sqrt{2}c^2\beta_M}
J_1(\sqrt{2}\beta_M \ts),
\label{eqn:int-check-A}
\eeqa
 independent of  $z_L$ and $z_R$.
The integral involving  ${\cal\check{B}}_\sigma$
in \fr{eqn:ave-wake-M-kappa} can be similarly
evaluated using \fr{eqn:bessel-int-formula} with $\nu=2$
\beqa
&&\int_{z_L}^{z_R}dz\,{\cal\check{B}}_{\beta_M}
\left(\frac{z+\ts}{c},z\right)
=\frac{\beta_M\check{\kappa}_0}{2c}\int_{z_L}^{z_R} dz\,
\Lambda_{\beta_M}\left(\frac{z+\ts}{c},z\right)\non\\
&&\quad =\frac{\beta_M\check{\kappa}_0}{2c}\int_{z_L}^{z_R} dz\,
\bigg\{
\frac{1}{2c\beta_M}\bigg(-1+J_0(\beta_M\sqrt{2\ts z+\ts^2}\,)
\bigg)\non\\
&&\qquad\qquad\qquad
+\frac{2z\ts +\ts^2}{2c\beta_M \ts^2}
J_2(\beta_M\sqrt{2\ts z+\ts^2}\,)
\bigg\}\non\\
&&\quad=
\frac{\check{\kappa}_0}{4c^2}
\bigg[
-z' +\frac{\sqrt{2}}{\beta_M\sqrt{\ts}}
\bigg\{ \sqrt{z'}J_1(\beta_M\sqrt{2\ts z'})\non\\
&&\qquad\qquad
+\frac{2z^{\prime 3/2}}{\ts}J_3(\beta_M \sqrt{2\ts z'})\bigg\}
\bigg]_{z'=z_L+\ts/2}^{z'=z_R+\ts/2}.
\label{eqn:int-check-B}
\eeqa

Finally  from \fr{eqn:ave-wake-M-edge},
\fr{eqn:int-check-A} one has
$$
\overline{{\cal W}_{\parallel M,edges}^{(r_0, \theta_0)}}(\epsilon,\ts\,)
=
\epsilon\frac{\check{\kappa}_0}{\sqrt{2}\beta_M}
(\breve{l}_M^{(r_0,\theta_0)}-\breve{p}_M^{(r_0,\theta_0)}
-\breve{s}_M^{(r_0,\theta_0)})
J_1(\sqrt{2}\beta_M \ts)+{\cal O}(\epsilon^2),
$$
and from \fr{eqn:ave-wake-M-kappa}, \fr{eqn:int-check-B}
\beqa
&&
\overline{{\cal W}_{\parallel M,\check{\kappa}_0}^{(r_0, \theta_0)}}
(\epsilon,\ts\, )
=-\epsilon\frac{\check{\kappa}_0}{4}
(\breve{l}_M^{(r_0,\theta_0)}-\breve{p}_M^{(r_0,\theta_0)})
\bigg[
-(z_R-z_L) +\frac{\sqrt{2}}{\beta_M\sqrt{\ts}}\non\\
&&\quad
\times\bigg\{
\sqrt{z_R+\frac{\ts}{2}}\,
J_1\left(\beta_M\sqrt{2\ts\left(z_R+\frac{\ts}{2}\right)}\,\right)
\non\\
&&\qquad
+\frac{2\left(z_R+\frac{\ts}{2}\right)^{3/2}}{\ts}
J_3\left(\beta_M\sqrt{2\ts \left(z_R+\frac{\ts}{2}\right)}\right)\non\\
&&\qquad
-\sqrt{z_L+\frac{\ts}{2}}\,
J_1\left(\beta_M\sqrt{2\ts \left(z_L+\frac{\ts}{2}\right)}\,\right)
\non\\
&&\qquad
-\frac{2\left(z_L+\frac{\ts}{2}\right)^{3/2}}{\ts}
J_3\left(\beta_M\sqrt{2\ts \left(z_L+\frac{\ts}{2}\right)}\,\right)
\bigg\}
\bigg]
+{\cal O}(\epsilon^2),
\non
%
\eeqa
where, from \fr{eqn:lM},\fr{eqn:pM},\fr{eqn:sM}
we introduce the abbreviations
\beqa
&&\breve{l}_M^{(r_0,\theta_0)}:=\frac{\NN{M}^2}{c^2 Q_{tot}}
l_M^{(r_0,\theta_0)}=2\mu c^2{\sum_N}'
\frac{1}{\NN{N}^2}
\bigg(-E^{\Phi}_{M,N}
+\frac{C^{\Phi}_{M,N}-D^{\Phi}_{M,N}}{\beta_N^2}
\bigg)\non\\
&&\qquad\qquad \times
J_n\left(x_{q(n)}\frac{r_0}{a}\right)e^{-in\theta_0},\non\\
&&\breve{p}_M^{(r_0,\theta_0)}:=\frac{\NN{M}^2}{c^2Q_{tot}}
p_M^{(r_0,\theta_0)}=\mu c^2r_0\cos\theta_0
J_m\left(x_{p(m)}\frac{r_0}{a}\right)e^{-im\theta_0},\non\\
&&\breve{s}_M^{(r_0,\theta_0)}:=\frac{\NN{M}^2}{c^2 Q_{tot}}
s_M^{(r_0,\theta_0)}=\mu c^2{\sum_N}'
\frac{1}{\NN{N}^2}
\bigg(-E^{\Phi}_{M,N}
+\frac{C^{\Phi}_{M,N}-D^{\Phi}_{M,N}}{\beta_N^2}
\bigg)\non\\
&&\qquad\qquad
\times J_n\left(x_{q(n)}\frac{r_0}{a}\right)e^{-in\theta_0}.\non
\eeqa
It is worth noting that the expressions for
the wake potentials are independent of $Q_{tot}$.

With the following dimensionless variables for some length $L$
$$
\widehat{\kappa}_0:=L\check{\kappa}_0,\quad
\widehat{s}:=\frac{\ts}{L},\quad
\widehat{\beta}_M:=L\beta_M,\quad
\widehat{z}_R:=\frac{z_R}{L},\quad
\widehat{z}_L:=\frac{z_L}{L},
$$
one may introduce the dimensionless quantities
\beqa
\zeta_{M,1}(\widehat{s})&:=&\frac{\widehat{\kappa}_0}{\widehat{\beta}_M}
J_1(\sqrt{2}\widehat{\beta}_M \widehat{s}),\non\\
\zeta_{M,2}(\widehat{s})&:=&\widehat{\kappa}_0\bigg[
(\widehat{z}_R-\widehat{z}_L)-
\frac{\sqrt{2}}{\widehat{\beta}_M\sqrt{\widehat{s}}}\non\\
&&\quad
\times\bigg\{
\sqrt{\widehat{z}_R+\frac{\widehat{s}}{2}}\,
J_1\left(\beta_M\sqrt{2\widehat{s}\left(\widehat{z}_R+\frac{\widehat{s}}{2}
\right)}\,\right)
\non\\
&&\qquad
+\frac{2\left(\widehat{z}_R+\frac{\widehat{s}}{2}\right)^{3/2}}{\widehat{s}}
J_3\left(\widehat{\beta}_M\sqrt{2\widehat{s}\left(\widehat{z}_R+\frac{\widehat{s}}{2}
\right)}\right)\non\\
&&\qquad
-\sqrt{\widehat{z}_L+\frac{\widehat{s}}{2}}\,
J_1\left(
\widehat{\beta}_M\sqrt{2\widehat{s}\left(
\widehat{z}_L+\frac{\widehat{s}}{2}\right)}\,\right)
\non\\
&&\qquad
-\frac{2\left(\widehat{z}_L
+\frac{\widehat{s}}{2}\right)^{3/2}}{\widehat{s}}
J_3\left(\widehat{\beta}_M\sqrt{2\widehat{s} \left(\widehat{z}_L
+\frac{\widehat{s}}{2}\right)}\,\right)
\bigg\}
\bigg],
\non
\eeqa
in terms of which
\beqa
\overline{{\cal W}_{\parallel M,edges}^{(r_0, \theta_0)}}(\epsilon,\ts\,)
&=&\frac{\epsilon}{\sqrt{2}}\,\zeta_{M,1}(\widehat{s})\,
(\breve{l}_M^{(r_0,\theta_0)}-\breve{p}_M^{(r_0,\theta_0)}
-\breve{s}_M^{(r_0,\theta_0)}),
\non\\
\overline{{\cal W}_{\parallel M,\check{\kappa}_0}^{(r_0, \theta_0)}}(\epsilon,\ts\,)
&=&
\frac{\epsilon}{4}\zeta_{M,2}\,(\widehat{s})\,
(\breve{l}_M^{(r_0,\theta_0)}-\breve{p}_M^{(r_0,\theta_0)}).\non
\eeqa
Natural choices for $L$ include $L=a$ or $L=z_R-z_L$.
In Fig.\ref{fig:wake-int-range} we plot $\zeta_{M,1}$ and $\zeta_{M,2}$
for the choice
$$
\widehat{\kappa}_0=1,\quad
\widehat{\beta}_M=1,\quad
\widehat{z}_R=2,\quad
\widehat{z}_L=1.\quad
$$
In the regime $\widehat{s}\gg 1$,
$\zeta_{M,2}$  tends to
$\widehat{\kappa}_0(\widehat{z}_R-\widehat{z}_L)$.
\begin{figure}[h]
\unitlength 1mm
\begin{picture}(160,70)
\put(10,0){\includegraphics[width=10.0cm]{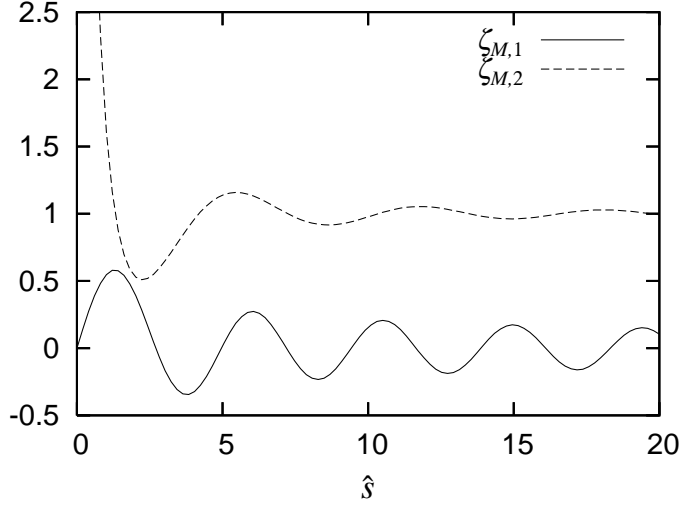}}
\end{picture}
\caption{
Dimensionless profiles for contributions to ${\cal W}_{\parallel\quad
  M}^{(r_0,\theta_0)}(\ts\,)$ to ${\cal O}(\epsilon^2)$.
}
\label{fig:wake-int-range}
\end{figure}

\section{Smooth Convected Localized Bunches
  with Fixed Total Charge}
In the previous section attention was concentrated on the ultrarelativistic  limit.
In this section, by contrast with \S\ref{sec:traveling-charge-source1},
%
%
we construct a moving source model {\it with finite total
charge $Q_{tot}$},
and a smooth charge density satisfying \fr{eqn:const-e-charge} moving at less than the speed of light.
 Large numbers of charged particles moving with a common
axial velocity
 may be modelled  by a localised  smooth distribution of electric charge
 with a prescribed
{\it convective axial velocity field  with constant longitudinal speed
 $v<c_0$ (independent of the local curvature $\kappa(z)$),
 charge density $\rho(z-vt,r,\theta)$ and current density
 components}
 $J_r=J_\theta=0,\, J_0(z-vt,r,\theta)=v\,\rho(z-vt,r,\theta)$.
The localized charge density profile is thereby maintained as a function of the
 arc-length parameter $z$ as it travels along the beam pipe.
Note that in this source  model   $J_0^{(0)}=J_0$ and all higher
orders are taken zero.


We assume here that $\rho$ can be expressed as
$$
\rho(\epsilon,t,z,x_1,x_2)=Q(\epsilon,t)
\rho^{\perp}(x_1,x_2)\rho^{\parallel}(z-vt),
$$
where $v$ is given ($v\leq c$),
$\rho^{\perp}(x_1,x_2)$ and $\rho^{\parallel}(z-vt)$ are
arbitrary smooth functions subject to
$$
\int_\DD \rho^{\perp}(x_1,x_2)\hashat 1=1,\qquad
\int_{-\infty}^{\infty}dz\, \rho^{\parallel}(z-vt)=1,
$$
and
$$
Q(\epsilon,t):=\frac{Q_{tot}}{1-\epsilon\langle\rho^{\parallel}\kappa_0\rangle_z(t)
\langle x_1\rho^{\perp}\rangle_{\DD}}=
Q_{tot}+\epsilon Q_{tot}\langle\rho^{\parallel}\kappa_0\rangle_z(t)
\langle x_1\rho^{\perp}\rangle_{\DD}+{\cal O}(\epsilon^2),
$$
with
\beqa
\langle\rho^{\parallel}\kappa_0\rangle_z(t)
&:=&\int_{-\infty}^{\infty}dz\, \rho^{\parallel}(z-vt)\kappa_0(z),\non\\
\langle x_1\rho^{\perp}\rangle_{\DD}
&:=&
\int_\DD x_1\rho^{\perp}(x_1,x_2)\hashat 1.
\non
\eeqa
Correspondingly, we define
\beqa
\rho^{(0)}(z-vt,x_1,x_2)&:=&
Q_{tot}\rho^{\perp}(x_1,x_2)\rho^{\parallel}(z-vt),\non\\
\rho^{(1)}(t,z,x_1,x_2)&:=&
Q_{tot}\langle\rho^{\parallel}\kappa_0\rangle_z (t)
\langle x_1\rho^{\perp}\rangle_{\DD}
\rho^{\perp}(x_1,x_2)\rho^{\parallel}(z-vt).\label{EXTRA}
\non
\eeqa
The currents $J_0^{(0)}$ and $J_0^{(1)}$ are defined by
\beqa
J_0^{(0)}(z-vt,r,\theta)&:=&v\rho^{(0)}(z-vt,x_1,x_2),\non\\
J_0^{(1)}(t,z,r,\theta)&:=&v\rho^{(1)}(t,z,x_1,x_2).\non
\eeqa
\subsection{Lowest Order Fields from Moving
Smooth Convected Localized Charge}
The equations for $\gamma_N^{H(0)}$ and $\gamma_N^{E(0)}$
are given by \fr{eqn:Sol-gammaH:0} and \fr{eqn:Sol-gammaE:0}.
The causal solutions are given by
\beqa
\gamma_N^{H(0)}(t,z)&=&{\cal H}_{\alpha_N}[\gamma_N^{H(0)init}] (t,z),
\non\\
\gamma_N^{E(0)}(t,z)&=&{\cal H}_{\beta_N}[\gamma_N^{E(0)init}] (t,z)
-\frac{\mu c^2}{\NN{N}^2}(c^2-v^2){\cal I}_{\beta_N}
[\overline{\rho_N^{(0)\prime}}] (t,z),\non
\eeqa
where  $\rho_N^{(0)}$ is given by
\fr{eqn:def-rho-j-N} and
\beq
{\rho_N^{(0)\,\prime}}(t,z)=Q_{tot}\rho^{\parallel\prime}(z-vt)
\int_\DD\rho^{\perp}(x_1,x_2)
{\Phi_N}\hashat 1.
\label{eqn:rho-bar-N-0-loc}
\eeq
In the ultra-relativistic limit, $v\rightarrow c$,
the second term in
$\gamma_N^{E(0)}(t,z)$ tends to zero.

\subsection{First Order Fields from Moving Convected Localized Charge}

The computation of the longitudinal fields follows along the lines detailed above for the point source.
The  distributional source is simply replaced by the smooth sources (\ref{EXTRA})  at each respective order.
For a general such source one obtains a system of fully coupled modal equations.

The causal solution to $\gamma_M^{H(1)}$ for $\gamma_M^{H(1)}$ can be written as
$$
\gamma_M^{H(1)}(t,z)=
 {\cal I}_{\alpha_M}[\widehat{K}_M^{H(1)}](t,z)
+{\cal I}_{\alpha_M}[\widehat{L}_M^{H(1)}](t,z).
$$
As before the functional form of
$\widehat{K}_M^{H(1)}(t,z)$  given in terms of $\gamma_N^{H(0)}$
and $\gamma_N^{E(0)}$ is the same as for  $K_M^{H(1)}(t,z)$
defined in \fr{eqn:K-M-H:1}
but now,
$\widehat{L}_M^{H(1)}(t,z)$ is calculated to be
$$
\widehat{L}_M^{H(1)}(t,z):=
2\frac{c^2v}{\NN{M}^2}\kappa_0(z){\sum_N}'\frac{G_{M,N}^{\bar{\Psi},\Phi}}{\NN{N}^2\beta_N^2}
\overline{\rho_N^{(0)\prime}}(t,z).
$$

Similarly the causal solution for $\gamma_M^{E(1)}$ is written
\beqa
\gamma_M^{E(1)}(t,z)&=&
 {\cal I}_{\beta_M}[\widehat{K}_M^{E(1)}](t,z)
+{\cal I}_{\beta_M}[\widehat{L}_M^{E(1)}](t,z)
+{\cal I}_{\beta_M}[\widehat{P}_M^{E(1)}](t,z)\non\\
&&
+{\cal I}_{\beta_M}[\widehat{R}_M^{E(1)}](t,z)
+{\cal I}_{\beta_M}[\widehat{S}_M^{E(1)}](t,z).
\label{eqn:gamma-M-E-1-localized}
\eeqa
with  the functional form of
$\widehat{K}_M^{E(1)}(t,z)$ written in terms of $\gamma_N^{H(0)}$ and
$\gamma_N^{E(0)}$
analogous to that of  $K_M^{E(1)}(t,z)$ given
in \fr{eqn:K-M-E-1}
and
\beqa
\widehat{L}_M^{E(1)}(t,z)&:=&
\frac{\mu c^2 (c^2+v^2)}{\NN{M}^2}
\kappa_0(z){\sum_N}'
\frac{1}{\NN{N}^2}\non\\
&&
\times\bigg\{
-E_{M,N}^{\Phi}
+\frac{C_{M,N}^{\Phi}-D_{M,N}^{\Phi}}{\beta_N^2}
\bigg\}\overline{\rho_N^{(0)\prime}}(t,z),\non\\
\widehat{P}_M^{E(1)}(t,z)&:=&
-\frac{\mu c^2 v}{\NN{M}^2}\overline{\dot\rho_M^{(1)}}(t,z),
\non\\
\widehat{R}_M^{E(1)}(t,z)&:=&
-\frac{\mu c^4}{\NN{M}^2}\int_\DD
\bigg\{\rho^{(1)\prime}-\bigg(
\rho^{(0)}\kappa_0'(z)+\rho^{(0)\prime}\kappa_0(z)\bigg)x_1
\bigg\}\overline{\Phi_M}\hashat 1,\non\\
\widehat{S}_M^{E(1)}(t,z)&:=&
\frac{\mu c^4}{\NN{M}^2}\kappa_0'(z)
{\sum_N}'\frac{1}{\NN{N}^2}
\bigg(-E_{M,N}^{\Phi}+\frac{C_{M,N}^{\Phi}-D_{M,N}^{\Phi}}{\beta_N^2}
\bigg)\overline{\rho^{(0)}_N}(t,z).
\non
\eeqa
where
$$
{\dot\rho_M^{(1)}}= Q_{tot}\langle x_1\rho^{\perp}\rangle_\DD
\frac{\partial}{\partial t}\bigg\{
\langle\rho^{\parallel}\kappa_0\rangle_z(t)\rho^{\parallel}(z-vt)
\bigg\}
\int_\DD \rho^{\perp}(x_1,x_2){\Phi_M}
\hashat 1,
$$
and
\beqa
&&
\int_\DD
\bigg\{\rho^{(1)\prime}-\bigg(
\rho^{(0)}\kappa_0'(z)\rho^{(0)\prime}\kappa_0(z)
\bigg)x_1\bigg\}
{\Phi_M}
\hashat 1\non\\
&&=
Q_{tot}\langle\rho^{\parallel}\kappa_0\rangle_z(t)
\langle x_1\rho^{\perp}\rangle_\DD\rho^{\parallel\,\prime}(z-vt)
\int_\DD\rho^{\perp}(x_1,x_2){\Phi_M}\hashat 1
\non\\
&&
\quad -Q_{tot}
\bigg\{\kappa_0'(z)\rho^{\parallel}(z-vt)+
\kappa_0(z)\rho^{\parallel\,\prime}(z-vt)\bigg\}
\int_\DD x_1\rho^{\perp}(x_1,x_2){\Phi_M}\hashat 1.
\non
\eeqa

\subsubsection{Axially Symmetric Smooth Convected Localized
Charge Distribution}
If the transverse
distribution depends only on $r$,
expressions for the electromagnetic $1$-forms simplify.
The source is axially symmetric if
\beq
\rho^{\perp}(x_1,x_2)={\cal R}(r),
\label{eqn:axial-dist}
\eeq
where ${\cal R}(r)$ is a smooth function satisfying
\beq
\int_0^{a}dr\, r{\cal R}(r)=\frac{1}{2\pi}.
\label{eqn:axial-dist-norm}
\eeq
It follows that
\beq
{\rho_N^{(0)\,\prime}}(t,z)=2\pi Q_{tot}\delta_{n,0}
\rho^{\parallel\prime}(z-vt)
\int_0^a dr\, r{\cal R}(r)J_0\left(x_{q(0)}\frac{r}{a}\right).
\label{eqn:symmetric-dist-source}
\eeq
The  expressions for $\gamma_M^{H(1)}$
and $\gamma_M^{E(1)}$   simplify and explicit expressions for the overlap coefficients can be computed in terms of  transverse projections of the radial profile ${\cal R}(r)$. We consider this case in simplifying the situation.
From \fr{eqn:symmetric-dist-source}, it follows immediately that
\beqa
\widehat{L}_M^{H(1)}(t,z)&=&
(\delta_{m,1}+\delta_{m,-1})
\frac{4\pi c^2 v}{\NN{M}^2} Q_{tot}\kappa_0(z)
\rho^{\parallel\,\prime}(z-vt)\non\\
&&
\times
\sum_{q(0)\in\mathbb{N}}
\frac{G_{M,0,q(0)}^{\bar{\Psi},\Phi}}{\NN{0,q(0)}^2\beta_{0,q(0)}^2}
\int_0^a dr\, r{\cal R}(r)J_0\left(x_{q(0)}\frac{r}{a}\right),
\non
\eeqa
and
\beqa
&&\widehat{L}_M^{E(1)}(t,z)=
(\delta_{m,1}+\delta_{m,-1})\frac{2\pi\mu c^2(c^2+v^2) Q_{tot}}{\NN{M}^2}
\kappa_0(z)\rho^{\parallel\,\prime}(z-vt)\non\\
&&
\times\sum_{q(0)\in\mathbb{N}}
\frac{1}{\NN{0,q(0)}^2}
\bigg\{
-E^{\Phi}_{M,0,q(0)}
+\frac{C^{\Phi}_{M,0,q(0)}-D^{\Phi}_{M,0,q(0)}}{\beta_{0,q(0)}^2}
\bigg\}\non\\
&&\qquad
\times\int_0^a dr\, r{\cal R}(r)J_0\left(x_{q(0)}\frac{r}{a}\right),
\non
\eeqa
Then, since
\beq
\langle x_1\rho^{\perp}\rangle_\DD=0,
\label{eqn:symmetric-dist-x1-rho-perp}
\eeq
one has
$$
\overline{\dot{\rho}^{(1)}_M}(t,z)=0.
$$
and therefore,
$$
\widehat{P}_M^{E(1)}(t,z)=0.
$$
Finally, using \fr{eqn:symmetric-dist-x1-rho-perp} and
$$
\int_\DD x_1\rho^{\perp}\overline{\Phi_M}\hashat 1=
\pi(\delta_{m,-1}+\delta_{m,1})\int_0^a dr\, r^2{\cal R}(r)
J_m\left(x_{p(m)}\frac{r}{a}\right),
$$
one has
\beqa
&&\int_\DD
\bigg\{\rho^{(1)\prime}-\bigg(\rho^{(0)}\kappa_0'(z)+
\rho^{(0)\prime}\kappa_0(z)\bigg)x_1\bigg\}
\overline{\Phi_M}
\hashat 1\non\\
&&=-Q_{tot}\pi(\delta_{m,-1}+\delta_{m,1})
\bigg(\kappa_0(z)\rho^{\parallel\,\prime}(z-vt)
+\kappa_0'(z)\rho^{\parallel}(z-vt)\bigg)\non\\
&&\qquad\times\int_0^a dr\, r^2{\cal R}(r)
J_m\left(x_{p(m)}\frac{r}{a}\right).
\non
\eeqa
It follows that
\beqa
&&
\widehat{R}_M^{E(1)}(t,z)=
\frac{\mu c^4}{\NN{M}^2}Q_{tot}\pi(\delta_{m,-1}+\delta_{m,1})
\non\\
&&\times
\bigg(\kappa_0(z)\rho^{\parallel\,\prime}(z-vt)
+\kappa_0'(z)\rho^{\parallel}(z-vt)\bigg)
\int_0^a dr\, r^2{\cal R}(r)
J_m\left(x_{p(m)}\frac{r}{a}\right).
\non
\eeqa
and
\beqa
&&\widehat{S}_M^{E(1)}(t,z)=(\delta_{m,1}+\delta_{m,-1})
\frac{2\pi Q_{tot}}{\NN{M}^2}\mu c^4\kappa_0'(z)\non\\
&&
\qquad\times\sum_{q(0)\in\mathbb{N}}\frac{1}{\NN{0,q(0)}^2}
\bigg(-E_{M,0,q(0)}^{\Phi}
+\frac{C_{M,0,q(0)}^{\Phi}-D_{M,0,q(0)}^{\Phi}}{\beta_{0,q(0)}^2}
\bigg).\non
\eeqa
In summary, for a smooth  axially symmetric charge distribution,
$\gamma_M^{H(1)}$ and $\gamma_M^{E(1)}$ can be written  as
$$
\gamma_M^{H(1)}(t,z)={\cal I}_{\alpha_M}[\widehat{K}_M^{H(1)}](t,z)
+(\delta_{m,-1}+\delta_{m,1}){\cal I}_{\beta_M}[\widehat{L}_M^{H(1)}](t,z),
$$
and
\beqa
\gamma_M^{E(1)}(t,z)&=&{\cal I}_{\beta_M}[\widehat{K}_M^{E(1)}](t,z)\non\\
&&+(\delta_{m,-1}+\delta_{m,1})
{\cal I}_{\beta_M}[\widehat{L}_M^{E(1)}
+\widehat{R}_M^{E(1)}
+\widehat{S}_M^{E(1)}](t,z),\non
\eeqa
using
$$
(\delta_{m,-1}+\delta_{m,1})^2=(\delta_{m,-1}+\delta_{m,1}).
$$

\section{Conclusions}
This paper offers an analytic perturbative approach to the computation
of electromagnetic fields generated by a variety of charged sources
moving with prescribed motions in a perfectly conducting beam pipe of
radius $a$ with planar curvature $\kappa(z)$. Results are given  in
terms of  expressions involving powers of $\vert a\kappa(z)\vert \ll 1$
and $|a^2\kappa^{\prime}(z)|$.
It has included a discussion of ultra-relativistic longitudinal wake potentials from which pipe impedances induced by $\kappa(z)\neq 0$ can be calculated. This has been explicitly illustrated for pipes with piecewise constant curvature modelling pipes with straight segments linked by
  circular arcs of (arbitrary) finite length.

  There are a number of extensions that follow from this work. They include the effects of non-planarity and non-circular cross-sections, both varying with length along the pipe, resistive pipe boundary conditions, generalisations to dielectric channels and the computation of fields satisfying periodic boundary conditions in cyclic machines. Although more challenging the use of the geometric perturbation technique presented here is immediately applicable to these problems and will be discussed elsewhere.

\section*{Appendix A}
\label{sec:app_A}
From the expressions for the orthonormal coframes (\ref{coframes})
in adapted coordinates it is straightforward to derive the
decompositions:
\beqa
&&
\D\hash(\D z\wedge\D\Psi_M)=
-\alpha_M^2(1+\epsilon\kappa_0(z)
r\cos\theta)\Psi_M{\hashat}1\non\\
&&\quad -\epsilon\kappa_0(z)({\hashat}\D\Psi_M)
\wedge(\cos\theta\D r-r\sin\theta\D\theta)\non\\
&&\qquad
-\epsilon \kappa_0'(z)r\cos\theta(\hashat\D\Psi_M)\wedge\D z
+{\cal O}(\epsilon^2),\non
\eeqa
\beqa
&&
\D\hash(\D z\wedge\D\Phi_N)=
-\beta_N^2(1+\epsilon\kappa_0(z) r\cos\theta)\Phi_N{\hashat}1\non\\
&&\quad -\epsilon\kappa_0(z)({\hashat}\D\Phi_N) \wedge(\cos\theta\D
r-r\sin\theta\D\theta)\non\\
&&\qquad
-\epsilon\kappa_0'(z)r\cos\theta(\hashat\D\Phi_N)\wedge\D z
+{\cal O}(\epsilon^2),\non
\eeqa
\beqa
\D\hash\D\Psi_M&=& -(1-\epsilon\kappa_0(z)
r\cos\theta)\alpha_M^2\Psi_M({\hashat}1)\wedge\D z\non\\
&&
\quad -\epsilon\kappa_0(z)(\hashat\D\Psi_M)\wedge (\cos\theta\D
z\wedge\D r -r\sin\theta\D z\wedge\D\theta),\non
\eeqa
\beqa
\D\hash\D\Phi_N&=& -(1-\epsilon\kappa_0(z)
r\cos\theta)\beta_N^2\Phi_N({\hashat}1)\wedge\D z\non\\
&&\quad -\epsilon\kappa_0(z)(\hashat\D\Phi_N)\wedge (\cos\theta\D
z\wedge\D r -r\sin\theta\D z\wedge\D\theta).\non
\eeqa

Furthermore if $\psi(r,\theta)$ represents either $\Phi_N(r,\theta)$
or $\Psi_M(r,\theta)$ one has
$$
\hashat\D\psi=
(\partial_r\psi)r\D\theta -\frac{1}{r}(\partial_\theta\psi)\D r,
$$
$$
\hash\D\psi= (1-\epsilon\kappa_0(z)
r\cos\theta)({\hashat}\D\psi)\wedge\D z,
$$
$$
\hash(\D z\wedge\D\psi)= \frac{{\hashat}\D\psi}{1-\epsilon\kappa_0(z)
r\cos\theta},
$$ and
$$
\hash\D z=\frac{1}{1-\epsilon\kappa_0(z) r\cos\theta}{\hashat}1.
$$

In terms of the symbol
$$
\Xi_{M,N}^k({\cal F,G},x,y) :=\int_0^a dr  r^k
{\cal F}_m\left(x_{p(m)}\frac{r}{a}\right) {\cal G}_n\left(y_{q(n)}\frac{r}{a}\right),
$$
the following overlap coefficients are defined:
\beqa
C_{M,N}^{\Psi}&:=&\int_\DD \cos\theta
\overline{\Psi_M}\frac{\partial\Psi_N}{\partial r}
\hashat 1\non\\
&=&\int_0^{2\pi} d\theta\frac{e^{i\theta}+e^{-i\theta}}{2}
e^{i(n-m)\theta}\int_0^a dr r J_m\left(x'_{p(m)}\frac{r}{a}\right)
\frac{x'_{q(n)}}{a}J'_n\left(x'_{q(n)}\frac{r}{a}\right)\non\\
&=&\pi
(\delta_{n-m+1,0}+\delta_{n-m-1,0})\frac{x'_{q(n)}}{a}\int_0^a dr r
J_m\left(x'_{p(m)}\frac{r}{a}\right)J'_{n}\left(x'_{q(n)}\frac{r}{a}\right)\non\\
&=& \pi (\delta_{n,m-1}+\delta_{n,m+1}) \frac{x'_{q(n)}}{a}
\Xi_{M,N}^1(J,J',x',x').
\non
\eeqa
\beqa
D_{M,N}^{\Psi}&:=&\int_\DD
\frac{\sin\theta}{r}\overline{\Psi_M}(\partial_\theta\Psi_N)
\hashat 1
= \pi n(\delta_{n,m-1}-\delta_{n,m+1})
\Xi_{M,N}^0(J,J,x',x'),\non\\
E_{M,N}^{\Psi}&:=&\int_\DD
r\cos\theta\overline{\Psi_M}\Psi_N{\hashat}1=
\pi(\delta_{n,m-1}+\delta_{n,m+1})
\Xi_{M,N}^2(J,J,x',x'),\non\\
F_{M,N}^{\Psi}&:=&\int_\DD r\cos\theta
\D\overline{\Psi_M}\wedge{\hashat}\D\Psi_N
=\pi(\delta_{n,m-1}+\delta_{n,m+1})\non\\
&&\times \bigg(
\frac{x'_{p(m)}}{a}\frac{x'_{q(n)}}{a}
\Xi_{M,N}^2(J',J',x',x')+mn\Xi_{M,N}^0(J,J,x',x')
\bigg),\non\\
C_{M,N}^{\Phi}&:=&\int_\DD
\cos\theta\overline{\Phi_M}(\partial_r\Phi_N) \hashat 1 =
\pi(\delta_{n,m-1}+\delta_{n,m+1})\frac{x_{q(n)}}{a}
\Xi_{M,N}^1(J,J',x,x),\non\\
D_{M,N}^{\Phi}&:=&\int_\DD
\frac{\sin\theta}{r}\overline{\Phi_M}(\partial_\theta\Phi_N)
\hashat 1 = \pi n (\delta_{n,m-1}-\delta_{n,m+1})
\Xi_{M,N}^0(J,J,x,x),\non\\
E_{M,N}^{\Phi}&:=&\int_\DD
r\cos\theta\overline{\Phi_M}\Phi_N{\hashat}1=
\pi(\delta_{n,m-1}+\delta_{n,m+1})
\Xi_{M,N}^2(J,J,x,x),\non\\
F_{M,N}^{\Phi}&:=&\int_\DD r\cos\theta
\D\overline{\Phi_M}\wedge{\hashat}\D\Phi_N
=\pi(\delta_{n,m-1}+\delta_{n,m+1})\non\\
&&\times \bigg(
\frac{x_{p(m)}}{a}\frac{x_{q(n)}}{a}
\Xi_{M,N}^2(J',J',x,x)+mn\Xi_{M;N}^0(J,J,x,x)
\bigg).\non \eeqa
 \beqa G_{M,N}^{\bar{\Psi},\Phi}&:=&\int_\DD
r\cos\theta
\D\overline{\Psi_M}\wedge\D\Phi_N\non\\
&=&i\pi(\delta_{n,m-1}+\delta_{n,m+1})\bigg( n\frac{x'_{p(m)}}{a}
\Xi_{M,N}^1(J',J,x',x)\non\\
&&+m\frac{x_{q(n)}}{a}\Xi_{M,N}^1(J,J',x',x)
\bigg), \non \eeqa and \beqa
G_{M,N}^{\bar{\Phi},\Psi}&:=&\int_\DD r\cos\theta
\D\overline{\Phi_M}\wedge\D\Psi_N\non\\
&=&i\pi(\delta_{n,m-1}+\delta_{n,m+1})\bigg( n\frac{x_{p(m)}}{a}
\Xi_{M,N}^1(J',J,x,x')\non\\
&& +m\frac{x'_{q(n)}}{a}
\Xi_{M,N}^1(J,J',x,x')\bigg).\non
\eeqa




\section*{Appendix B}
\begin{figure}[h]
\unitlength 1mm
\begin{picture}(160,100)
\put(0,50){\includegraphics[width=12.5cm]{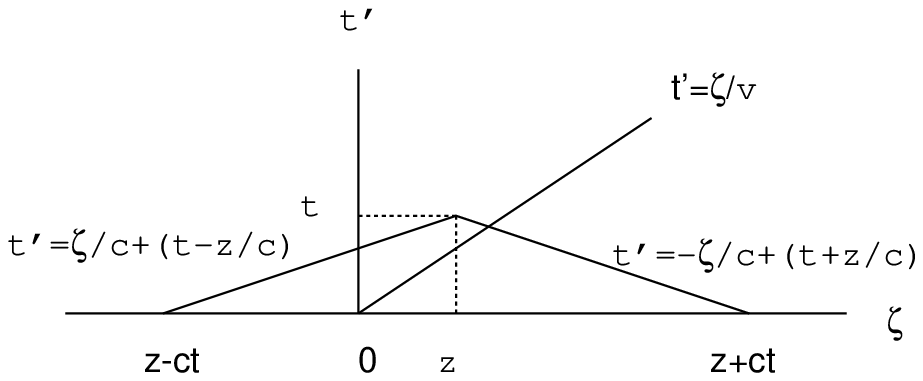}}
\put(0,0){\includegraphics[width=12.0cm]{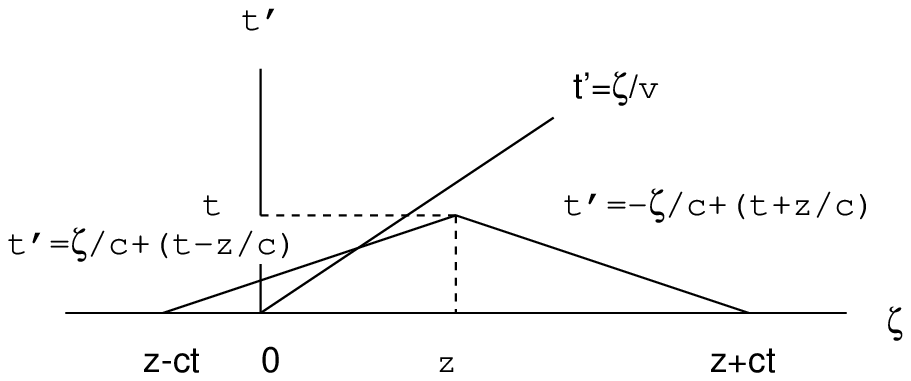}}
\put(10,90){(a)}
\put(10,35){(b)}
\end{picture}
\caption{Integration ranges for $\Upsilon_N(t,z)$. (a): $-ct<z\leq
vt$, (b): $vt\leq z<ct $. } \label{fig:integration}
\end{figure}

In this appendix the  derivation of the expressions for
$\Upsilon_N(t,z)$ in \fr{eqn:upsilon-before},
\fr{eqn:upsilon-before-vt}, \fr{eqn:upsilon-after-vt},
\fr{eqn:upsilon-after},
and ${\cal A}_\sigma(t,z)$, ${\cal B}_\sigma(t,z)$ in
\fr{eqn:AB-before}, \fr{eqn:AB-before-vt},
\fr{eqn:AB-after-vt}, \fr{eqn:AB-after}
is outlined.
 Since
\beq
\int_{\Sigma}dz
f(z)\delta'(z-z_0)=-f'(z_0),
\label{eqn:integration-delta'}
\eeq
\beq
\int_{\Sigma}dz f(z)\delta(z-z_0)=f(z_0),
\label{eqn:integration-delta}
\eeq
for a smooth function $f$ and $z_0$ contained within $\Sigma\subset\mathbb{R}$,
double integrals involving smooth functions and $\delta$
or $\delta'$ can be reduced to a single integral.
For $\Upsilon_N$,  ${\cal A}_\sigma$ and ${\cal B}_\sigma$, the
triangular integration range defined in the $(t',\zeta)$ plane by
$0\leq t'\leq t$ and $z-c(t-t')\leq \zeta\leq z+c(t-t')$ is reduced
to a line segment specified by the support, $t'=\zeta/v$,
of the distribution $\delta(\zeta - vt)$. The slope of this line can
only vary between $\frac{1}{c}$ and infinity. The value of this
slope determines which side of the bounding triangle the line
intersects: see Fig.\ref{fig:integration}(a) and
Fig.\ref{fig:integration}(b). For $v>0$ if $z> vt $ the line segment
intersects the line $t^\prime=(t- \frac{z}{c}) + \frac{\zeta}{c}$ at
$( \zeta=\frac{z-ct}{1-\frac{c}{v}}, t'=\frac{z-ct}{v-c} )$.
If $z<vt $ the line segment intersects the line $t'=(t+ \frac{z}{c}) -
\frac{\zeta}{c}$  at $(\zeta=\frac{z+ct}{1+\frac{c}{v}},
t'=\frac{z+ct}{v+c} )$. When either
$$
z\leq -ct,\qquad \mbox{or}\qquad z\geq ct,
$$
there is no intersection of the line $t'=\zeta/v$ with the triangle
and so  $\Upsilon_N=0$ yielding \fr{eqn:upsilon-before} and
\fr{eqn:upsilon-after}. When
$$
-ct<z<ct,
$$
there is an intersection of the line segment and the triangle, as
shown in Fig.\ref{fig:integration}(a), since $1/v>1/c$.

If
$$
{\cal K} (t,z):=\int_0^t dt'\int_{z-c(t-t')}^{z+c(t-t')}
d\zeta \delta^\prime(\zeta - vt)
f(t,z | t',\zeta),
$$
and
$$
{\cal J} (t,z):=\int_0^t dt'\int_{z-c(t-t')}^{z+c(t-t')} d\zeta
\delta(\zeta- vt) f(t,z | t',\zeta),
$$
it follows that if $-ct< z \leq vt$
$$
{\cal J}= \int_0^{t'_+(t,z)}\,dt' f(t,z | t',vt')
$$
and
\beq
{\cal K}= -\int_0^{t'_+(t,z)}\, dt'\,
\frac{\partial f}{\partial \zeta}\bigg|_{\zeta=vt'}
(t,z | t', \zeta),
\label{dog}
\eeq
where
$$
t'_+(t,z):=\frac{t+z/c}{1+v/c}.
$$
Similarly if $vt \leq z < ct$
$$
{\cal J}= \int_0^{t'_-(t,z)}\,dt'
f(t,z | t',vt')
$$
and
\beq
{\cal K}= -\int_0^{t'_-(t,z)}\,dt'
\frac{\partial f}{\partial \zeta}\bigg|_{\zeta=vt'}
f(t,z | t', \zeta),
\label{bigk}
\eeq
where
$$
t'_-(t,z):=\frac{t-z/c}{1-v/c}.
$$

Using the Bessel relation $J_0'(z)=-J_1(z)$ the integrals
\fr{eqn:upsilon-before-vt} and \fr{eqn:upsilon-after-vt} follow
immediately  from \fr{dog} and \fr{bigk}.
Similarly the integral representations of
${\cal A}_{\sigma}(t,z)$ and ${\cal B}_{\sigma}(t,z)$
follow from those for ${\cal K}$ and ${\cal J}$.

\begin{acknowledgments}
The authors are grateful to C Bane, G Bassi, J Ellison,  
D Jaroszynski, R Li,  M Poole, G Stupakov and  R Warnock  for valuable
discussions and to the Cockcroft Institute and EPSRC for financial
support for this research which is part of the Alpha-X collaboration.
\end{acknowledgments}
\pagebreak[4]


\end{document}